\documentclass[fleqn,usenatbib]{mnras}

\usepackage{newtxtext,newtxmath}

\usepackage[T1]{fontenc}
\usepackage{ae,aecompl}

\usepackage{graphicx}	
\usepackage{amsmath}	
\usepackage{amssymb}	
\usepackage{textcomp}
\usepackage{booktabs}
\usepackage[dvipsnames]{xcolor}
\usepackage{tabularx}
\usepackage{longtable}
\usepackage{multirow}
\usepackage{pdflscape} 
\usepackage{csvsimple}	

\title{S2COSMOS: Evolution of Gas Mass with Redshift Using Dust Emission}
\author[J. S. Millard et al.]{Jenifer S. Millard$^{1}$\thanks{E-mail: jenifer.millard@astro.cf.ac.uk},
Stephen A. Eales$^{1}$,
M.W.L. Smith$^{1}$,
H.L. Gomez$^{1}$,
K. Ma{\l}ek$^{2,3}$,
\newauthor
J.M. Simpson$^{4}$,
Y. Peng$^{5}$,
M. Sawicki$^{6}$,
R. A. Beeston$^{1}$,
Andrew Bunker$^{7}$,
Y. Ao$^{8}$,
\newauthor
A. Babul$^{9}$,
L.C. Ho$^{5,10}$,
Ho Seong Hwang$^{11}$,
M. J. Micha{\l}owski$^{12}$,
N. Scoville$^{13}$,
\newauthor
H. Shim$^{14}$,
Y. Toba$^{15,16,17}$
\\
$^{1}$School of Physics and Astronomy, Cardiff University, The Parade, Cardiff, CF24 3AA, UK\\
$^{2}$Aix Marseille Univ. CNRS, CNES, LAM, Marseille, France\\
$^{3}$National Centre for Nuclear Research, ul. Hoza 69, 00-681 Warszawa, Poland\\
$^{4}$EACOA fellow: Academia Sinica Institute of Astronomy and Astrophysics, No. 1, Sec. 4, Roosevelt Rd., Taipei 10617, Taiwan \\
$^{5}$The Kavli Institute for Astronomy and Astrophysics, Peking University, 5 Yiheyuan Road, Haidian District, Beijing 100871, PR China \\
$^{6}$Institute for Computational Astrophysics and Department of Astronomy and Physics, Saint Mary's University, Halifax, Nova Scotia, \\ B3H 3C3, Canada\\
$^{7}$Sub-dept. of Astrophysics, Department of Physics, University of Oxford, Denys Wilkinson Building, Keble Road, Oxford OX1 3RH, UK \\
$^{8}$Purple Mountain Observatory and Key Laboratory for Radio Astronomy, Chinese Academy of Sciences, 8 Yuanhua Road, Nanjing 210034, \\ PR China \\
$^{9}$Department of Physics and Astronomy, University of Victoria, 3800 Finnerty Road, Victoria V8P 1A1, Canada\\
$^{10}$Department of Astronomy, School of Physics, Peking University, 5 Yiheyuan Road, Haidian District, Beijing 100871, PR China \\
$^{11}$Korea Astronomy and Space Science Institute, 776 Daedeokdae-ro, Yuseong-gu, Daejeon 34055, Republic of Korea\\
$^{12}$Astronomical Observatory Institute, Faculty of Physics, Adam Mickiewicz University, ul. S{\l}oneczna 36, 60-286, Pozna\'{n}, Poland\\
$^{13}$California Institute of Technology, MC 249-17, 1200 East California Boulevard, Pasadena, CA 91125, USA\\
$^{14}$Department of Earth Science Education, Kyungpook National University, 80 Daehak-ro, Buk-gu, Daegu 41566, Republic of Korea\\
$^{15}$Department of Astronomy, Kyoto University, Kitashirakawa-Oiwake-cho, Sakyo-ku, Kyoto 606-8502, Japan\\
$^{16}$Academia Sinica Institute of Astronomy and Astrophysics, No. 1, Sec. 4, Roosevelt Rd., Taipei 10617, Taiwan \\
$^{17}$Research Center for Space and Cosmic Evolution, Ehime University, 2-5 Bunkyo-cho, Matsuyama, Ehime 790-8577, Japan\\
}

\date{Accepted 2020 March 2. Received 2019 October 10; in original form ZZZ}

\pubyear{2020}

\begin{document}
\label{firstpage}
\pagerange{\pageref{firstpage}--\pageref{lastpage}}
\maketitle

\begin{abstract}
We investigate the evolution of the gas mass fraction for galaxies in the COSMOS field using submillimetre emission from dust at 850$\mu$m. We use stacking methodologies on the 850\,$\mu$m S2COSMOS map to derive the gas mass fraction of galaxies out to high redshifts, $0 \le z \le 5$, for galaxies with stellar masses of $10^{9.5} < M_* ~(\rm M_{\odot}) < 10^{11.75}$.  
In comparison to previous literature studies we extend to higher redshifts, include more normal star-forming galaxies (on the main sequence), and also investigate the evolution of the gas mass fraction split by star-forming and passive galaxy populations.  We find our stacking results broadly agree with scaling relations in the literature. We find tentative evidence for a peak in the gas mass fraction of galaxies at around $z \sim 2.5-3$, just before the peak of the star formation history of the Universe. We find that passive galaxies are particularly devoid of gas, compared to the star-forming population. We find that even at high redshifts, high stellar mass galaxies still contain significant amounts of gas.
\end{abstract}




\section{Introduction}
Some of the biggest questions facing astronomy today are inspired by galaxies; specifically, we desire to know the origin and content of these complex structures, and how their content evolves over time. Of particular interest is how galaxies form stars, and this itself motivates investigations into the gas content of galaxies, since stars form from the gravitational collapse of gas clouds.  

For star-forming galaxies, there exists a relationship between the stellar mass of a galaxy and its star formation rate (SFR). Star-forming galaxies are said to lie on the Main Sequence (MS) (e.g. \citealt{Daddi2007}; \citealt{Elbaz2007}; \citealt{Noeske2007}; \citealt{Santini2009}; \citealt{Peng2010}; \citealt{Karim2011}; \citealt{Rodighiero2011}; \citealt{Whitaker2012}; \citealt{Sawicki2012}; \citealt{Steinhardt2014}; \citealt{Lee2015}; \citealt{Schreiber2015}; \citealt{Tomczak2016}); and studies have shown that this relationship evolves over time (e.g. \citealt{Daddi2007}; \citealt{Whitaker2012}; \citealt{Lee2015}; \citealt{Tomczak2016}). Observations of both nearby galaxies and galaxies at high redshift, $z$, have shown that as we progress through different epochs, the MS evolves to lower SFRs; galaxies in the nearby universe have SFRs some 20 times lower than those at $z \sim$ 2 for the same stellar mass, consistent with the peak of star formation activity in the universe at $z \sim$ 2 (e.g. \citealt{Daddi2007}; \citealt{Karim2011}; \citealt{Whitaker2012}; \citealt{Madau2014}; \citealt{Lee2015}). 

To understand these observations, we must focus our studies on the fuel for star formation activity, the interstellar medium (ISM). Observations and theory indicate that the evolution of the MS is rooted in the higher molecular gas content of galaxies at higher redshifts (\citealt{Dunne2011}; \citealt{Genzel2015}; \citealt{Scoville2016}; \citealt{Scoville2017}; \citealt{Imara2018}; \citealt{Tacconi2018}). As one looks back in time, one would expect galaxies to contain more of the fuel for star formation and less of the product of star formation. 

Even though knowing the gas content of galaxies is key to understanding their evolution, estimating the gas content of high-$z$ galaxies is non-trivial. The atomic phase of the ISM can be reliably estimated using the 21 cm line, but telescopes currently in operation are not sensitive enough to detect this at high-$z$ \citep{Catinella2010}. Gas in the molecular phase can be estimated using CO line transitions as a tracer (e.g. \citealt{Solomon2005}; \citealt{Coppin2009}; \citealt{Tacconi2010}; \citealt{Casey2011}; \citealt{Bothwell2013}; \citealt{Carilli2013}; \citealt{Tacconi2013}; \citealt{Combes2018}) but this method is costly in terms of telescope time. An added complication is that the conversion factor used to transform CO detections into mass estimates of molecular hydrogen is notoriously uncertain, particularly at low metallicities, and there are also indications that it can vary across the galactic disc (e.g. \citealt{Bolatto2013}; \citealt{Bethermin2015}; \citealt{Genzel2015}; \citealt{Scoville2016}). Further, the high-$z$ galaxies observed using CO transitions are often non-typical galaxies including massive star-forming sub-millimeter galaxies, galaxies that host AGN, and lensed sources (e.g. \citealt{Greve2005}; \citealt{Bothwell2013}; \citealt{Carilli2013}; \citealt{Riechers2013}; \citealt{Canameras2018}; \citealt{Harrington2018}; \citealt{Rivera2018}).

Over the past few years, an alternative approach to measure the molecular gas content of galaxies has been gaining momentum: using the optically thin dust continuum emission detected at a single sub-millimetere (sub-mm) wavelength as a tracer of the gas (\citealt{Eales2012}; \citealt{Scoville2014}; \citealt{Scoville2016}; \citealt{Scoville2017}). The far-infrared (FIR) or sub-mm emission from galaxies generally originates from two major sources: firstly, the re-emission of UV and optical starlight that has been absorbed by dust grains, and secondly, from active galactic nuclei (AGN) \citep{Scoville2016}. Dust is the primary source of FIR/sub-mm emission in most galaxies; it emits radiation as a modified blackbody and at long wavelengths, on the Rayleigh-Jeans (RJ) tail, this dust emission is usually optically thin.
With knowledge of the emissivity of dust per unit mass and the gas-to-dust ratio, we can use the emission from dust to trace the mass of gas in galaxies. 
\cite{Scoville2016} and \cite{Scoville2017} used samples of galaxies that have both 850$\mu$m measurements from the Atacama Large Millimeter/sub-millimeter Array (ALMA) and molecular gas masses estimated using CO observations, to derive an empirical relation to estimate gas masses using optically thin emission from dust. The advantage of this method lies in the opportunity it provides to quickly derive molecular gas masses for large numbers of galaxies at high redshifts, opening up the possibility of probing the gas mass fraction in galaxies at particularly interesting epochs, and over a large fraction of the history of the universe.  This method is particularly timely; the {\it Herschel Space Observatory} (hereafter {\it Herschel}, \citealt{Pilbratt2010}) has provided sub-mm measurements for hundreds of thousands of galaxies. ALMA can also be used to provide sub-mm continuum measurements to estimate the mass of the ISM in high-z galaxies (e.g. \citealt{Scoville2016}).

Recent studies comparing gas masses estimated using sub-mm dust continuum emission to gas masses estimated using more traditional methods, such as CO line emission, have added support to the validity of this method. \cite{Genzel2015} simulated the sub-mm emission produced by a population of galaxies, based on stacked {\it Herschel} FIR data from \cite{Magnelli2014}, and scaling relations developed using this dataset. They compared the known molecular gas masses to those determined using the simulated sub-mm emission and found that gas masses were successfully estimated within 0.35 dex of the true value, with most of the scatter caused by uncertainties in the dust temperature. \cite{Tacconi2018} built upon the work of \cite{Genzel2015}, incorporating new CO data into their studies (e.g. \citealt{DeCarli2016}; \citealt{Saintonge2017}), additional stacked data from {\it Herschel} (e.g. \citealt{Santini2014}; \citealt{Bethermin2015}), and further sub-mm emission observations (e.g. \citealt{Scoville2016}). They found that no matter the method used to determine the gas masses, the results all converged to the same scaling relations.

Although using dust as a tracer of the gas masses may be efficient in terms of integration time required to detect dust emission (e.g. \citealt{Genzel2015}; \citealt{Scoville2016}; \citealt{Scoville2017}; \citealt{Tacconi2018}), unfortunately sub-mm telescopes suffer from poor resolution. Indeed, the resolution of 850$\mu$m images produced by the Submillimetre Common-User Bolometer Array 2 (SCUBA-2) on the James Clerk Maxwell Telescope (JCMT) (\citealt{Holland2013}) is typically around $13^{\prime \prime}$ (\citealt{Dempsey2013}). This makes it difficult to measure accurate sub-mm measurements for individual galaxies; multiple optical sources will often lie within one beamsize. However, stacking methodologies can be employed to combat the poor resolution; we lose information on individual galaxies, but gain information on the galactic population as a whole.

A recent study examining the evolution of dust emission in COSMOS galaxies using stacking methodologies was performed by \cite{Bethermin2015}. Like other \textit{Herschel}-based SED stacking analyses (e.g. \citealt{Genzel2015};  \citealt{Tacconi2018}), average dust masses of binned sources were inferred by fitting spectral energy distributions (SEDs) to average stacked long-wavelength fluxes - specifically in that study, fluxes ranging from the mid-infrared to millimetre wavelengths. The fitted SEDs were constructed using dust emissivity models (specifically, the models from \citealt{DraineLi2007}), wherein dust temperatures, and therefore dust masses, are luminosity-weighted. Molecular gas masses were then estimated using the derived dust masses, and an assumed gas-to-dust mass ratio (specifically, a metallicity dependent gas-to-dust ratio from \citealt{Leroy2011}).

\cite{Bethermin2015} examined galaxies with stellar masses (>3 $\times 10^{10} M_{\odot}$) and redshifts $z \leq$ 4.
In this paper, we probe lower stellar masses, redshifts $>4$\footnote{Note that the redshift data used in this study (\citealt{Davies2015}; \citealt{Andrews2017}; \citealt{Driver2018}; see also Sections \ref{Sec:DriverMAGPHYS} and \ref{Sec:accuracyphotz}) is a combination of spectroscopic and photometric redshifts, with more sources having photometric redshifts as we progress further back through cosmic time. A discussion on the uncertainties of assigned redshifts for sources used in this study can be found in Section \ref{Sec:accuracyphotz}. We caution the reader that although we do probe to higher redshifts, these redshifts are photometric, and therefore are associated with a higher uncertainty compared to redshifts derived from spectroscopic observations.}, and examine the validity of deriving gas masses using a single sub-mm measurement at 850$\mu$m for large samples of galaxies. There is also a major difference in method between our analysis and that of \cite{Bethermin2015} because we use mass-weighted rather than luminosity-weighted temperatures. We divide galaxies in the COSMOS field into bins of stellar mass and redshift, and use stacking to calculate the average dust continuum emission for galaxies in different bins. We make use of relations developed by \cite{Scoville2016} and \cite{Scoville2017} to estimate average gas masses and gas fractions for the binned galactic population. We examine the evolution of the gas mass fraction of galaxies over cosmic time in different stellar mass bins, beyond the peak of star formation activity and compare these results to scaling relations in the literature. We also split our sample into passive and star-forming galaxies, as a diagnostic tool for understanding the results of our stacking analysis on the collective population.

We use the cosmological parameters from {\it Planck} (\citealt{Planck2015}) and make use of {\tt astropy.cosmology} (\citealt{Astropy2013}, \citealt{Astropy2018}) assuming {\tt FlatLambdaCDM} and $H_0 = 67.7 {\rm km \ Mpc^{-1} s^{-1}}$,  $\Omega_{M_{0}} = 0.307$ and $\Omega_{B_{0}} = 0.0486$.

\section{The COSMOS Field: Data and Source Catalogues}

\subsection{Submillimetre Images of the COSMOS Field}
SCUBA-2 is a 10,000 pixel bolometer camera installed at the JCMT, which operates in the sub-millimetre (sub-mm) wavelength regime \citep{Holland2013}. Specifically, it surveys the sky at 450$\mu$m and 850$\mu$m. SCUBA-2 has been used to target the COSMOS (Cosmic Evolution Survey) field \citep{Scoville2007}, a 2 deg$^2$ area of sky centred at RA = 10:00:28.60 and Dec = +02:12:21.00 (J2000). We make use of the 850$\mu$m map from the SCUBA-2 COSMOS survey (S2COSMOS, \citealp{Simpson2019}), the deepest and most sensitive sub-mm image of the COSMOS field to-date. This map incorporates archival data from the SCUBA-2 Cosmology Legacy Survey (S2CLS, \citealt{Geach2017}; \citealt{Michalowski2017}) and new data from S2COSMOS, providing a complete and homogeneous map of the COSMOS field at 850$\mu$m.

The median instrumental noise level over the central 1.6deg$^2$ region of the S2COSMOS survey is $\sigma_{850\mu {\rm m}}$ = 1.2mJy beam$^{-1}$. There is also further coverage of 1deg$^2$, which has median instrumental noise levels of $\sigma_{850\mu {\rm m}}$ = 1.7mJy beam$^{-1}$. Confusion noise is less than the instrumental noise, and is estimated to be $\sigma_c \sim 0.4$mJy beam$^{-1}$ \citep{Simpson2019}. We note that, unless otherwise stated, we make use of the matched-filtered map, which is more sensitive to point source emission. For further details, we refer the reader to \cite{Simpson2019}. 

We will later stack on positions of galaxies in the 850$\mu$m S2COSMOS map to determine the average sub-mm properties of the population. We therefore require a source catalogue to provide the locations of COSMOS galaxies; our choice of catalogue is discussed in Section \ref{Sec:input_cats}.

\subsection{COSMOS galaxies: input catalogues} \label{Sec:input_cats}
In this study, we wish to make use of stacking techniques to determine statistical information about the gas properties of galaxies as they evolve through cosmic time. Stacking will allow us to push to lower stellar masses where one cannot detect an individual object, and potentially provide a more unbiased estimate of the population. For example, it gives us the opportunity to probe gas masses across galaxies with a wider range of star forming properties than could necessarily be achieved by studying individual objects.

To study the evolution of gas mass, we divide our sources into bins of stellar mass ($M_*$) and redshift. A common way of estimating the stellar mass of galaxies is to use Spectral Energy Distribution (SED) fitting routines. Since the COSMOS field is one of the most well studied areas of sky, with data covering almost the entire electromagnetic spectrum, photometric catalogues have been created for COSMOS sources, and subsequently exploited using many SED fitting routines. 

In Appendix \ref{App:Acrosscheck}, we compare the source catalogue from \cite{Driver2018} made using the SED fitting code {\sc magphys} \citep{daCunha2008} to one created using the SED fitting routine CIGALE (\citealt{Noll2009}, \citealt{Boquien2019}) as part of the {\it Herschel} Extragalactic Legacy Project (HELP) database (\citealt{Vaccari2016}; \citealt{Malek2018}; \citealt{Malek2019short}; \citealt{Shirley2019}; Oliver et al. {\it in prep.}). This analysis shows that our choice of source catalogue is not likely to significantly impact the results of this study, and choosing the CIGALE catalogue over the Driver/{\sc magphys} catalogue would not change our conclusions. We ultimately choose to use the {\sc magphys} dataset for the rest of this study, since it is limited to the central regions of the map (Figure \ref{fig:SCUBA2_map}) where the noise levels are lowest \citep{Simpson2019}.

In this study, we compile our final source list using the Driver/{\sc magphys} catalogue and the COSMOS2015 source catalogue of \cite{Laigle2016} - see Sections \ref{Sec:DriverMAGPHYS} and \ref{Sec:COSMOS2015} for a description of these, respectively. Section \ref{Sec:finalcat} describes our final source catalogue used in the stacking analysis.

\subsubsection{Driver/MAGPHYS catalogue of region} \label{Sec:DriverMAGPHYS}
{\sc magphys} \citep{daCunha2008} is an SED fitting code that fits pre-determined libraries of physically motivated model SEDs to panchromatic photometry data, returning probabilistic estimations of various physical parameters of the sources in question. This includes, for example, stellar masses and star formation histories. The stellar emission is based on synthetic spectra from \cite{BruzualCharlot2003}, wherein a \cite{Chabrier2003} stellar initial mass function (IMF) is assumed. Dust attenuation follows the model present by \cite{CharlotFall2000}, where starlight can be attenuated by dust in both spherically symmetric birth clouds and in the ambient ISM.  {\sc magphys} employs an energy balance between the Ultraviolet-to-Near-Infrared (UV-NIR) and the Mid-Infrared-to-Far-Infrared (MIR-FIR) components; the UV-optical light attenuated by dust is re-emitted in the FIR. The dust emission responsible for the MIR to FIR originates from three sources: polycyclic aromatic hydrocarbons, small dust grains and large grains. Grains have temperatures of 30-60\,K in birth clouds, and another cooler component with temperatures of 15-25\,K is also modelled in the diffuse clouds with dust emissivity index $\beta$ of 1.5 and 2.0 in the warm and cold components. The best fit model is determined using a $\chi^2$ minimization technique, and, subsequently, best-fit values for each parameter are returned. In addition to this, probability distribution functions (PDFs) are generated for each parameter, from the summation of $e^{-\chi^2 / 2}$ over all models. The PDFs detail the most likely value for a given parameter. Here, we make use of the 50th-percentile values for any given parameter, since these are more representative of the range of models that fit the photometry than the single best-fit model.

We consider the {\sc magphys} source catalogue, presented in \cite{Driver2018}, which makes use of photometry provided by \cite{Andrews2017} using 22 filters (FUV, NUV, {\it ugriz}YJHK, IRAC1234, MIPS24/70, PACS100/160, SPIRE250/350/500); note that not all sources have fluxes for all of the filters, see \cite{Driver2018} for more details. This photometric catalogue is based on G10-COSMOS, a small region (1 deg$^2$) within the COSMOS field.  \cite{Driver2018} use a {\it i}-band < 25 mag limited catalogue, based on results from a Source Extractor \citep{Bertin1996} analysis of {\it i}-band Subaru observations. The final photometric catalogue was created using the {\sc lambdar} routine (\citealt{Wright2016}; \citealt{Andrews2017}). {\sc lambdar} produces aperture-matched photometry, and, for coarser-resolution long-wavelength maps (such as the {\it Herschel} maps), deblends the flux from multiple sources by sharing the flux between overlapping apertures. The final {\sc magphys} source catalogue used by \cite{Driver2018} is that presented in \citet{Andrews2017}, with some additional minor adjustments to the  selection process for assigning FIR fluxes to sources using {\sc lambdar}. This process resulted in an extension to the number of sources with associated FIR data (see Appendix A of \citealt{Driver2018}) compared to those presented in \cite{Andrews2017}. We make use of the photometric catalogue {\tt G10CosmosLAMBDARCatv06}, containing 185,907 sources. 

In their {\sc magphys} fitting work, \cite{Driver2018} make use of the BC03 stellar libraries and a modified version of {\sc magphys} that includes the derivation of model FIR fluxes based on photon energy, as opposed to photon number, and the use of the latest PACS and SPIRE filter curves.  We make use of their {\sc magphys} catalogue {\tt MagPhysG10v05}, which contains 173,399 sources. When cross-matched to {\tt G10CosmosLAMBDARCatv06}, we find a match for every {\sc magphys}  source. 

Redshifts in the {\sc magphys} catalogue are sourced from an updated catalogue from \cite{Davies2015}. Where possible, spectroscopic redshifts are used; these are obtained via an independent extraction of spectroscopic redshifts from the zCOSMOS-Bright sample\footnote{See \cite{Davies2015} and \cite{Andrews2017} for full details of their bespoke pipeline.}, combined with additional redshifts from PRIMUS, VVDS and SDSS (\citealt{Lilly2007zbright}; \citealt{Cool2013}; \citealt{LeFevre2013}; \citealt{Ahn2014}). Where there are multiple spectroscopic redshifts for a given source, the most robust one is chosen. If spectroscopic redshifts are unavailable, photometric redshifts are sourced from COSMOS2015 \citep{Laigle2016}. The {\sc magphys} sample contains 21,494 sources with reliable spectroscopic redshifts; the remaining 151,905 have photometric redshifts.  The source catalogue for {\sc magphys} is considered redshift complete down to the magnitude limit ($i$ < 25 mag).

\subsubsection{COSMOS2015} \label{Sec:COSMOS2015}
The COSMOS2015 catalogue (\citealt{Laigle2016}) contains half a million NIR (Near Infrared) selected objects over the COSMOS field, with photometry covering wavelengths from the X-ray range through to the radio. The source detection image is a combination of YJHK$_s$ from UltraVISTA DR2 \citep{McCracken2012} and WIRCAM \citep{McCracken2010}, and z$^{++}$ images from Subaru Suprime-Cam (\citealt{Taniguchi2007}; \citealt{Taniguchi2015}). COSMOS2015 is 90\% complete to a stellar mass of 10$^{10} M_{\odot}$ out to $z$ = 4. Photometric redshifts errors are small; $\sigma_{\Delta z / (1+z)}$ = 0.007 for $z < 3$, and $\sigma_{\Delta z / (1+z)}$ = 0.021 for 3 < $z$ < 6.
The photometric redshift errors were estimated using the normalized median absolute deviation (\citealt{Hoaglin1983}; \citealt{Laigle2016}), which calculates the dispersion of the photometric redshifts, as compared to the spectroscopic redshifts. Spectroscopic redshifts were compiled from multiple surveys (see \citealt{Laigle2016} for details), and only the highly reliable 97\% confidence level spectroscopic redshifts were used \citep{Lilly2007zbright}.

\begin{figure}
	\includegraphics[width=\columnwidth]{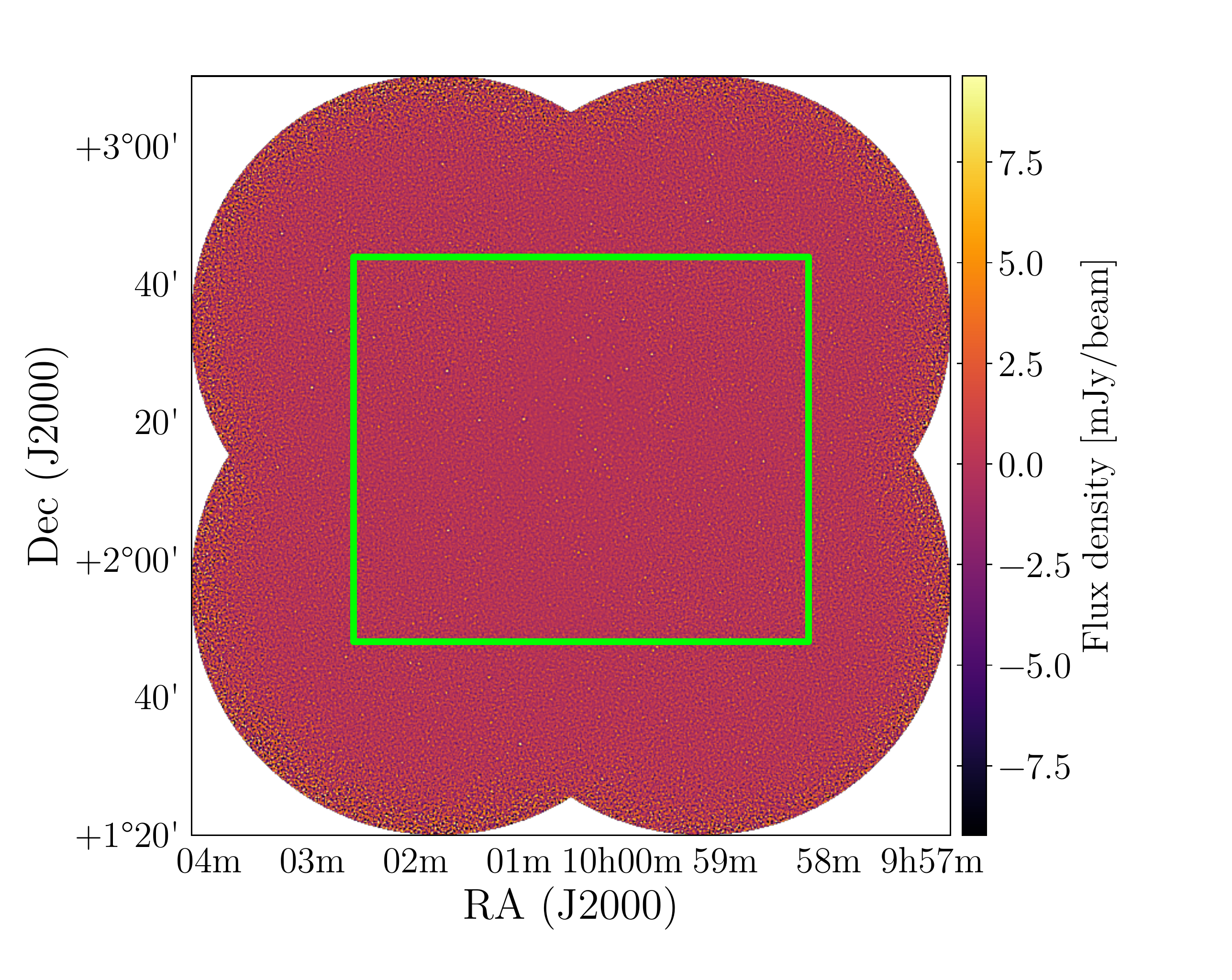}
    \caption{The extent of {\sc magphys} sources in the COSMOS field (lime square), covering approximately 1 deg$^2$, across the 850$\mu$m match-filtered SCUBA-2 map.}
    \label{fig:SCUBA2_map}
\end{figure}

\subsection{COSMOS galaxies: final catalogue used in this work} \label{Sec:finalcat}

We cross-match the {\tt MagPhysG10v05} catalogue to the publicly available COSMOS2015 catalogue \citep{Laigle2016} on RA and Dec. We do this to ensure catalogue completeness, and to exploit the additional photometry provided in the COSMOS2015 survey (Section~\ref{sec:remove AGN}). This reduces our {\sc magphys} sample to 155,858 sources. The extent of the {\sc magphys} sources across the SCUBA-2 850$\mu$m map is shown in Figure \ref{fig:SCUBA2_map}. The sources discarded in the matching process are evenly distributed across the COSMOS field, with the exception of the locations of bright stars in the field, more sources are removed around these than the field average (Figure \ref{fig:radec_notin_COSMOS2015}). We attribute this effect to more aggressive bright star masking in the COSMOS2015 catalogue compared to the {\sc magphys} catalogue. The stellar mass and redshift distribution of removed sources are shown in Figure \ref{fig:mstarz_notin_COSMOS2015}. Many of the sources are relatively nearby, with $z$ < 1, where our number counts are highest (see Figure \ref{fig:z_mstar_bins}). 

The final {\sc magphys}-COSMOS2015 sample contains 21,212 sources with reliable spectroscopic redshifts; the remaining 134,646 have photometric redshifts.  
We next filter out the {\sc magphys}-COSMOS2015 matched galaxies with log$(M_*/M_{\odot}) < 9.5$ (see Section \ref{Sec:stackinggas}), leaving 64,684 sources, of which 13,955 have reliable spectroscopic redshifts.  

\subsubsection{Removing Active Galactic Nuclei}
\label{sec:remove AGN}
AGN (Active Galactic Nuclei) emit radiation across the electromagnetic spectrum. By definition, this means that the emission from AGN contaminates the galactic emission used to estimate physical properties, such as stellar mass and star-formation rates, particularly in the infrared \citep{Ciesla2015}. Disentangling the galactic infrared emission from the AGN infrared emission is difficult. But, if the emission from AGN is not properly accounted for, galactic stellar mass estimates and star-formation rates can suffer systematic uncertainties of up to 50\% \citep{Ciesla2015}. Accurately estimating physical properties of galaxies hosting AGN using SED-fitting methods is non-trivial. As such, it is important that we remove AGN from our sample.

We first remove potential AGN from our {\sc magphys} sample by following the prescription described in \citet{Driver2018} with an additional step. AGN are identified using a combination of IR, radio and X-ray data. Unless otherwise stated, we use photometry from {\tt G10CosmosLAMBDARCatv06}. As in \citet{Driver2018}, we remove sources with stellar masses greater than $10^{12}M_{\odot}$. We remove AGN based on their NIR (3.6-8\,$\mu$m) colours using the criteria from \citet{Donley2012} (Equations 1 and 2 therein). We remove sources that are radio-loud using the criteria from \citet{Seymour2008}, where log$\rm _{10}(S_{1.4GHz}/S_{Ks}) > 1.5$ and log$\rm _{10}(S_{24\mu m} / S_{1.4GHz}) < 0.0$. The 1.4GHz radio fluxes are obtained from COSMOS2015 (using the {\tt FLUXPEAK} photometry). We also reject any source that has a non-zero flux in any of the {\it XMM-Newton} bands in COSMOS2015 \citep{Laigle2016}.

We additionally make use of the {\it Chandra}-COSMOS Legacy Survey (CCLS)\footnote{\url{https://irsa.ipac.caltech.edu/data/COSMOS/tables/chandra/}} \citep{Civano2016,Marchesi2016}, a catalogue of 4016 X-ray point sources across $\sim$2.2 deg$^2$ of the COSMOS field. We identify AGN in this catalogue by selecting sources that have a reliable optical counterpart, a spectroscopic or photometric redshift, and are not flagged as stars \citep{Suh2019}. Using these criteria, we classify 3713 sources in the CCLS as AGN. Subsequently, we cross-match this reduced CCLS catalogue to the {\sc magphys}-COSMOS2015 catalogue, identifying 715 sources that have a {\it Chandra} counterpart. We additionally reject any sources that have a flux recorded for any of the three {\it Chandra} bands.

In combining the prescription from \citet{Driver2018} and the AGN flagged in the CCLS, we remove 1026 sources that are potential AGNs, this provides a final sample of 63,658 galaxies. For stacking purposes, we do not filter the catalogue based on the goodness-of-fit threshold $\chi^2_{\rm thr,MAG}$ level (see Appendix \ref{App:Achi2} for details), this ensures that we sample the complete catalogue of sources for our stellar mass cut-off. 

\subsubsection{Precision of photometric redshifts in the final sample} \label{Sec:accuracyphotz}
We now consider the precision of photometric redshifts for sources in our final sample. We identify sources with reliable spectroscopic redshifts and use their corresponding COSMOS2015 photometric redshifts (removing sources that have catastrophic failures for photometric redshift), giving us a sub-sample of 12,332 sources to consider, out to a maximum spectroscopic redshift of 3.47, with a median spectroscopic redshift of 0.68. We estimate the precision of the photometric redshifts by following the prescription in \cite{Laigle2016}. We calculate the normalized median absolute deviation \citep{Hoaglin1983}, $\sigma$:
\begin{equation}
    \sigma = 1.48 \times {\rm median} \left( \vert z_p - z_s \vert / \left( 1 + z_s \right) \right)
\end{equation}
where $z_s$ are the spectroscopic redshifts, and $z_p$ are the photometric redshifts. We find $\sigma$ = 0.0104, a value roughly 50 per cent higher than the corresponding value calculated in \cite{Laigle2016}, for a similar redshift range. Overall, the dispersion of the photometric redshifts is low, giving us confidence that the photometric redshifts are accurate. Therefore, although many of our sources do not have reliable spectroscopic redshifts, we do not anticipate that this will impact significantly on our results.

\section{Deriving Submillimetre Fluxes and Gas Masses: Our Stacking Analysis}
\subsection{Submillimetre Fluxes from Stacking} \label{Sec:stackingflux}
Individually, most of the galaxies in our sample have 850$\mu$m fluxes below the noise level of the S2COSMOS map \citep{Simpson2019}. To circumvent this issue, we make use of well-established stacking methodologies, wherein we co-add the emission from many similar sources to determine an average flux, which is representative of the sub-population in question. We lose information about individual sources, but gain information on the sample of galaxies as a whole. 

\begin{figure}
	\includegraphics[width=\columnwidth]{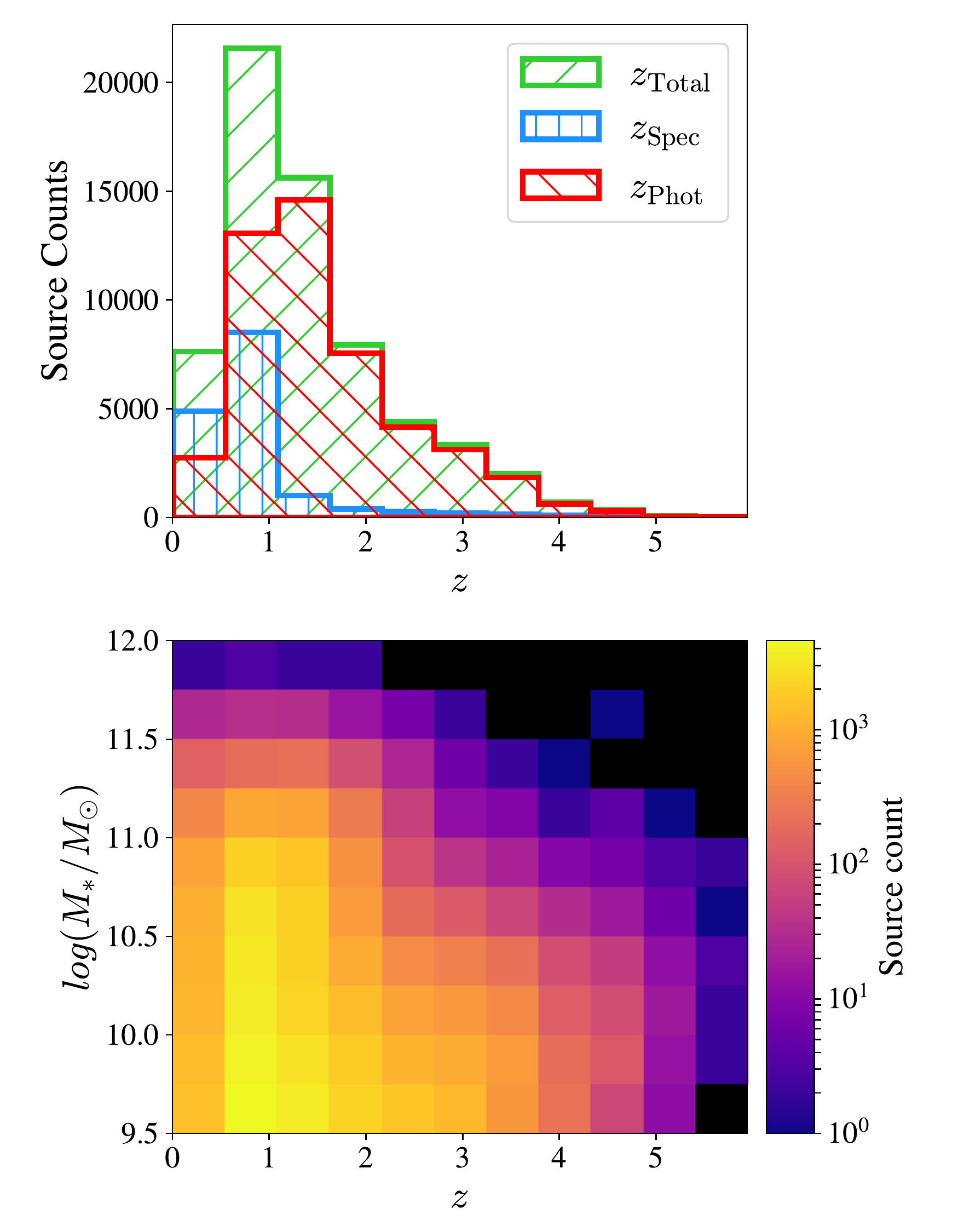}
    \caption{{\it Upper:} A histogram of the redshifts for sources in the COSMOS field with log$(M_*/M_{\odot}) > 9.5$.  These sources are used in the stacking analysis to determine ISM masses. The $z$ bin resolution is the same as that displayed in the lower figure. Green: all the sources. Blue: sources with spectroscopic redshifts in the {\sc magphys} catalogue. Red: sources with photometric redshifts in the {\sc magphys} catalogue. {\it Lower:} A 2D histogram illustrating the distribution of {\sc magphys} sources in the COSMOS field with log$(M_*/M_{\odot}) > 9.5$, in $(M_* - z)$ space.  Black denotes that there are no sources in the bin.}
    \label{fig:z_mstar_bins}
\end{figure}

We stack using the COSMOS2015 (\citealt{Laigle2016}) positions for the sources in our catalogue of 63,658 galaxies. We split the sources into regularly spaced ${\rm log} (M_*)$ and $z$ bins.  The distribution of sources in this parameter space is displayed in Figure \ref{fig:z_mstar_bins} (our lower limit on stellar mass is log$(M_*/M_{\odot}) \geq 9.5$). 

For each ($M_*-z$) bin, small cutouts of the 850$\mu$m S2COSMOS data and error maps \citep{Simpson2019} are made, centred on each individual source. The average `stamp', $ \langle S_{s,850\mu {\rm m}} \rangle$, for sources in a bin is calculated using the inverse variance-weighted (IVW) mean of each of the individual cutouts:
\begin{equation}
    \langle S_{s,850\mu {\rm m}} \rangle = \frac{\Sigma_i S_i/\sigma_i^2}{\Sigma_i 1/\sigma_i^2}
	\label{eq:fluxav}
\end{equation}
where $S_i$ is the data map cutout for the $i$th source, and $\sigma_i$ is the error map cutout for the $i$th source. The average flux, $\langle S_{850\mu {\rm m}} \rangle$, for a given bin is then taken to be the central pixel value of the final coadded stamp (where the pixel scale of S2COSMOS is 2$^{\prime \prime}$), since the maps are in units of mJy/beam. Figure \ref{fig:goodstack} (top panel) shows an example of a stacked cutout for a ($M_*-z$) bin with a clear, strong detection of 850$\mu$m flux. In this stack, the emission is uniformly and centrally concentrated, as expected, since the galaxies are all assumed to be point sources in the SCUBA-2 COSMOS map. Conversely, Figure \ref{fig:goodstack} (bottom panel) is an example where we do not observe a strong detection of 850$\mu$m flux. In this stack, there is no central peak of 850$\mu$m emission. This implies that the galaxies used in the stack do not have significant 850$\mu$m emission. See Table \ref{tab:bin_numbers} for details of the specified ($M_*-z$) bin shown in Figure \ref{fig:goodstack}.

\subsubsection{Error Estimation using Monte Carlo Simulations} \label{sec:MCsim}
To estimate the errors on the average fluxes for each bin, we make use of Monte Carlo (MC) simulations to generate random source positions within the $850\mu$m S2COSMOS map. This gives us an estimation of the significance of the stacked signal determined from our source catalogue. Since the average value of the S2COSMOS map is zero, randomly selected positions should exhibit a Gaussian flux distribution, centred on zero. In other words, on average, flux values from random positions within the data map should not be significant. The width of the average distribution of randomly selected flux values is an estimate of the error on our average fluxes.

\begin{figure*}
	\includegraphics[width=\textwidth]{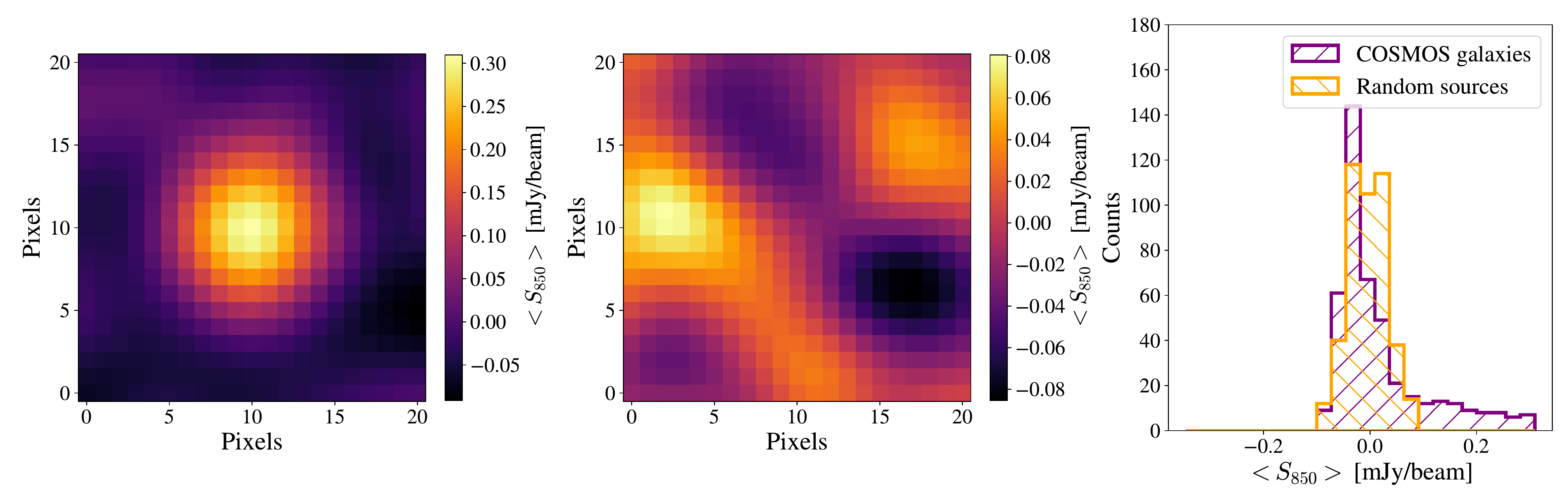}
		\includegraphics[width=\textwidth]{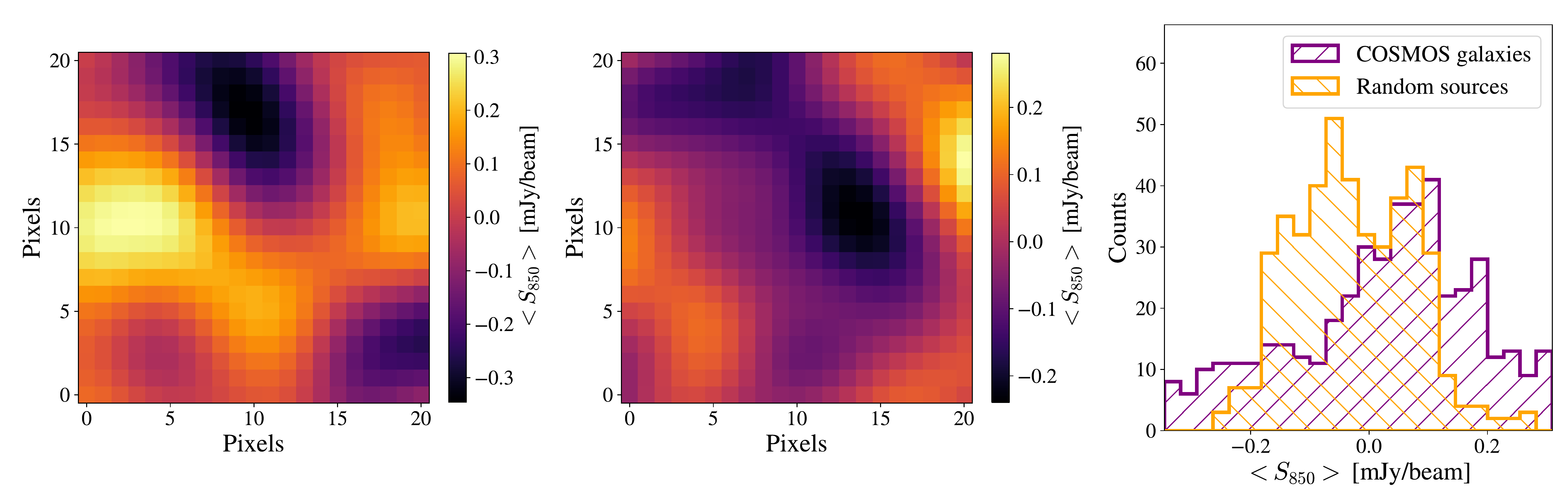}
    \caption{Examples of the 850$\mu$m IVW stacked stamps for $<z_{\rm bin}>$ = 0.82 and log($M_* / \rm M_{\odot})$ bin = 10.75-11.00 ({\it top}) and $<z_{\rm bin}>$ = 0.28 and log($M_* / \rm M_{\odot})$ bin = 11.25-11.50 ({\it bottom}).  In the {\it left} to {\it right} panels, we show: {\it left:} the 850$\mu$m IVW stacked stamp resulting from stacking at the location of COSMOS sources ($N=2033$ sources in top panel, $N=145$ sources in the bottom panel). In the top left panel, we see a clear detection of 850$\mu$m flux present in the centre of the stack. In the bottom left panel, we do not detect the source in the 850$\mu$m stack. {\it Middle:} An example of the resulting IVW stack from random positions within the extent of the original {\sc magphys} catalogue using the same number of sources as in the left panel.  There is no clear detection of emission from these mock sources, as expected. {\it Right:} Histograms of the pixel values for the IVW stack of our galaxy sample (purple) and mock galaxies (orange). In the top panel, the stacked galaxies show a deviation from a Gaussian distribution, with additional pixels with high positive values, correlating with the strong detection of 850$\mu$m flux displayed in the stacked stamp in the left panel. The pixels in the stack of mock sources are approximately Gaussian and centred on zero.}
    \label{fig:goodstack}
\end{figure*}

We generate 1000 artificial catalogues of random sources, with the maximum extent of the RA and Dec distribution of the sources limited to that of the input source  catalogue. For each ($M_*-z$) bin, we consider the number of sources in this bin for the original {\sc magphys}-COSMOS2015 catalogue, and randomly extract, without replacement, this many sources from the first artificial catalogue. We then stack these sources, using the 850$\mu$m S2COSMOS data map and noise map, as we did for the {\sc magphys}-COSMOS2015 catalogue of galaxies. Examples of  random stacks are shown in the middle panels of Figure \ref{fig:goodstack}. In the IVW stacks of randomly selected positions, we see no significant emission detected in the centre of the stacks. 

For each ($M_* - z$) bin, 1000 mock stacked stamps are generated, and the central pixel values for each are stored. Subsequently, for each ($M_* - z$) bin, a histogram is made of the central pixel values. We find that the histograms of the central pixel values for stacked stamps generated using the artificial catalogues are approximately Gaussian distributed, centred on zero; Appendix \ref{App:MChists} contains illustrative examples of these distributions. The error on the average fluxes for each bin is taken to be the width of these distributions (i.e. the 16th- and 84th-percentiles).

\subsubsection{Biases in the stacking: SIMSTACK}
Even with coarse resolution, the standard IVW method of stacking sources does not produce biased fluxes due to multiple sources being within one beamsize as long as the mean of the map is zero and the galaxies are not clustered. We have confirmed this using our MC simulations, which produce a mean signal consistent with zero. However, if the galaxies in the {\it initial} sample are clustered, the coarse angular resolution will cause a positive bias to the stacked signal, and result in an artificial boost to the stacked fluxes. To check the impact of galaxy clustering on our stacking results, we make use of the Python implementation of {\sc simstack} \citep{Viero2013}, a stacking and deblending method that attempts to account for any flux boosting introduced by galaxy clustering. We apply {\sc simstack} to the non match-filtered S2COSMOS 850$\mu$m map \citep{Simpson2019} and our {\sc magphys}-COSMOS2015 galaxy catalogue, and the sources are split into the same $M_{*}$ and $z$ bins as for the IVW stacking.  We use the non-matched filtered map because SIMSTACK carries out a convolution with a Gaussian beam within its code. We use the standard Python implementation of {\sc simstack} \citep{Viero2013}, with one small change: we alter the FWHM of the SCUBA-2 PSF to be 13$^{\prime \prime}$, reflecting the latest results from SCUBA-2 calibration tests in \cite{Dempsey2013}.

\subsubsection{Results}
\begin{figure*}
	\includegraphics[width=\textwidth]{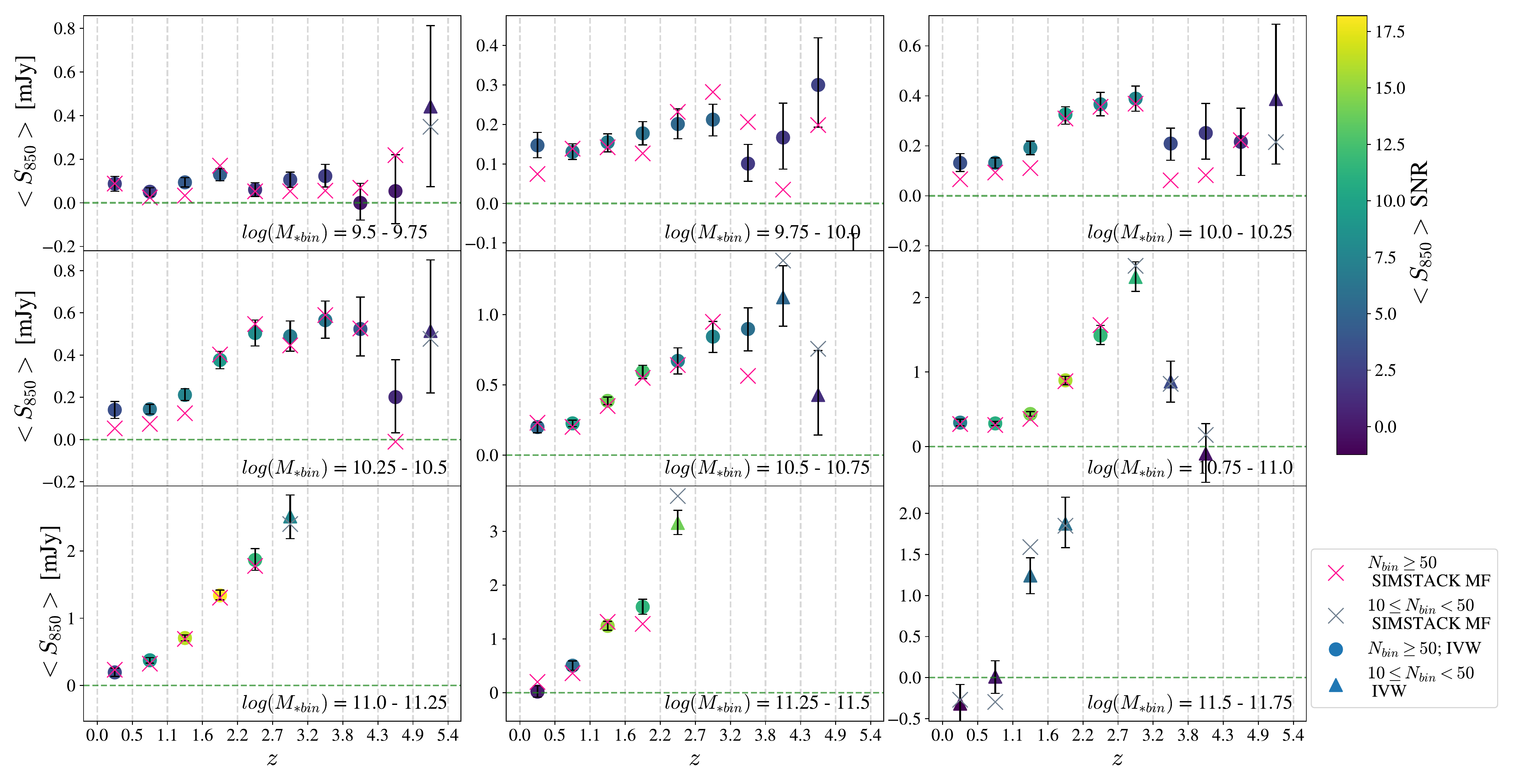}
    \caption{The resulting stacked fluxes from the Inverse-Variance Weighting (IVW) method and SIMSTACK. Filled circles are the IVW fluxes for stacks containing at least 50 sources. Filled triangles are the IVW fluxes for stacks containing at least 10 sources, but less than 50 sources. The colour of the points represents the SNR of the stacked flux for that given point. Crosses are the stacked fluxes determined by using SIMSTACK: pink means there are at least 50 sources in the stack, grey means there are at least 10 sources, but less than 50 sources in the stack. The stellar mass bin that each subplot represents is labelled in the bottom right of the respective subplot. A horizontal green dashed line denotes $<S_{850}> = 0$. Vertical grey dashed lines mark borders of redshift bins.}
    \label{fig:stackedflux}
\end{figure*}

The results from stacking the binned {\sc magphys}-COSMOS2015 sources on the 850$\mu$m SCUBA-2 map are presented in Figure \ref{fig:stackedflux}. The number of sources used in each bin is displayed in Table \ref{tab:bin_numbers}. 

For all $M_*$ bins, we see that the fluxes calculated from both stacking methods (Figure \ref{fig:stackedflux}) follow the same general trend; there is little evidence for flux boosting caused by biases in our original catalogue. Curiously, and more notably at higher redshifts and in low SNR $(M_* - z)$ bins, some of the fluxes from {\sc SIMSTACK} are higher than those calculated using the IVW stacking method. For stellar mass bins $10.0 \leq$ log$(M_{* \rm bin}/M_{\odot}) \leq 11.0$, we see a peak in the mean $850\mu$m flux from dust emission at around $z \sim 2.5-3.5$, just beyond the peak of star formation (SF) in the history of the universe, which is at $z \sim 2$ (e.g. \citealt{Madau2014}). Lower stellar mass bins are too noisy to draw similar conclusions for. Higher stellar mass bins may follow a similar trend, but the lack of high mass galaxies at high redshifts makes it impossible to draw meaningful conclusions about the location of the peak flux in the highest stellar mass bins.

The fluxes for most stacks have good SNRs (Figure \ref{fig:stackedflux}), but there are a handful of negative stacked fluxes, indicating non-detections. In these instances, upper limits of ISM masses are calculated based on the 3$\sigma$ flux errors, determined from Monte Carlo simulations.

The similarity of the results from SIMSTACK, which makes a correction for clustering, and our stacking results suggests that galaxy clustering has little effect on the stacked fluxes\footnote{Note that the PSF FWHM of the SCUBA-2 beam 13$^{\prime \prime}$ at $z = 0.1, 0.5, 1, 2, 3, 4$ translates to a physical scale on the sky, $\eta$, of $\approx 25, 82, 108, 112, 103, 93$ kpc, respectively \citep{Wright2006}.}.
In contrast, James Simpson ({\it priv. comm.)} found evidence of flux boosting of up to 20\% at 850$\mu$m when using a modified version of SIMSTACK \citep{Simpson2019} that accounts for holes in galactic catalogues caused by bright star masking, and includes background modelling. The flux boosting they observed did not evolve over redshift, so although our stacked flux values may be higher by 20\%, there is no evidence to suggest any trends in $f_{ISM}$ would change. We also note that there are other caveats that are likely to have more of an impact on our results - these are discussed later in Section \ref{Sec:caveats}. Moving forward, we focus mostly on the results from IVW stacking. 

\subsection{Deriving Gas Masses and Gas Fractions using Submillimetre Fluxes} \label{Sec:stackinggas}
Currently, our knowledge of gas-to-dust ratios and dust emissivities is limited to too few galaxies to be of practical use for statistical evolution studies of the universe. As such, we follow the methods presented in \cite{Eales2012}, \cite{Scoville2016} and \cite{Scoville2017}, which allow us to extrapolate the existing data, and calculate gas masses for thousands of galaxies using a calibrated conversion factor, $\alpha_{850}$, and the specific luminosity of sources at 850$\mu$m, $L_{850}$. 

As detailed in \cite{Scoville2016} and \cite{Scoville2017}, there is a direct relation between the CO(1-0) luminosity and $L_{850}$. By making use of a standard Galactic CO(1-0) conversion factor, \cite{Scoville2016} and \cite{Scoville2017} show that the RJ emission from dust can be used to calculate ISM masses:
\begin{equation}
\begin{split}
    M_{\rm ISM} = \ & 1.78 \ S_{\nu_{\rm obs}} [mJy] \ (1+z)^{-4.8} \left(\frac{\nu_{850}}{\nu_{\rm obs}}\right)^{3.8} (d_L)^2 \\
    & \times \left \{ \frac{6.7 \times 10^{19}}{\alpha_{850}} \right \} \frac {\Gamma_0}{\Gamma_{\rm RJ}} \quad 10^{10}M_{\odot}
	\label{eq:ISMmasses}
\end{split}
\end{equation}
where $S_{\nu_{\rm obs}}$ is the flux density of the source at the observation wavelength, $z$ is the redshift of the source, $\nu_{\rm obs}$ is the observation frequency, $d_L$ is the luminosity distance of the source, in Gpc, $\alpha_{850} = 6.7 \pm 1.7 \times 10^{19} {\rm erg} s^{-1} {\rm Hz}^{-1} M_{\odot}^{-1}$, and $\Gamma_0 / \Gamma_{\rm RJ}$ accounts for the deviation of the Planck function in the rest frame from the RJ form:
\begin{equation}
    \Gamma_{\rm RJ} (T_d, \nu_{\rm obs}, z) = \frac{h \nu_{\rm obs} (1+z) / k T_d} {e^{h \nu_{\rm obs} (1+z) / k T_d} - 1}
	\label{eq:RJcorr}
\end{equation}
where $h$ is the Planck constant, $k$ is the Boltzmann constant and $T_d$ is the temperature of the dust, in Kelvin. In Equation \ref{eq:ISMmasses}, $\Gamma_0$ is calculated for $z = 0$ and $T_d$ = 25\,K, as used to calibrate $\alpha_{850}$. In these calculations, we assume $T_d$ = 25\,K, as in \cite{Scoville2016}, and we refer the interested reader to this paper for full calibration details.

Note that there is an important difference in our analysis here from the stacking analysis of \cite{Bethermin2015}. Bethermin et al. estimated ISM masses using dust temperatures estimated from the SEDs formed from the stacked flux densities at a large number of far-IR and submm wavebands. It is well-known (\citealt{EalesWW1989}; \citealt{DunneEales2001}) that this procedure produces luminosity-weighted dust temperatures which are higher than the temperature of most of the dust because warm dust is more luminous than cold dust. Ideally, we would use mass-weighted dust temperatures. Since we do not have any direct information about the mass-weighted dust temperatures of the galaxies in our sample, we follow Scoville et al. in assuming that the mass-weighted dust temperature is 25\,K. We note that this is consistent with the mass-weighted dust temperatures estimated for both a low-redshift sample \citep{DunneEales2001} and a sample of bright high-redshift submm sources \citep{Pearson2013}. We also note that theoretical simulations \citep{Liang2019} suggest that the mass-weighted dust temperature does not evolve much with redshift.

Since in \cite{Scoville2016} and \cite{Scoville2017} the equations for calculating ISM masses are calibrated against molecular hydrogen mass measurements conducted using CO line emission, we consider the gas traced by dust to be molecular hydrogen, rather than a combination of molecular and atomic gas phases.

Subsequently, we define the gas fraction, $f_{\rm ISM}$, as:
\begin{equation} \label{eq:massfrac}
    f_{\rm ISM} \equiv \frac{M_{\rm ISM}}{M_* + M_{\rm ISM}}
\end{equation}
where, for each $M_*$ bin, the value of $M_*$ in Equation \ref{eq:massfrac} is determined from a bootstrap analysis of the {\sc magphys} stellar masses from individual sources in the bin.  In brief, for a given ($M_*-z)$ bin we randomly select, with replacement, $X$ sources, where $X$ is the number of sources in the bin.   For the selected sources, we perturb each galaxy's stellar mass within their error (using a Gaussian centred on the 50th-percentile stellar mass, with width as the average of the 16th- and 84th-percentiles from {\sc magphys}).
5.6\,per\,cent of our sample have identical values returned by {\sc magphys} for the 16th-, 50th- and 84th-percentiles ie there is no error provided from the PDF of the stellar mass for these sources, or the PDF is extremely narrow. The fraction of sources in a given $(M_* - z)$ bin with no stellar mass error varies as a function of stellar mass. In the highest stellar mass bins the fraction increases to around 20-50\%. Similarly, the highest redshift bins have slightly higher fractions of galaxies without stellar masses errors, typically up to 30\%.  We attribute the narrow PDFs/lack of error on stellar mass returned by {\sc magphys} for these sources due to the error being smaller than the {\sc magphys} parameter grid-spacing used to build up the PDF. This may be caused by over-constrained fitting. For these sources, we allocate them a fractional error based on the average fractional error of sources in their associated bin in the bootstrap estimation.  

For each iteration of the bootstrap analysis, we store the median of the perturbed stellar masses, and repeat 10,000 times. We use the 16th-, 50th-, and 84th-percentiles of the median bootstrapped stellar masses as the stellar mass and error in each bin. The errors obtained from the bootstrap analysis are negligible in comparison to the errors on our stacked fluxes. On a bin-by-bin basis, the fractional stellar mass error is at least 30 times lower than the fractional flux error. Thus, we do not consider errors in the stellar mass in determining $f_{\rm ISM}$. The assumed stellar masses for a given bin are displayed in Table \ref{tab:bin_numbers}.

$M_{\rm ISM}$ is calculated using Equations \ref{eq:ISMmasses} and \ref{eq:RJcorr}. For each $(M_*-z)$ bin, we derive the gas fraction using the average stacked fluxes at 850$\mu$m as the values for $S_{\nu_{\rm obs}}$ in Equation \ref{eq:ISMmasses}. $d_L$ is determined using the centre of the $z$ bin in question.

We note that Equation \ref{eq:ISMmasses} was developed with calibration samples limited to $M_* > 5 \times 10^{10} M_{\odot}$, to avoid contamination by sources which are likely to have below solar metallicity, or where there may be an abundance of gas without CO \citep{Scoville2016}. In this study, we choose to probe masses below this limit, but caution the reader to keep these limitations in mind when considering the results of this work.

\begin{figure*}
	\includegraphics[width=\textwidth, trim = 0cm 0.5cm 0cm 0cm, clip]{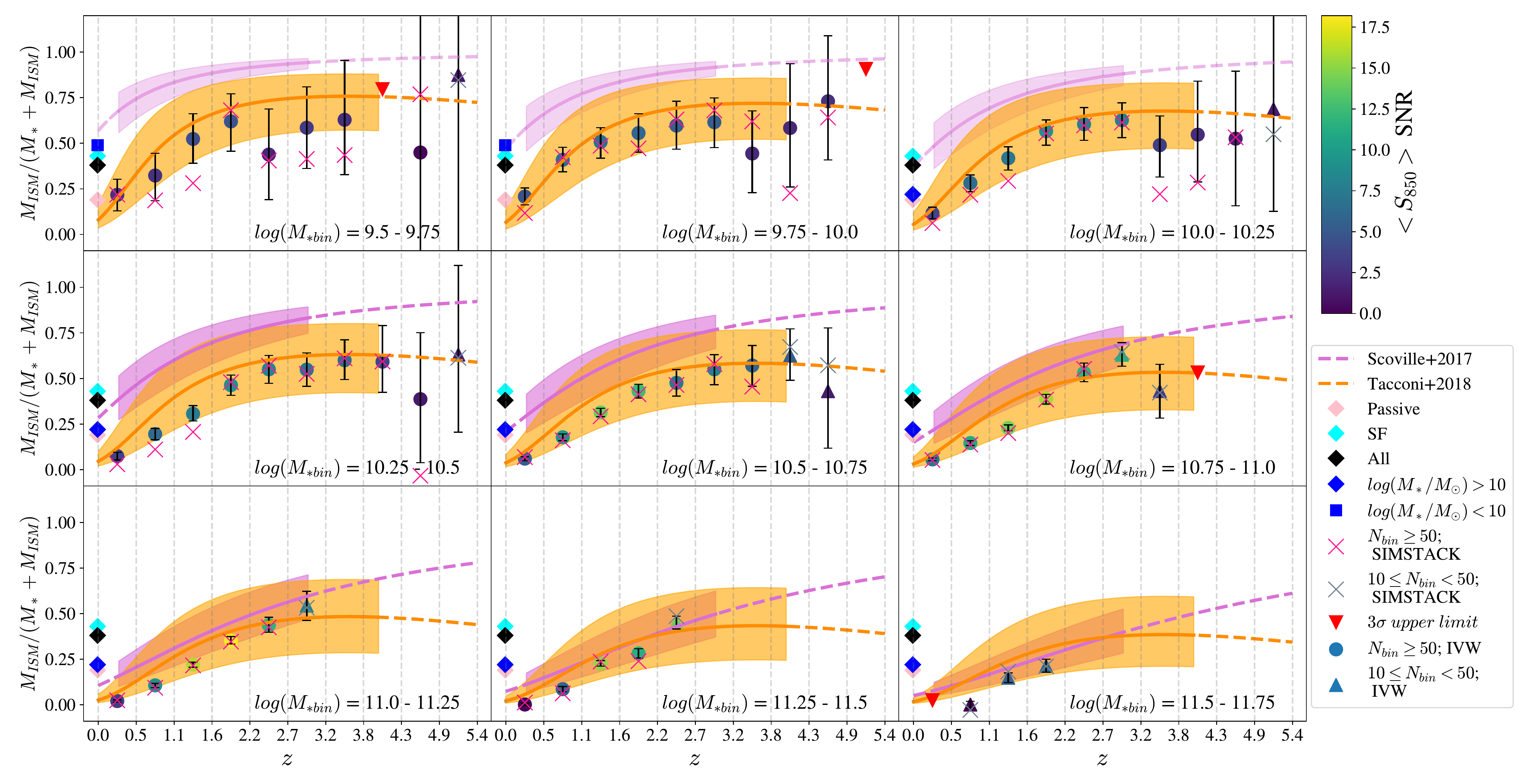}
    \caption{
    The resulting stacked ISM mass fractions with redshift, based on 850$\mu$m fluxes in the COSMOS field. We compare both the Inverse-Variance Weighting (IVW) method and SIMSTACK. Filled circles are the IVW data for stacks with N $\geq$ 50. Filled triangles are the IVW data for stacks with 10 $\leq$ N $<$ 50. The colour of the points represents the SNR of the stacked flux for that given point. Red triangles denote negative stacked fluxes; here, the 3$\sigma$ error value is used to calculate an upper limit to the ISM mass fraction. The stellar mass bin that each subplot represents is labeled in the bottom right of the respective subplot. Vertical grey dashed lines mark borders of redshift bins. The redshifts used in calculating $M_{\rm ISM}$ are assumed to be the centre of the redshift bins. The purple and orange lines are the scaling relations from \citealt{Scoville2017} and \citealt{Tacconi2018} (shown as the dashed lines where the relationship extends beyond the redshift limit of their samples), for galaxies on the Main Sequence (MS). The shaded regions show the extent of the scaling relations (over the redshift ranges used in \citealt{Scoville2017} and \citealt{Tacconi2018}) for galaxies 5 times above and below the MS  i.e. $0.2 \leq ({\rm sSFR}/{\rm sSFR}_{\rm MS}) \leq 5$ (the shaded region is fainter where the mass limit extends beyond their sample masses). The average gas fraction based on 640 local galaxies (black diamond) is also shown at redshift zero (Pieter De Vis, {\it priv. comm.}), defined as $M_{\rm gas}/ (M_{\rm gas} + M_*)$ (see main text). We also split the local gas fraction into late type and early type galaxies (490 LTGs and 150 ETGs, the cyan and pink diamonds respectively) and by stellar mass (373 galaxies with ${\rm log}$ $M_* <10$ and 267 galaxies with ${\rm log}$ $M_* >10$, denoted by the blue square and diamond, respectively).
    }
    \label{fig:stackedmassfrac}
\end{figure*}

\section{The evolution of gas fraction with redshift}
Even with small differences between some of the stacked fluxes calculating using the IVW method and SIMSTACK, Figure \ref{fig:stackedmassfrac} illustrates that we still see the same evolution in the ISM mass fraction over cosmic time. 
The largest difference between the two stacking methods is for the lowest stellar mass bin, where the gas mass fractions for several redshift bins determined using the fluxes from SIMSTACK are lower than using fluxes from IVW stacking. However, the fluxes mostly agree within statistical errors. This is also the stellar mass bin with the largest errors on the fluxes, and lies outside of the stellar mass range for which the relation developed in \cite{Scoville2016} and \cite{Scoville2017} has been calibrated. As such, the small discrepancies displayed here are not too concerning.
Generally, in low stellar mass bins ($\rm log(M_*/M_{\odot}) < 10.5$), we see the ISM mass fraction in galaxies increasing up to a maximum at around $z \sim 2.5 - 3$, and then plateauing out, with some small peaks and dips as we move to higher $z$. This trend is hard to see for the lowest stellar mass bin, but, within errors, we believe that this bin displays a similar trend to the other low stellar mass bins. For higher stellar mass bins ($\rm log(M_*/M_{\odot}) \geq 10.5$), the ISM mass fraction in galaxies increases with redshift, without clearly reaching a maximum, even out to $z \sim 3$. We see that as one progresses to higher stellar mass bins, the overall gas mass fraction decreases. The most massive galaxies in the nearby universe are particularly gas-poor, as also seen in local galaxy surveys e.g. \citet{Saintonge2011,DeVis2017a}, and previous studies of gas evolution with redshift \citep{Tacconi2013,Scoville2017,Tacconi2018}.

\subsection{Comparison to literature scaling relations}
We compare our stacking results to scaling relations from \cite{Scoville2017} and \cite{Tacconi2018} (Figure \ref{fig:stackedmassfrac}, and Equations \ref{eq:scoville_scaling} and \ref{eq:tacc_scaling}) and with observations of the gas fraction derived for local galaxies at $z=0$.  The latter are taken from Pieter De Vis ({\it priv. comm.}) where the observed gas fraction for 640 galaxies are averaged\footnote{If $\rm H_2$ measurements are not available, they are estimated using a $\rm H_2/HI$ scaling relation derived for DustPedia galaxies \citep{Casasola2020}}. The gas mass estimates are taken from  the DustPedia survey \citep{Davies2017}, HAPLESS, the dust-selected sample from a blind {\it Herschel} survey (\citealt{Clark2015}), and the HI selected sample HIGH \citep{DeVis2017a,DeVis2017b}. The gas mass fraction for these observations is defined as $M_{\rm gas}/ (M_{\rm gas} + M_*)$ where $M_{\rm gas} = (1+M_{\rm H_2}/M_{\rm HI})x_i$ and $x_i$ is a metallicity-dependent factor to account for He \citep{Clark2016,DeVis2019}. We further split the observed gas fractions of local galaxies into late type and early type galaxies, and by stellar mass.  The observed gas fractions can act as a benchmark to the scaling relations and the stacked values derived here, and are shown in Figure~\ref{fig:stackedmassfrac}.

The scaling relation from \cite{Scoville2017} is based on a reduced sample of 575 galaxies detected with ALMA Bands 6 and 7, with SFRs from the MS up to 50 times the MS, a redshift distribution of $z=0.3-3$, and stellar masses above $3 \times 10^{10} M_{\odot}$. ISM masses are calculated using 850$\mu$m fluxes from the RJ tail of dust emission (Section~\ref{Sec:stackinggas}). In that scaling relation, it is assumed that $\beta = 2$:
\begin{equation*}
    f_{\rm ISM} \equiv \frac{M_{\rm ISM}}{M_* + M_{\rm ISM}}
\end{equation*}
\begin{multline}
     \phantom{f_{\rm gas}} = \left\{ 1 + 1.41(1+z)^{-1.84} \right. \times ({\rm sSFR/sSFR}_{\rm MS})^{-0.32} \\
     \left. \times \left(M_*/10^{10}M_{\odot}\right)^{0.70} \right\} ^{-1}
	\label{eq:scoville_scaling}
\end{multline}
where $({\rm sSFR/sSFR}_{\rm MS})$ is the ratio of the specific star-formation rate $({\rm sSFR})$ compared to the specific star-formation rate on the MS $({\rm sSFR}_{\rm MS})$.   

\cite{Tacconi2018} present a few different scaling relations for the ratio of molecular gas to stellar mass ($\mu_{\rm mol}$) with redshift, based on 1444 star-forming galaxies between $z=0-4$. The SFRs of their sample range from $\sim$20 times lower than the MS, to up to $\sim$150 times above, probing the stellar mass range log$(M_*/M_{\odot}) = 9.0-11.8$. Molecular gas masses in \citet{Tacconi2018} are estimated from CO line fluxes, far-infrared (FIR) spectral energy distributions (SEDs) and mm photometry. They define a `best' scaling relation based on their combined samples\footnote{We have taken the best fit relation from \citet{Tacconi2018} for the $\beta=2$ case, where $\beta$ is a parameter introduced in their work that accounts for the redshift evolution in specific star formation rates.} for $\mu_{\rm mol}$ (their Table 3 and Equation 6) of:
\begin{multline}
     {\rm log} \frac{M_{\rm mol}}{M_*} = 0.12 - 3.62({\rm log}(1+z) - 0.66)^2 \\ \phantom{log \frac{M_{\rm mol}}{M_*}55} + 0.53{\rm log(sSFR/sSFR}_{\rm MS}) \\ - 0.35({\rm log}M_* - 10.7) + 0.11{\rm log}(R_e/R_{e,0})
	\label{eq:tacc_scaling}
\end{multline}
where $M_{\rm mol}$ is the total mass of molecular gas (including a 36\,per\,cent mass fraction of helium and a correction for CO-dark molecular clouds).   $R_{e,0}$ is defined as the mean effective radius of the star-forming population as a function of $z$ and $M_*$ \citep{vanderWel2014}. Since in this study we derived our stacked fluxes from the central pixel values of stacked stamps, measured in Jy/beam, we simply set $(R_e/R_{e,0}) = 1$\footnote{\citet{Tacconi2018} show that the evolution of molecular gas fraction only very weakly depends on size. Although, we note that \cite{Tacconi2018} use rest-frame optical sizes, rather than sizes of the distribution of gas - measurements of which are extremely uncertain, and will remain so until there are significant sub-mm samples, hopefully obtained with ALMA.}. \citet{Tacconi2018} suggest that one can approximate the gas mass by $\mu_{\rm mol} \sim \mu_{\rm gas}$ and $f_{\rm mol} \sim f_{\rm gas}$ for $z>0.4$, where $f_{\rm mol} = (1+ M_*/M_{\rm mol})^{-1}$.  (\cite{Scoville2017} and \cite{Tacconi2018} quote errors on the parameters in their scaling relations, though when comparing these relations to our results, we do not consider their errors here.) When comparing to the scaling relations, we consider the relations for galaxies on the MS, such that the ratio $\rm (sSFR)/(sSFR_{MS}) = 1$ (i.e. we normalize by the sSFR value on the MS).

We note here that although the \citet{Scoville2017} relation for $f_{\rm ISM}$ (Equation~\ref{eq:scoville_scaling}) is said to apply to both the HI and $\rm H_2$ components, the mm-method described there (and in Section~\ref{Sec:stackinggas}) is only calibrated against CO and $L_{850}$. It is possible then that this method may only be strictly valid when using dust emission to trace the molecular component of the gas mass and not the total cool ISM mass.  In this scenario, the \citet{Scoville2017} and \citet{Tacconi2018} relations are therefore both tracing the molecular gas component and should be comparable, except that \citet{Tacconi2018} has made a further correction for the metallicity dependence of CO factors and gas-to-dust ratios.

These scaling relations are shown in Figure~\ref{fig:stackedmassfrac}. We take the centre of the stellar mass bins quoted in the lower right of each subplot as the assumed stellar mass value.  We find our stacking results are more in agreement with the scaling relation from \cite{Tacconi2018} in the lowest stellar mass bins, but as we move to the highest stellar mass bins, we see that our stacking data agrees with both relations, which are comparable with each other. The scaling relations follow a similar trend to our data: the higher the stellar mass of a galaxy, the lower the ISM mass fraction \citep{Saintonge2011,Saintonge2017}.  The \citet{Tacconi2018} relation also shows the plateauing  in gas mass fraction over cosmic time, as shown in our stacked data points.

Figure~\ref{fig:stackedmassfrac} shows that our stacked data and the \citet{Tacconi2018} relations lie well below the average gas (HI + H$_2$) fractions derived using local galaxy samples at $z=0$, though the stacked data most closely agrees with the average gas fractions derived locally for high-mass ($f_{\rm gas}|_{z=0} \sim 0.22$), and passive galaxies samples (based on early type galaxies $f_{\rm gas}|_{z=0} \sim 0.19$). 

The \citet{Scoville2017} trends agree well with the observed $z=0$ gas fractions in the low $z$ bins, particularly when comparing with the average local gas fraction split by stellar mass (blue symbols in Figure~\ref{fig:stackedmassfrac} where $f_{\rm gas}|_{z=0} \sim 0.49$ and $f_{\rm gas}|_{z=0} \sim 0.22$ for low and high stellar mass samples respectively). At first glance this may imply that the \citet{Scoville2017} gas relations are indeed valid for dust emission tracing both the atomic and molecular gas component of galaxies since it matches well the observed properties that include both HI and $\rm H_2$. However, we note that the average gas fractions for $z=0$ star forming galaxies (based on the average properties of late type galaxies, $f_{\rm gas}|_{z=0} \sim 0.43$) are also comparable with \citet{Scoville2017} at low redshifts, so the agreement with local galaxy values could simply be due to their sample containing a higher fraction of star-forming systems compared to this work and the \citet{Tacconi2018} sample. We noted earlier that the measurements of gas mass from \citet{Tacconi2018} and potentially the \citet{Scoville2017} method used here may only be valid in tracing the molecular mass component i.e. our $f_{\rm ISM}$ may in fact only trace $f_{\rm mol}$ with redshift. It is not entirely surprising then that the local observed gas fractions are higher, because these also include the contribution of atomic gas, which is known to be significant at $z=0$ \citep{Lagos2011, Saintonge2011}, whereas the calibration relationships that we are using only include molecular gas. The scaling relations based on molecular gas will always be biased low compared to the total gas fractions measured in the lowest redshift bins.

The difference in the gas fractions for ${\rm log}~ M_*<10.75$ between our work/the scaling relations of \citet{Tacconi2018} and \citet{Scoville2017} may be due to two effects: (i) metallicity and/or (ii) the sample selections.  The \citet{Scoville2017} gas fractions were defined using ALMA mm fluxes without any metallicity corrections (indeed, they caution against using their relationship at ${\rm log} M_* < 10.3$ for this reason). 
We note that one can test the metallicity effect since \citet{Tacconi2018} has already presented scaling relations for molecular-to-stellar mass for gas masses derived using molecular gas from CO, FIR SED fitting techniques and the long-wavelength mm photometry method, from \citet{Scoville2013,Scoville2016,Scoville2017}, having corrected for the metallicity dependence of the gas-to-dust ratio (assuming that the dust-to-gas ratio is nearly linearly correlated with metallicity). 
Their relationships for these three methods for determining the gas mass (their Table 3) suggest that even when correcting for metallicity, the gas fractions derived from mm photometry are always higher at stellar masses ${\rm log} M_* < 11$ and redshifts $z<3$, compared to gas fractions derived using CO or FIR SED fitting methods. 
Beyond this redshift, and at higher stellar masses, they tend to have slightly lower gas fractions than seen in the best fit scaling relations from \citet{Tacconi2018} and our work, though the differences are small in this stellar mass regime. The fact that the trends are very similar no matter how the gas fraction is determined at higher stellar masses (ie mm photometry or from CO observations), may, at first glance, suggest that the metallicity correction applied by \citet{Tacconi2018} is not valid in the lowest redshift and stellar mass bins, and causes the gas fractions to be overestimated in this regime. 
However, \citet{Tacconi2018} applies the same metallicity dependence to gas masses derived from dust masses based on FIR SED fitting, which produces scaling relations similar to CO and our stacked data.  Thus the differences can not entirely be due to metallicity corrections. 
Although, like \citet{Scoville2017}, we have not corrected for metallicity dependence in gas-to-dust ratios, our data is in close agreement with the \citet{Tacconi2018} best fit relationship.  This suggests that the largest driver in the differences here is not the metallicity correction, or method used to derive the gas masses, but rather the sample selection. By stacking on a submillimetre map based on optical/NIR source catalogues, we are less likely to sample dust rich galaxies compared to \citet{Scoville2017} and the combined samples from \citet{Tacconi2018}. It also means that our gas fractions in Figure~\ref{fig:stackedmassfrac} may be biased low if we are including passive galaxies or less star forming systems, compared to eg \cite{Scoville2017}, or the galaxies in the local samples, since it is well known that galaxies below the main sequence have lower gas fractions (e.g. \citealt{Saintonge2012}). This could partly explain the offset at $z=0$ seen between our stacked $f_{\rm ISM}$ and the higher range of gas fractions observed in local galaxies ($f_{\rm gas} \sim 0.4-0.5$) since these are dominated by star forming and low mass galaxies. We will return to this in the next Section.

\subsection{Star-forming and Passive Galaxies}

\begin{figure*}
	\includegraphics[width=\textwidth, trim = 0cm 0.4cm 0cm 0cm, clip]{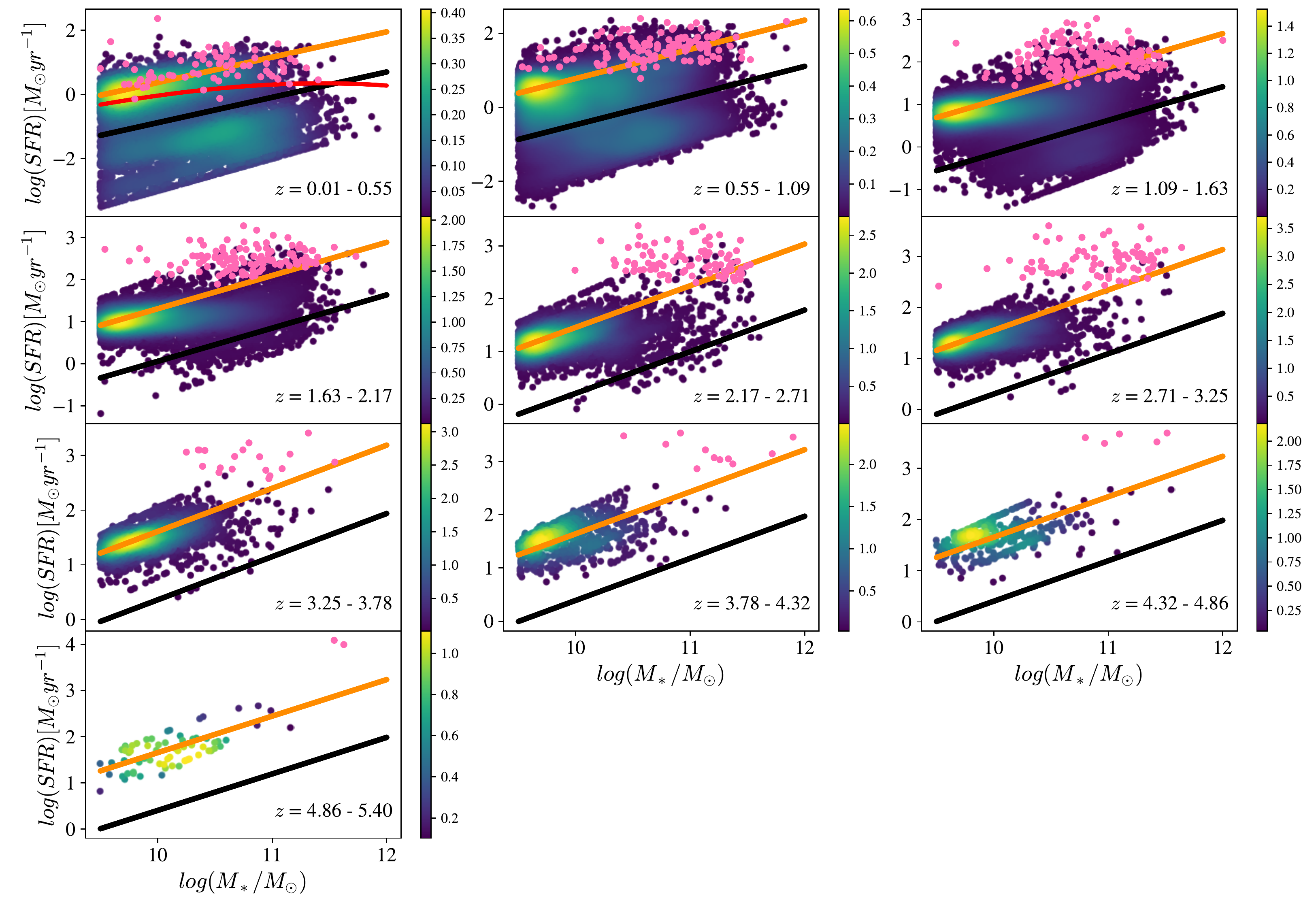}
    \caption{The star formation rate (SFR) versus stellar mass for the galaxy catalogue used in this work, split into redshift bins. SFRs and stellar masses are taken from the {\sc magphys} fits to the SEDs. The colour bar shows the number density of galaxies in the sample, the orange line is the so-called Main Sequence line of \citet{Sargent2014} which evolves with redshift. The red line in the lowest redshift bin is the MS trend from \citet{Saintonge2016}. The black line denotes where we split our sample into 'star-forming' and 'passive' sources. The black line also evolves with redshift in a similar manner to the orange line; there is a fixed offset between the two.  The properties of galaxies that make up the sample of \citet{Scoville2017} are also shown in pink. We attribute the striping in the distribution of sources at low SFRs in the lowest redshift bins to sources with high $\chi^2$ (poor fits).}
    \label{fig:starformingsplit}
\end{figure*}

Here we test whether the stacked gas fractions in Figure~\ref{fig:stackedmassfrac} are biased low at low redshifts due to passive galaxies, compared to the observed local gas fractions and the mm dust photometry from \citet{Scoville2017}.  Figure~\ref{fig:starformingsplit} compares the star formation rates (SFRs, derived using {\sc magphys}) with the stellar masses, $M_*$, in redshift bins.  We can see that at $z<1.6$, there is clear evidence of two populations of galaxies. A tight correlation between SFR and stellar mass exists for the more star forming galaxies: this is the so-called main sequence (MS) relation. Figure~\ref{fig:starformingsplit} shows the MS relation from \citet{Sargent2014} which evolves with redshift (see also \citealt{Noeske2007,Daddi2007}), with scatter around the line of approximately 0.3\,dex \citep{Peng2010,Sargent2014}. (The $z=0$ MS from \citet{Saintonge2016} is also shown). Most of our sample lies along the MS indicative of normal star forming systems, but a significant fraction of sources have lower star formation rates for a given stellar mass, by approximately two orders of magnitude. These are quiescent/passive sources i.e. red and dead galaxies, with potentially some galaxies with intermediate star forming properties, so-called green valley sources. This passive population quickly disappears from our sample at redshifts greater than 1.6, where our sample becomes dominated by main sequence galaxies at higher redshifts. Figure~\ref{fig:starformingsplit} also demonstrates the lack of high mass systems in our higher redshift bins.

As we are interested in whether a galaxy in our sample is more-or-less star forming, rather than the absolute value of SFR, we simply split our sample into two sets: a star-forming and a passive set based on a split in SFR-$M_*$ space, as indicated by the black line in Figure~\ref{fig:starformingsplit}. This split region is offset by -1.25\,dex compared to the MS relation of \citet{Sargent2014}. Although this offset is somewhat arbitrary, it is sufficient for this test.

Figure~\ref{fig:starformingsplit} compares the SFR-$M_*$ plane with the \citet{Scoville2017} mm sample. We can see that their sample contains more massive systems in each redshift bin and, except for the lowest redshift bin, these tend to sit above the MS trend from \citet{Sargent2014}. Furthermore, in every redshift bin we see that the \citet{Scoville2017} galaxies are on average more star forming than our sample. Thus the sample selection can explain the differences in our observed scaling relations. By stacking on a catalogue of optical/NIR sources in a deep submillimetre field, we have improved on number counts in each redshift range, probed down to lower stellar masses out to $z=5$, and include galaxies with lower star-formation rates.  We will return to the issue of selection effects in Section~\ref{Sec:caveats}.

\begin{figure*}
	\includegraphics[width=\textwidth, trim = 0cm 0.4cm 0cm 0cm, clip]{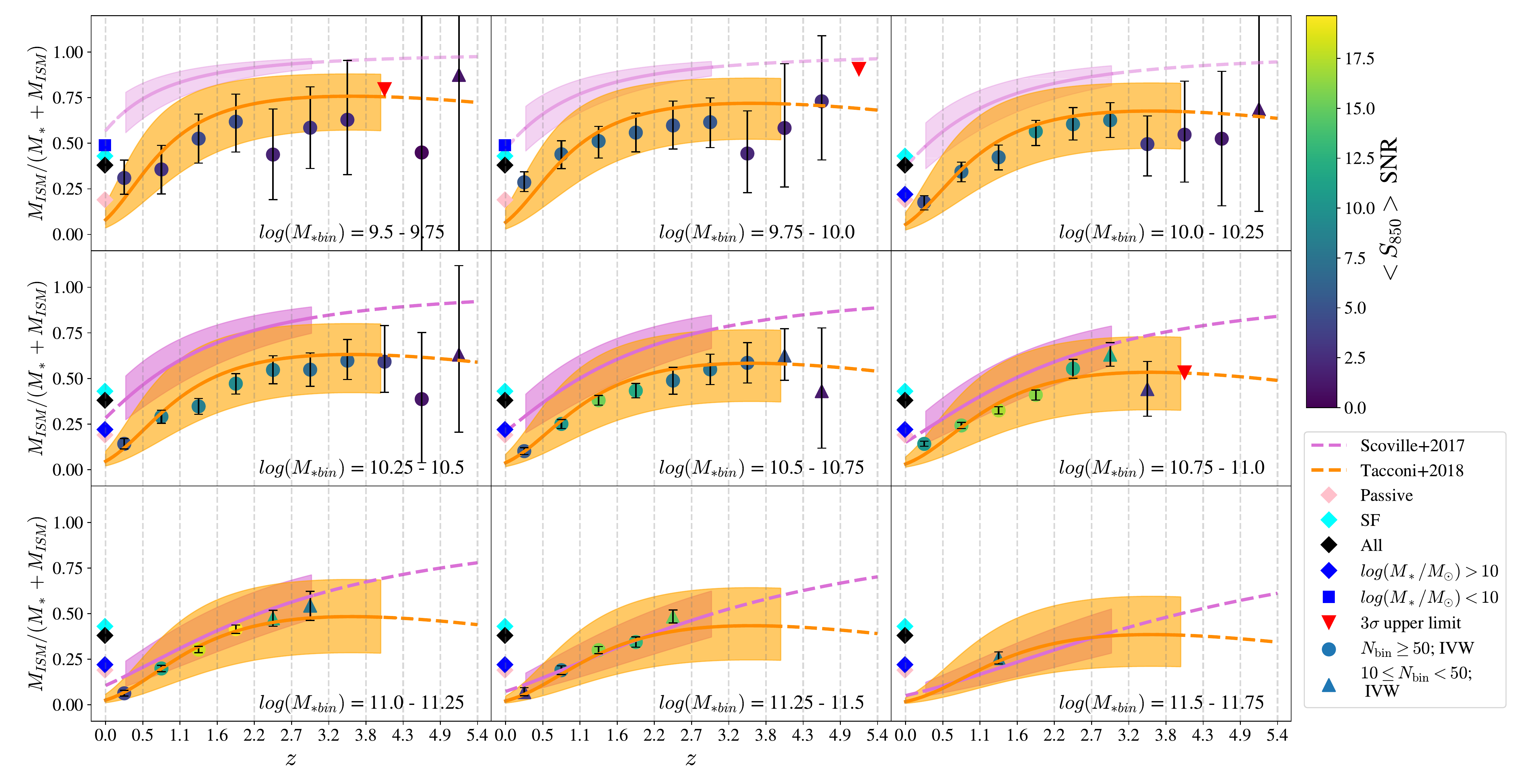}
	\includegraphics[width=\textwidth, trim = 0cm 0.4cm 0cm 0cm, clip]{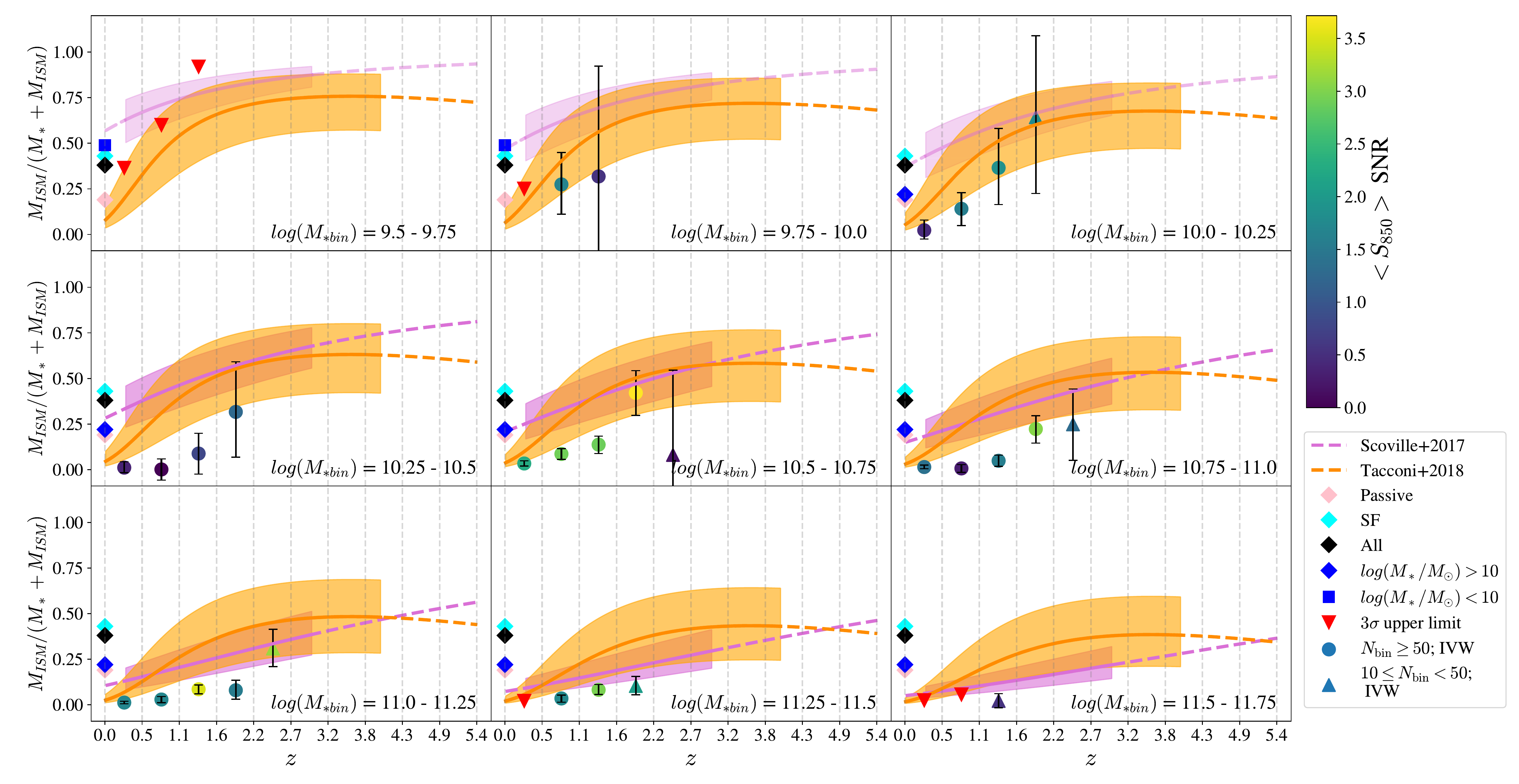}
    \caption{As in Figure~\ref{fig:stackedmassfrac}, here we show the resulting stacked ISM mass fractions with redshift, based on 850$\mu$m fluxes in the COSMOS field, but now split into star-forming (top) and passive (bottom) sets. Filled circles are the IVW data for stacks with N $\geq$ 50. Filled triangles are the IVW data for stacks with 10 $\leq$ N $<$ 50. The colour of the points represents the SNR of the stacked flux for that given point. Red triangles denote negative stacked fluxes; here, the 3$\sigma$ error value is used to calculate an upper limit to the ISM mass fraction. The purple and orange lines are the scaling relations from \citealt{Scoville2017} and \citealt{Tacconi2018} (shown as the dashed lines where the relationship extends beyond the redshift limit of their samples), for galaxies on the Main Sequence (MS).The average gas fraction based on 640 local galaxies (black diamond) is also shown at redshift zero (Pieter De Vis, {\it priv. comm.}), defined as $M_{\rm gas}/ (M_{\rm gas} + M_*)$ (see main text). We also split the local gas fraction into late type and early type galaxies (490 LTGs and 150 ETGs, the cyan and pink diamonds respectively) and by stellar mass (373 galaxies with ${\rm log}$ $M_* <10$ and 267 galaxies with ${\rm log}$ $M_* >10$, denoted by the blue square and diamond, respectively).}
    \label{fig:starformingsplit_fism}
\end{figure*}

Figure~\ref{fig:starformingsplit_fism} shows the stacked gas mass fractions for the passive and star-forming galaxies.  The SNR is very low for the passive galaxies and the number counts too are low at higher redshifts and low stellar masses.  As such, we do not draw any conclusions from this except that passive galaxies do not have much 850$\mu$m emission and therefore not much gas (as expected), and that the inclusion of these galaxies in our low $z$ sample does indeed bias our gas fraction scaling relation low in Figure~\ref{fig:stackedmassfrac}. We see still that the gas mass fractions derived from stacking on the COSMOS map for our star forming and passive samples are lower than those derived for local samples of galaxies (even when split into LTGs and ETGs). This likely originates from the fact that the tracer we use here to measure gas mass does not include the atomic phase, whereas the local measurements do.

\section{Discussion and Caveats} \label{Sec:caveats}
Figure \ref{fig:stackedmassfrac} clearly illustrates that out to around $z \sim 3$, the gas mass fraction of galaxies increases. Beyond this point, gas mass fractions seem to approximately plateau. However, there are number of technical concerns we need to consider before we take these results too seriously.  

We have four $(M_*-z)$ bins with non-detections. The non-detection at $z \sim 5.1$ may simply be due to the low number of galaxies in that bin. As shown in Table \ref{tab:bin_numbers}, all the stellar mass bins at this redshift have low numbers of galaxies (if they even have enough sources to warrant a stacking analysis at all), and all the measured fluxes, and calculated gas mass fractions, have large errors. Therefore, it is unsurprising that we have a non-detection at this redshift in one of our stellar mass bins. 
The non-detection at $z \sim 4$ in the lowest stellar mass bin cannot be simply attributed to a low number of galaxies in that bin - indeed, there are a significant number of galaxies in this $(M_* -z)$ bin (see Table \ref{tab:bin_numbers}). Measurements at this redshift in the other low mass bins have large errors, even with similarly high source counts. Also, in this stellar mass bin, the flux measurements at $z \sim 3.5$ and $z \sim 4.6$ have large errors, even if they do have a positive flux detection. Therefore, it is not surprising that we have a non-detection here too. 
Table \ref{tab:bin_numbers} shows that the non-detection at $z \sim 4$ in stellar mass bin log$(M_{*,bin})$ = 10.75 - 11.0 can be attributed to a low number of galaxies in this bin, particularly as the non-detection is for the highest redshift bin in this stellar mass bin. We simply lack enough high mass sources at high redshift to enable a detection of sub-mm emission from dust above the noise levels of the map. At first, the non-detection at $z \sim 0.3$ may seem unusual. But, when one considers the results of splitting the sources into star-forming and passive galaxies (Figure \ref{fig:starformingsplit_fism}), we see that the galaxies in this bin are mostly passive galaxies, which are gas poor. It is therefore unsurprising that we do not detected significant 850$\mu$m dust emission from galaxies in this $(M_* - z)$ bin.

Inevitably, despite the faintness of the {\sc magphys} catalogue magnitude limit ($i$ < 25\,mags), the sample we consider in this study is not complete at high redshifts, roughly $z>1$ for the low-mass bins and $z>2$ for the high-mass bins. Without having much deeper magnitude-limited samples, which only exist for small area fields such as GOODS-North and GOODS-South, we cannot quantify the biases involved. However, the two obvious biases are that we will be missing highly-dust-obscured galaxies such as SMGs (\citealt{Lang2019}; \citealt{Stach2019}), which will be fainter because of the obscuration, and passive galaxies, which will be fainter in the rest-frame UV because of the lack of young stars. We expect a  bias in the high-redshift bins toward star-forming galaxies since at a given stellar mass and redshift, these are brighter than passive galaxies and will be preferentially included in magnitude-limited samples.

Another technical issue we need to consider is the effect of the errors on the photometric redshifts. Since there are far more low-redshift galaxies in our sample than high-redshift galaxies, the errors on the photometric redshifts will produce a  larger fraction of low-redshift interlopers in the high-redshift bins than high-redshift interlopers in the low-redshift bins. Low-redshift interlopers into high redshift stacks will bias the fluxes lower; see Figure \ref{fig:stackedflux}, where the average fluxes measured for the low redshift stacks are less than those measured for the high redshift stacks. Photometric redshift error will therefore tend to reduce the high redshift signal. However, as established in Section \ref{Sec:accuracyphotz}, although photometric redshift uncertainties increase at higher redshifts, particularly above z > 3, the photometric redshifts used in this study are reasonably accurate, so we do not anticipate low redshift interlopers into high redshift bins to be a significant issue affecting these results.

\begin{figure}
	\includegraphics[width=\columnwidth, trim = 0cm 0cm 0.8cm 0.8cm, clip]{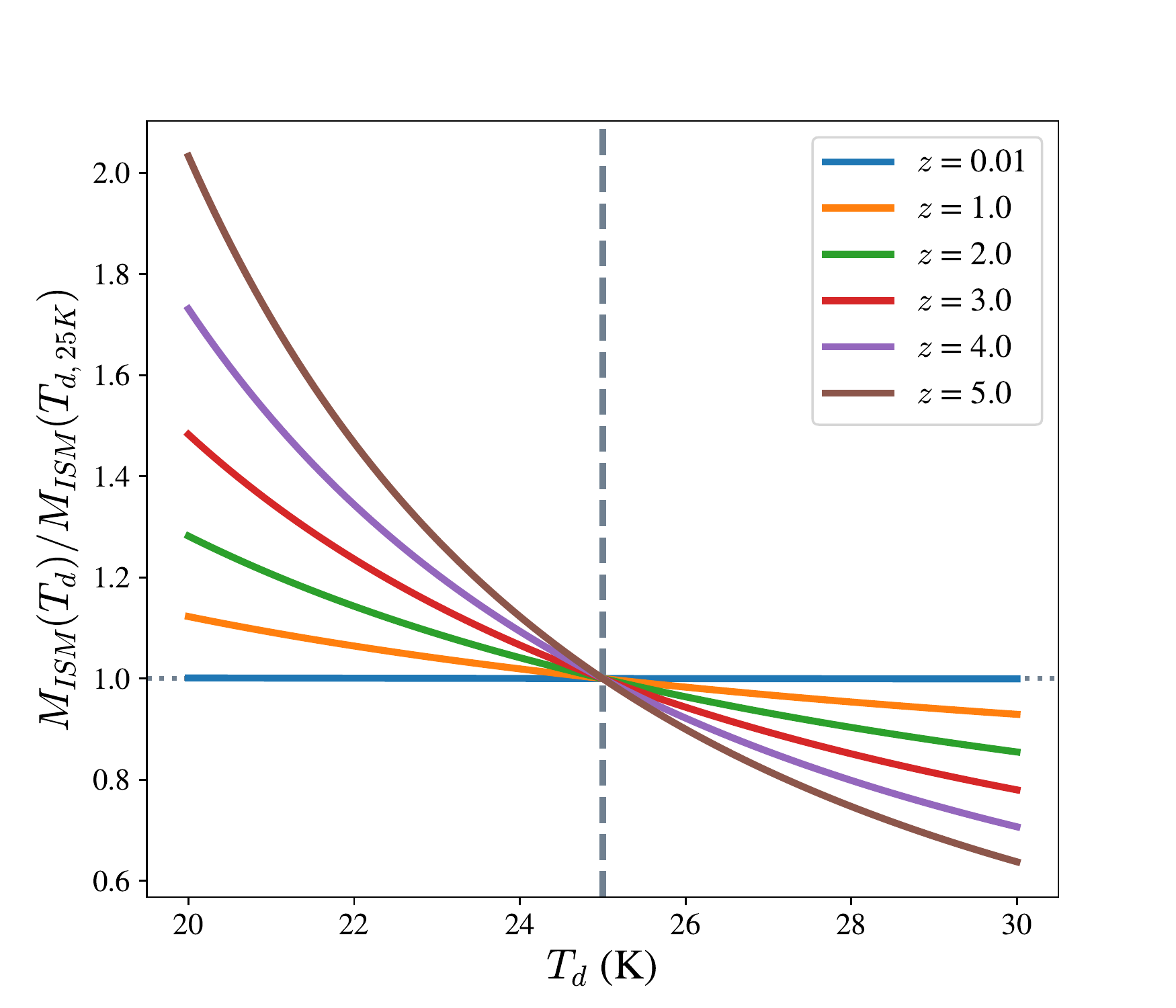}
    \caption{The relation between assumed mass-weighted dust temperature, $T_d$, and calculated ISM mass, $\rm M_{ISM}$, using emission observed at 850$\mu$m, for different discrete redshifts.}
    \label{fig:temp_variation}
\end{figure}

A fundamental assumption we have made is that the dust is at a constant temperature of $T_d$ = 25\,K.
A higher dust temperature would lead to a lower gas fraction calculated using Equations \ref{eq:ISMmasses} and \ref{eq:RJcorr}; Figure \ref{fig:temp_variation} illustrates how the calculated ISM mass for an artificial source changes with assumed dust temperature for different discrete redshifts, compared with the values calculated at 25\,K, when using dust emission detected at 850$\mu$m. The curved shape of the lines originates from the observation wavelength approaching the peak of the blackbody emission. Even at $z=3$, a change in mass-weighted dust temperature to 30\,K would only change the ISM mass by a factor of $\sim$20\%. Beyond this redshift, the reduction in ISM mass increases - however, the fluxes measured for sources at these higher redshifts are extremely uncertain. The flux uncertainties are, at the least, comparable to the uncertainty introduced in this method by the dust temperature assumption. We conclude that even if the dust temperature assumption of 25\,K is an underestimate of the mass-weighted dust temperature of galaxies, particularly at higher redshifts, the results of this study are robust against the error this assumption introduces - we will still see similar trends in the evolution of the gas masses fraction over cosmic time.

It is worth noting that in this work we are considering \textit{mass-weighted} dust temperature, which represents the temperature of the bulk of the ISM. Large variations in this temperature between galaxies, even at different epochs, are unlikely, since this mass-weighted dust temperature depends on the mean radiation energy density to the power of $\sim$1/6 \citep{Scoville2016}. Significant variations in the ISM environment over cosmic time would be required to instigate large mass-weighted dust temperature changes, and thus far, there is little evidence to support such variations. 

Recent simulations of $z = 2-6$ galaxies \citep{Liang2019} found that the mass-weighted dust temperature evolves little over these redshifts, and their results support adopting a value of 25\,K to estimate the gas mass of high redshift galaxies using long-wavelength emission on the RJ, as in this study.

A caveat affecting the determination of gas mass fraction from 850$\mu$m dust emission is the assumed CO-to-H$_2$ conversion factor (used when deriving Equation \ref{eq:ISMmasses}). \cite{Scoville2016} use a single value of $\alpha_{\textrm{CO}}$ based on a standard Galactic conversion factor $X_{\textrm{CO}} = 3 \times 10^{20}$ cm$^{-2}$(K km s$^{-1}$)$^{-1}$, and include a factor of 1.36 to account for the mass of heavy elements e.g. Helium. The exact value of $X_{\textrm{CO}}$ is debated, with values between 1 and $4 \times 10^{20}$ cm$^{-2}$(K km s$^{-1}$)$^{-1}$ often used in general galactic or extragalactic studies (\citealt{Bolatto2013}; and references therein).  Within this range, higher $X_{\textrm{CO}}$ values are appropriate for lower metallicity regions, and lower $X_{\textrm{CO}}$ values for starburst galaxies. Had \cite{Scoville2016} chosen to use a lower value of $X_{\textrm{CO}}$, the gas mass fractions determined here would be lower; similarly, a higher value of $X_{\textrm{CO}}$ would see our gas mass fractions increase. The trends from \cite{Scoville2017} (Figure \ref{fig:stackedmassfrac}) would also scale in a similar manner, since they use the same equations to derive their scaling relations. Since all our gas mass fraction calculations would be affected, our overall trends would not change, just the relative normalisation.

A final big caveat is that, like \cite{Scoville2016, Scoville2017}, we have made the assumption that the dust-to-gas ratio is independent of redshift. There is a lot of evidence that in the local universe, and above a transition metallicity (12 + log(O/H) $\simeq$ 8.0), the dust-to-gas ratio is proportional to the metallicity \citep{James2002, Sandstrom2013, RemyRuyer2014}. If this proportionality extends to all redshifts, we could, in principle, use metallicity measurements to correct the ISM measurements. This can have a large effect. For example, \cite{Tacconi2018} did decide to make corrections for metallicity. Their Equation 4 gives a relationship between metallicity and stellar mass and redshift. If we apply this equation to a galaxy at $z=3$ with a stellar mass of $log_{10}(M_*) = 11.0$ and make the assumption of the same proportionality between metallicity and dust-to-gas ratio that
exists at low redshift, we calculate that our ISM mass estimate would increase by
41\%, which would increase the gas fraction plotted in Figure \ref{fig:stackedmassfrac} by $\simeq$20\%. However, we are suspicious of making these corrections because of the assumptions required,
the uncertainties in the metallicity measurements, and the fact that we would have to apply these corrections to extreme high-redshift objects with large gas fractions. In a second paper, we sidestep this problem completely by using our measurements to estimate dust masses rather than ISM masses, which we then compare with predictions from IllustrisTNG \textit{(in prep.)}.

None of these potential problems therefore seem big enough to invalidate our basic result that there is strong evolution in the gas fraction of galaxies.

The evolution of HI and $\rm H_2$ gas mass with redshift are predicted to be very different, with the former only mildly evolving with redshift and the latter evolving by a factor of $\sim 7$ between $z=0$ and $z=3$ \citep{Lagos2011}. Their model predictions suggest that the $\rm H_2$ slightly dominates over HI at $z=2-5$ with HI dominating the cold gas mass at lower redshifts. With our method it is only possible to investigate the evolution of the molecular phase of the ISM. It is reassuring, both for our method and for the simulations, that we too find strong evolution over the redshift range $0<z<3$.

Our study and those of \cite{Scoville2017} and \cite{Tacconi2018} all find strong evolution in the gas fraction of galaxies. The samples of galaxies were selected in very different ways, with ours being based on stellar mass while the sample of Scoville et al. was selected in the far-infrared. Given the different selection methods and the other differences in the methodology, the agreement between the results from all three papers seems to us quite good (Figure \ref{fig:stackedmassfrac}). It is clear that galaxies at high redshift do have a much higher fraction of gas than those nearby. Our analysis has been based on a stacking analysis on the SCUBA-2 images. \cite{Dudzeviciute2020} have recently used the individual sources detected in another deep SCUBA-2 image to estimate the ISM mass function (the space-density of galaxies as a function of ISM mass), using the same basic method as ours of using the submillimetre continuum emission to estimate the mass of the ISM in a galaxy. They find strong evolution out to $z \sim 3$, which is broadly in agreement with the strong evolution we see in Figure \ref{fig:stackedmassfrac}.

Our results suggest the gas fraction increases rapidly with redshift until $z \sim 2-3$ and is then roughly constant (Figure \ref{fig:stackedmassfrac}). The precise form of this relationship seems to depend on stellar mass, with the galaxies with lower stellar masses reaching this plateau at a lower redshift than the galaxies in the higher stellar mass bins. This apparent maximum in the gas fraction, which has a value of $\sim0.5$ may be genuine or it may reflect a fundamental problem of our method - that we are relying on dust to trace the gas. The existence of dust relies on the existence of metals, and it is possible that the true gas fraction was higher at higher redshifts but not enough dust had been formed to trace the gas. Some evidence to support lack of dust in high redshift galaxies comes from the extremely blue rest-frame UV continuum slopes observed in Lyman break galaxies at $z>5$ \citep{Stanway2005,Wilkins2011}.  Therefore, the plateau in gas mass fraction that we see at the highest redshifts may actually be the result of less dust at the highest redshifts and evolution in the dust-to-gas ratio rather than evolution in the gas fraction. 

This fundamental ambiguity in this method, as well as the difficulty in modelling the selection biases in our sample and in the samples of \cite{Scoville2017} and \cite{Tacconi2018}, means that it is difficult to compare these results with the predictions of hydrodynamic \citep{Dave2020} and semi-analytic models \citep{Lilly2013, Peng2014} . We therefore suspect this method has gone as far as it can. In a second paper (Millard et al. {\it in preparation}), we return to using the submm emssion to estimate the mass of dust in a high-redshift galaxy rather than the mass of the ISM, a method which produces results which can relatively easily be compared with the predictions of cosmological hydrodynamic simulations.

\section{Conclusions}
In this work, we have derived estimates of average gas mass fractions for a large sample of stellar mass selected galaxies using long-wavelength dust continuum emission at a single wavelength. We find relationships between gas fraction, stellar mass, and redshift similar to ones found in previous investigations, which have been based on samples that are biased with large star-formation rates or ISM masses. At low redshifts, we find gas mass fractions much lower than found in studies of low-redshift galaxy samples, probably because our method is calibrated against ISM measurements that only include the molecular phase. We find that at low stellar masses ($\rm log(M_*/M_{\odot}) \leq 10.5$), the gas fraction increases with redshift, reaching a plateau at z$\sim$ 2.5-3, just before the peak in star-formation rate density in the Universe. At higher stellar masses ($\rm log(M_*/M_{\odot}) \geq 10.5$), the gas mass fraction increases with redshift, without clearly reaching a maximum, even out to $z = 3$. At each redshift, the galaxies with higher stellar masses have lower gas fractions. We show that star-forming galaxies are much more gas rich than their passive counterparts. We show that the results are robust against mass-weighted dust temperature variations.

\section*{Acknowledgements}
We thank Luke Davies and Simon Driver for providing the {\sc magphys} dataset on which this study is based. We thank Pieter De Vis for providing average gas fraction measurements for local galaxies. We thank Ian Smail for his comments at several stages of paper development, and also the anonymous referee for their comments and suggestions - both parties ultimately improved this manuscript.

The James Clerk Maxwell Telescope is operated by the East Asian Observatory on behalf of The National Astronomical Observatory of Japan; Academia Sinica Institute of Astronomy and Astrophysics; the Korea Astronomy and Space Science Institute; the Operation, Maintenance and Upgrading Fund for Astronomical Telescopes and Facility Instruments, budgeted from the Ministry of Finance (MOF) of China and administrated by the Chinese Academy of Sciences (CAS), as well as the National Key R\&D Program of China (No. 2017YFA0402700). Additional funding support is provided by the Science and Technology Facilities Council of the United Kingdom and participating universities in the United Kingdom and Canada (ST/M007634/1, ST/M003019/1, ST/N005856/1). The James Clerk Maxwell Telescope has historically been operated by the Joint Astronomy Centre on behalf of the Science and Technology Facilities Council of the United Kingdom, the National Research Council of Canada and the Netherlands Organization for Scientific Research and data from observations undertaken during this period of operation is used in this manuscript. This research used the facilities of the Canadian Astronomy Data Centre operated by the National Research Council of Canada with the support of the Canadian Space Agency. The data used in this work were taken as part of Program ID M16AL002.

JSM, HLG and RAB acknowledge support from the European Research Council (ERC) in the form of Consolidator Grant {\sc CosmicDust} (ERC-2014-CoG-647939). KM has been supported by the National Science Centre (grant UMO-2013/09/D/ST9/04030). YP acknowledges National Key R\&D Program of China Grant 2016YFA0400702 and NSFC Grant No. 11773001. MJM acknowledges the support of the National Science Centre, Poland through the SONATA BIS grant 2018/30/E/ST9/00208.




\bibliographystyle{mnras}
\bibliography{gasref} 

\begin{thebibliography}{}
\makeatletter
\relax
\def\mn@urlcharsother{\let\do\@makeother \do\$\do\&\do\#\do\^\do\_\do\%\do\~}
\def\mn@doi{\begingroup\mn@urlcharsother \@ifnextchar [ {\mn@doi@}
  {\mn@doi@[]}}
\def\mn@doi@[#1]#2{\def\@tempa{#1}\ifx\@tempa\@empty \href
  {http://dx.doi.org/#2} {doi:#2}\else \href {http://dx.doi.org/#2} {#1}\fi
  \endgroup}
\def\mn@eprint#1#2{\mn@eprint@#1:#2::\@nil}
\def\mn@eprint@arXiv#1{\href {http://arxiv.org/abs/#1} {{\tt arXiv:#1}}}
\def\mn@eprint@dblp#1{\href {http://dblp.uni-trier.de/rec/bibtex/#1.xml}
  {dblp:#1}}
\def\mn@eprint@#1:#2:#3:#4\@nil{\def\@tempa {#1}\def\@tempb {#2}\def\@tempc
  {#3}\ifx \@tempc \@empty \let \@tempc \@tempb \let \@tempb \@tempa \fi \ifx
  \@tempb \@empty \def\@tempb {arXiv}\fi \@ifundefined
  {mn@eprint@\@tempb}{\@tempb:\@tempc}{\expandafter \expandafter \csname
  mn@eprint@\@tempb\endcsname \expandafter{\@tempc}}}

\bibitem[\protect\citeauthoryear{{Abdurro'uf}}{{Abdurro'uf}}{2018}]{Abdurrouf2018}
{Abdurro'uf} Masayuki A.,  2018, \mn@doi [\mnras] {10.1093/mnras/sty1771},
  \href {https://ui.adsabs.harvard.edu/abs/2018MNRAS.479.5083A} {479, 5083}

\bibitem[\protect\citeauthoryear{{Ahn} et~al.,}{{Ahn} et~al.}{2014}]{Ahn2014}
{Ahn} C.~P.,  et~al., 2014, \mn@doi [The Astrophysical Journal Supplement
  Series] {10.1088/0067-0049/211/2/17}, \href
  {https://ui.adsabs.harvard.edu/\#abs/2014ApJS..211...17A} {211, 17}

\bibitem[\protect\citeauthoryear{{Albareti} et~al.,}{{Albareti}
  et~al.}{2017}]{SDSS13}
{Albareti} F.~D.,  et~al., 2017, \mn@doi [\apjs] {10.3847/1538-4365/aa8992},
  \href {http://adsabs.harvard.edu/abs/2017ApJS..233...25A} {233, 25}

\bibitem[\protect\citeauthoryear{{Andrews}, {Driver}, {Davies}, {Kafle},
  {Robotham}  \& {Wright}}{{Andrews} et~al.}{2017}]{Andrews2017}
{Andrews} S.~K.,  {Driver} S.~P.,  {Davies} L.~J.~M.,  {Kafle} P.~R.,
  {Robotham} A. S.~G.,   {Wright} A.~H.,  2017, \mn@doi [\mnras]
  {10.1093/mnras/stw2395}, \href
  {https://ui.adsabs.harvard.edu/\#abs/2017MNRAS.464.1569A} {464, 1569}

\bibitem[\protect\citeauthoryear{{Astropy Collaboration} et~al.,}{{Astropy
  Collaboration} et~al.}{2013}]{Astropy2013}
{Astropy Collaboration} et~al., 2013, \mn@doi [\aap]
  {10.1051/0004-6361/201322068}, \href
  {http://adsabs.harvard.edu/abs/2013A%26A...558A..33A} {558, A33}

\bibitem[\protect\citeauthoryear{{Bertin} \& {Arnouts}}{{Bertin} \&
  {Arnouts}}{1996}]{Bertin1996}
{Bertin} E.,  {Arnouts} S.,  1996, \mn@doi [\aaps] {10.1051/aas:1996164}, \href
  {https://ui.adsabs.harvard.edu/abs/1996A%26AS..117..393B} {117, 393}

\bibitem[\protect\citeauthoryear{{B{\'e}thermin} et~al.,}{{B{\'e}thermin}
  et~al.}{2015}]{Bethermin2015}
{B{\'e}thermin} M.,  et~al., 2015, \mn@doi [\aap]
  {10.1051/0004-6361/201425031}, \href
  {https://ui.adsabs.harvard.edu/abs/2015A&A...573A.113B} {573, A113}

\bibitem[\protect\citeauthoryear{{Bolatto}, {Wolfire}  \& {Leroy}}{{Bolatto}
  et~al.}{2013}]{Bolatto2013}
{Bolatto} A.~D.,  {Wolfire} M.,   {Leroy} A.~K.,  2013, \mn@doi [\araa]
  {10.1146/annurev-astro-082812-140944}, \href
  {https://ui.adsabs.harvard.edu/abs/2013ARA&A..51..207B} {51, 207}

\bibitem[\protect\citeauthoryear{{Boquien}, {Burgarella}, {Roehlly}, {Buat},
  {Ciesla}, {Corre}, {Inoue}  \& {Salas}}{{Boquien} et~al.}{2019}]{Boquien2019}
{Boquien} M.,  {Burgarella} D.,  {Roehlly} Y.,  {Buat} V.,  {Ciesla} L.,
  {Corre} D.,  {Inoue} A.~K.,   {Salas} H.,  2019, \mn@doi [\aap]
  {10.1051/0004-6361/201834156}, \href
  {https://ui.adsabs.harvard.edu/abs/2019A&A...622A.103B} {622, A103}

\bibitem[\protect\citeauthoryear{{Bothwell} et~al.,}{{Bothwell}
  et~al.}{2013}]{Bothwell2013}
{Bothwell} M.~S.,  et~al., 2013, \mn@doi [\mnras] {10.1093/mnras/sts562}, \href
  {https://ui.adsabs.harvard.edu/abs/2013MNRAS.429.3047B} {429, 3047}

\bibitem[\protect\citeauthoryear{{Bruzual} \& {Charlot}}{{Bruzual} \&
  {Charlot}}{2003}]{BruzualCharlot2003}
{Bruzual} G.,  {Charlot} S.,  2003, \mn@doi [\mnras]
  {10.1046/j.1365-8711.2003.06897.x}, \href
  {http://adsabs.harvard.edu/abs/2003MNRAS.344.1000B} {344, 1000}

\bibitem[\protect\citeauthoryear{{Ca{\~n}ameras} et~al.,}{{Ca{\~n}ameras}
  et~al.}{2018}]{Canameras2018}
{Ca{\~n}ameras} R.,  et~al., 2018, \mn@doi [\aap]
  {10.1051/0004-6361/201833625}, \href
  {https://ui.adsabs.harvard.edu/abs/2018A&A...620A..61C} {620, A61}

\bibitem[\protect\citeauthoryear{{Calistro Rivera} et~al.,}{{Calistro Rivera}
  et~al.}{2018}]{Rivera2018}
{Calistro Rivera} G.,  et~al., 2018, \mn@doi [\apj] {10.3847/1538-4357/aacffa},
  \href {https://ui.adsabs.harvard.edu/abs/2018ApJ...863...56C} {863, 56}

\bibitem[\protect\citeauthoryear{{Carilli} \& {Walter}}{{Carilli} \&
  {Walter}}{2013}]{Carilli2013}
{Carilli} C.~L.,  {Walter} F.,  2013, \mn@doi [\araa]
  {10.1146/annurev-astro-082812-140953}, \href
  {https://ui.adsabs.harvard.edu/abs/2013ARA&A..51..105C} {51, 105}

\bibitem[\protect\citeauthoryear{{Casasola} et~al.,}{{Casasola}
  et~al.}{2020}]{Casasola2020}
{Casasola} V.,  et~al., 2020, \mn@doi [\aap] {10.1051/0004-6361/201936665},
  \href {https://ui.adsabs.harvard.edu/abs/2020A&A...633A.100C} {633, A100}

\bibitem[\protect\citeauthoryear{{Casey} et~al.,}{{Casey}
  et~al.}{2011}]{Casey2011}
{Casey} C.~M.,  et~al., 2011, \mn@doi [\mnras]
  {10.1111/j.1365-2966.2011.18885.x}, \href
  {https://ui.adsabs.harvard.edu/abs/2011MNRAS.415.2723C} {415, 2723}

\bibitem[\protect\citeauthoryear{{Catinella} et~al.,}{{Catinella}
  et~al.}{2010}]{Catinella2010}
{Catinella} B.,  et~al., 2010, \mn@doi [\mnras]
  {10.1111/j.1365-2966.2009.16180.x}, \href
  {https://ui.adsabs.harvard.edu/abs/2010MNRAS.403..683C} {403, 683}

\bibitem[\protect\citeauthoryear{{Chabrier}}{{Chabrier}}{2003}]{Chabrier2003}
{Chabrier} G.,  2003, \mn@doi [\pasp] {10.1086/376392}, \href
  {http://adsabs.harvard.edu/abs/2003PASP..115..763C} {115, 763}

\bibitem[\protect\citeauthoryear{{Charlot} \& {Fall}}{{Charlot} \&
  {Fall}}{2000}]{CharlotFall2000}
{Charlot} S.,  {Fall} S.~M.,  2000, \mn@doi [\apj] {10.1086/309250}, \href
  {http://adsabs.harvard.edu/abs/2000ApJ...539..718C} {539, 718}

\bibitem[\protect\citeauthoryear{{Ciesla} et~al.,}{{Ciesla}
  et~al.}{2015}]{Ciesla2015}
{Ciesla} L.,  et~al., 2015, \mn@doi [\aap] {10.1051/0004-6361/201425252}, \href
  {https://ui.adsabs.harvard.edu/abs/2015A&A...576A..10C} {576, A10}

\bibitem[\protect\citeauthoryear{{Ciesla} et~al.,}{{Ciesla}
  et~al.}{2016}]{Ciesla2016}
{Ciesla} L.,  et~al., 2016, \mn@doi [\aap] {10.1051/0004-6361/201527107}, \href
  {https://ui.adsabs.harvard.edu/\#abs/2016A&A...585A..43C} {585, A43}

\bibitem[\protect\citeauthoryear{{Civano} et~al.,}{{Civano}
  et~al.}{2016}]{Civano2016}
{Civano} F.,  et~al., 2016, \mn@doi [\apj] {10.3847/0004-637X/819/1/62}, \href
  {https://ui.adsabs.harvard.edu/abs/2016ApJ...819...62C} {819, 62}

\bibitem[\protect\citeauthoryear{{Clark} et~al.,}{{Clark}
  et~al.}{2015}]{Clark2015}
{Clark} C.~J.~R.,  et~al., 2015, \mn@doi [\mnras] {10.1093/mnras/stv1276},
  \href {http://adsabs.harvard.edu/abs/2015MNRAS.452..397C} {452, 397}

\bibitem[\protect\citeauthoryear{{Clark}, {Schofield}, {Gomez}  \&
  {Davies}}{{Clark} et~al.}{2016}]{Clark2016}
{Clark} C.~J.~R.,  {Schofield} S.~P.,  {Gomez} H.~L.,   {Davies} J.~I.,  2016,
  \mn@doi [\mnras] {10.1093/mnras/stw647}, \href
  {http://cdsads.u-strasbg.fr/abs/2016MNRAS.459.1646C} {459, 1646}

\bibitem[\protect\citeauthoryear{{Combes}}{{Combes}}{2018}]{Combes2018}
{Combes} F.,  2018, \mn@doi [\aapr] {10.1007/s00159-018-0110-4}, \href
  {https://ui.adsabs.harvard.edu/abs/2018A&ARv..26....5C} {26, 5}

\bibitem[\protect\citeauthoryear{{Cool} et~al.,}{{Cool}
  et~al.}{2013}]{Cool2013}
{Cool} R.~J.,  et~al., 2013, \mn@doi [\apj] {10.1088/0004-637X/767/2/118},
  \href {https://ui.adsabs.harvard.edu/\#abs/2013ApJ...767..118C} {767, 118}

\bibitem[\protect\citeauthoryear{{Coppin} et~al.,}{{Coppin}
  et~al.}{2009}]{Coppin2009}
{Coppin} K.~E.~K.,  et~al., 2009, \mn@doi [\mnras]
  {10.1111/j.1365-2966.2009.14700.x}, \href
  {https://ui.adsabs.harvard.edu/abs/2009MNRAS.395.1905C} {395, 1905}

\bibitem[\protect\citeauthoryear{{Daddi} et~al.,}{{Daddi}
  et~al.}{2007}]{Daddi2007}
{Daddi} E.,  et~al., 2007, \mn@doi [\apj] {10.1086/521818}, \href
  {https://ui.adsabs.harvard.edu/abs/2007ApJ...670..156D} {670, 156}

\bibitem[\protect\citeauthoryear{{Dav{\'e}}, {Crain}, {Stevens}, {Narayanan},
  {Saintonge}, {Catinella}  \& {Cortese}}{{Dav{\'e}} et~al.}{2020}]{Dave2020}
{Dav{\'e}} R.,  {Crain} R.~A.,  {Stevens} A. R.~H.,  {Narayanan} D.,
  {Saintonge} A.,  {Catinella} B.,   {Cortese} L.,  2020, arXiv e-prints, \href
  {https://ui.adsabs.harvard.edu/abs/2020arXiv200207226D} {p. arXiv:2002.07226}

\bibitem[\protect\citeauthoryear{{Davies} et~al.,}{{Davies}
  et~al.}{2015}]{Davies2015}
{Davies} L.~J.~M.,  et~al., 2015, \mn@doi [\mnras] {10.1093/mnras/stu2515},
  \href {https://ui.adsabs.harvard.edu/\#abs/2015MNRAS.447.1014D} {447, 1014}

\bibitem[\protect\citeauthoryear{{Davies} et~al.,}{{Davies}
  et~al.}{2017}]{Davies2017}
{Davies} J.~I.,  et~al., 2017, \mn@doi [\pasp]
  {10.1088/1538-3873/129/974/044102}, \href
  {https://ui.adsabs.harvard.edu/abs/2017PASP..129d4102D} {129, 044102}

\bibitem[\protect\citeauthoryear{{De Vis} et~al.,}{{De Vis}
  et~al.}{2017a}]{DeVis2017a}
{De Vis} P.,  et~al., 2017a, \mn@doi [\mnras] {10.1093/mnras/stw2501}, \href
  {http://adsabs.harvard.edu/abs/2017MNRAS.464.4680D} {464, 4680}

\bibitem[\protect\citeauthoryear{{De Vis} et~al.,}{{De Vis}
  et~al.}{2017b}]{DeVis2017b}
{De Vis} P.,  et~al., 2017b, \mn@doi [\mnras] {10.1093/mnras/stx981}, \href
  {https://ui.adsabs.harvard.edu/abs/2017MNRAS.471.1743D} {471, 1743}

\bibitem[\protect\citeauthoryear{{De Vis} et~al.,}{{De Vis}
  et~al.}{2019}]{DeVis2019}
{De Vis} P.,  et~al., 2019, \mn@doi [\aap] {10.1051/0004-6361/201834444}, \href
  {https://ui.adsabs.harvard.edu/abs/2019A&A...623A...5D} {623, A5}

\bibitem[\protect\citeauthoryear{{Decarli} et~al.,}{{Decarli}
  et~al.}{2016}]{DeCarli2016}
{Decarli} R.,  et~al., 2016, \mn@doi [\apj] {10.3847/1538-4357/833/1/70}, \href
  {https://ui.adsabs.harvard.edu/abs/2016ApJ...833...70D} {833, 70}

\bibitem[\protect\citeauthoryear{{Dempsey} et~al.,}{{Dempsey}
  et~al.}{2013}]{Dempsey2013}
{Dempsey} J.~T.,  et~al., 2013, \mn@doi [\mnras] {10.1093/mnras/stt090}, \href
  {https://ui.adsabs.harvard.edu/\#abs/2013MNRAS.430.2534D} {430, 2534}

\bibitem[\protect\citeauthoryear{{Donley} et~al.,}{{Donley}
  et~al.}{2012}]{Donley2012}
{Donley} J.~L.,  et~al., 2012, \mn@doi [\apj] {10.1088/0004-637X/748/2/142},
  \href {https://ui.adsabs.harvard.edu/abs/2012ApJ...748..142D} {748, 142}

\bibitem[\protect\citeauthoryear{{Draine} \& {Li}}{{Draine} \&
  {Li}}{2007}]{DraineLi2007}
{Draine} B.~T.,  {Li} A.,  2007, \mn@doi [\apj] {10.1086/511055}, \href
  {https://ui.adsabs.harvard.edu/\#abs/2007ApJ...657..810D} {657, 810}

\bibitem[\protect\citeauthoryear{{Driver} et~al.,}{{Driver}
  et~al.}{2018}]{Driver2018}
{Driver} S.~P.,  et~al., 2018, \mn@doi [\mnras] {10.1093/mnras/stx2728}, \href
  {https://ui.adsabs.harvard.edu/\#abs/2018MNRAS.475.2891D} {475, 2891}

\bibitem[\protect\citeauthoryear{{Dudzevi{\v{c}}i{\={u}}t{\.{e}}}
  et~al.,}{{Dudzevi{\v{c}}i{\={u}}t{\.{e}}} et~al.}{2019}]{Dudzeviciute2020}
{Dudzevi{\v{c}}i{\={u}}t{\.{e}}} U.,  et~al., 2019, arXiv e-prints, \href
  {https://ui.adsabs.harvard.edu/abs/2019arXiv191007524D} {p. arXiv:1910.07524}

\bibitem[\protect\citeauthoryear{{Duncan} et~al.,}{{Duncan}
  et~al.}{2018a}]{Duncan2018a}
{Duncan} K.~J.,  et~al., 2018a, \mn@doi [\mnras] {10.1093/mnras/stx2536}, \href
  {https://ui.adsabs.harvard.edu/\#abs/2018MNRAS.473.2655D} {473, 2655}

\bibitem[\protect\citeauthoryear{{Duncan}, {Jarvis}, {Brown}  \&
  {R{\"o}ttgering}}{{Duncan} et~al.}{2018b}]{Duncan2018b}
{Duncan} K.~J.,  {Jarvis} M.~J.,  {Brown} M. J.~I.,   {R{\"o}ttgering} H.
  J.~A.,  2018b, \mn@doi [\mnras] {10.1093/mnras/sty940}, \href
  {https://ui.adsabs.harvard.edu/\#abs/2018MNRAS.477.5177D} {477, 5177}

\bibitem[\protect\citeauthoryear{{Dunne} \& {Eales}}{{Dunne} \&
  {Eales}}{2001}]{DunneEales2001}
{Dunne} L.,  {Eales} S.~A.,  2001, \mn@doi [\mnras]
  {10.1046/j.1365-8711.2001.04789.x}, \href
  {https://ui.adsabs.harvard.edu/abs/2001MNRAS.327..697D} {327, 697}

\bibitem[\protect\citeauthoryear{{Dunne} et~al.,}{{Dunne}
  et~al.}{2011}]{Dunne2011}
{Dunne} L.,  et~al., 2011, \mn@doi [\mnras] {10.1111/j.1365-2966.2011.19363.x},
  \href {https://ui.adsabs.harvard.edu/abs/2011MNRAS.417.1510D} {417, 1510}

\bibitem[\protect\citeauthoryear{{Eales}, {Wynn-Williams}  \& {Duncan}}{{Eales}
  et~al.}{1989}]{EalesWW1989}
{Eales} S.~A.,  {Wynn-Williams} C.~G.,   {Duncan} W.~D.,  1989, \mn@doi [\apj]
  {10.1086/167341}, \href
  {https://ui.adsabs.harvard.edu/abs/1989ApJ...339..859E} {339, 859}

\bibitem[\protect\citeauthoryear{{Eales} et~al.,}{{Eales}
  et~al.}{2012}]{Eales2012}
{Eales} S.,  et~al., 2012, \mn@doi [\apj] {10.1088/0004-637X/761/2/168}, \href
  {http://adsabs.harvard.edu/abs/2012ApJ...761..168E} {761, 168}

\bibitem[\protect\citeauthoryear{{Elbaz} et~al.,}{{Elbaz}
  et~al.}{2007}]{Elbaz2007}
{Elbaz} D.,  et~al., 2007, \mn@doi [\aap] {10.1051/0004-6361:20077525}, \href
  {https://ui.adsabs.harvard.edu/abs/2007A&A...468...33E} {468, 33}

\bibitem[\protect\citeauthoryear{{Fritz}, {Franceschini}  \&
  {Hatziminaoglou}}{{Fritz} et~al.}{2006}]{Fritz2006}
{Fritz} J.,  {Franceschini} A.,   {Hatziminaoglou} E.,  2006, \mn@doi [\mnras]
  {10.1111/j.1365-2966.2006.09866.x}, \href
  {https://ui.adsabs.harvard.edu/\#abs/2006MNRAS.366..767F} {366, 767}

\bibitem[\protect\citeauthoryear{{Geach} et~al.,}{{Geach}
  et~al.}{2017}]{Geach2017}
{Geach} J.~E.,  et~al., 2017, \mn@doi [\mnras] {10.1093/mnras/stw2721}, \href
  {https://ui.adsabs.harvard.edu/abs/2017MNRAS.465.1789G} {465, 1789}

\bibitem[\protect\citeauthoryear{{Genzel} et~al.,}{{Genzel}
  et~al.}{2015}]{Genzel2015}
{Genzel} R.,  et~al., 2015, \mn@doi [\apj] {10.1088/0004-637X/800/1/20}, \href
  {http://adsabs.harvard.edu/abs/2015ApJ...800...20G} {800, 20}

\bibitem[\protect\citeauthoryear{{Greve} et~al.,}{{Greve}
  et~al.}{2005}]{Greve2005}
{Greve} T.~R.,  et~al., 2005, \mn@doi [\mnras]
  {10.1111/j.1365-2966.2005.08979.x}, \href
  {https://ui.adsabs.harvard.edu/abs/2005MNRAS.359.1165G} {359, 1165}

\bibitem[\protect\citeauthoryear{{Harrington} et~al.,}{{Harrington}
  et~al.}{2018}]{Harrington2018}
{Harrington} K.~C.,  et~al., 2018, \mn@doi [\mnras] {10.1093/mnras/stx3043},
  \href {https://ui.adsabs.harvard.edu/abs/2018MNRAS.474.3866H} {474, 3866}

\bibitem[\protect\citeauthoryear{Hoaglin, Mosteller  \& Tukey}{Hoaglin
  et~al.}{1983}]{Hoaglin1983}
Hoaglin D.,  Mosteller F.,   Tukey J.,  1983, Understanding robust and
  exploratory data analysis.
Wiley series in probability and mathematical statistics: Applied probability
  and statistics, Wiley

\bibitem[\protect\citeauthoryear{{Holland} et~al.,}{{Holland}
  et~al.}{2013}]{Holland2013}
{Holland} W.~S.,  et~al., 2013, \mn@doi [\mnras] {10.1093/mnras/sts612}, \href
  {https://ui.adsabs.harvard.edu/abs/2013MNRAS.430.2513H} {430, 2513}

\bibitem[\protect\citeauthoryear{{Hunt} et~al.,}{{Hunt}
  et~al.}{2019}]{Hunt2019}
{Hunt} L.~K.,  et~al., 2019, \mn@doi [\aap] {10.1051/0004-6361/201834212},
  \href {https://ui.adsabs.harvard.edu/\#abs/2019A&A...621A..51H} {621, A51}

\bibitem[\protect\citeauthoryear{{Hurley} et~al.,}{{Hurley}
  et~al.}{2017}]{Hurley2017}
{Hurley} P.~D.,  et~al., 2017, \mn@doi [\mnras] {10.1093/mnras/stw2375}, \href
  {http://adsabs.harvard.edu/abs/2017MNRAS.464..885H} {464, 885}

\bibitem[\protect\citeauthoryear{{Imara}, {Loeb}, {Johnson}, {Conroy}  \&
  {Behroozi}}{{Imara} et~al.}{2018}]{Imara2018}
{Imara} N.,  {Loeb} A.,  {Johnson} B.~D.,  {Conroy} C.,   {Behroozi} P.,  2018,
  \mn@doi [\apj] {10.3847/1538-4357/aaa3f0}, \href
  {http://adsabs.harvard.edu/abs/2018ApJ...854...36I} {854, 36}

\bibitem[\protect\citeauthoryear{{James}, {Dunne}, {Eales}  \&
  {Edmunds}}{{James} et~al.}{2002}]{James2002}
{James} A.,  {Dunne} L.,  {Eales} S.,   {Edmunds} M.~G.,  2002, \mn@doi
  [\mnras] {10.1046/j.1365-8711.2002.05660.x}, \href
  {https://ui.adsabs.harvard.edu/abs/2002MNRAS.335..753J} {335, 753}

\bibitem[\protect\citeauthoryear{{Karim} et~al.,}{{Karim}
  et~al.}{2011}]{Karim2011}
{Karim} A.,  et~al., 2011, \mn@doi [\apj] {10.1088/0004-637X/730/2/61}, \href
  {https://ui.adsabs.harvard.edu/abs/2011ApJ...730...61K} {730, 61}

\bibitem[\protect\citeauthoryear{{Lagos}, {Baugh}, {Lacey}, {Benson}, {Kim}  \&
  {Power}}{{Lagos} et~al.}{2011}]{Lagos2011}
{Lagos} C.~D.~P.,  {Baugh} C.~M.,  {Lacey} C.~G.,  {Benson} A.~J.,  {Kim}
  H.-S.,   {Power} C.,  2011, \mn@doi [\mnras]
  {10.1111/j.1365-2966.2011.19583.x}, \href
  {https://ui.adsabs.harvard.edu/abs/2011MNRAS.418.1649L} {418, 1649}

\bibitem[\protect\citeauthoryear{{Laigle} et~al.,}{{Laigle}
  et~al.}{2016}]{Laigle2016}
{Laigle} C.,  et~al., 2016, \mn@doi [The Astrophysical Journal Supplement
  Series] {10.3847/0067-0049/224/2/24}, \href
  {https://ui.adsabs.harvard.edu/\#abs/2016ApJS..224...24L} {224, 24}

\bibitem[\protect\citeauthoryear{{Lang} et~al.,}{{Lang}
  et~al.}{2019}]{Lang2019}
{Lang} P.,  et~al., 2019, \mn@doi [\apj] {10.3847/1538-4357/ab1f77}, \href
  {https://ui.adsabs.harvard.edu/abs/2019ApJ...879...54L} {879, 54}

\bibitem[\protect\citeauthoryear{{Le F{\`e}vre} et~al.,}{{Le F{\`e}vre}
  et~al.}{2013}]{LeFevre2013}
{Le F{\`e}vre} O.,  et~al., 2013, \mn@doi [\aap] {10.1051/0004-6361/201322179},
  \href {https://ui.adsabs.harvard.edu/\#abs/2013A&A...559A..14L} {559, A14}

\bibitem[\protect\citeauthoryear{{Lee} et~al.,}{{Lee} et~al.}{2015}]{Lee2015}
{Lee} N.,  et~al., 2015, \mn@doi [\apj] {10.1088/0004-637X/801/2/80}, \href
  {https://ui.adsabs.harvard.edu/abs/2015ApJ...801...80L} {801, 80}

\bibitem[\protect\citeauthoryear{{Leroy} et~al.,}{{Leroy}
  et~al.}{2011}]{Leroy2011}
{Leroy} A.~K.,  et~al., 2011, \mn@doi [\apj] {10.1088/0004-637X/737/1/12},
  \href {https://ui.adsabs.harvard.edu/abs/2011ApJ...737...12L} {737, 12}

\bibitem[\protect\citeauthoryear{{Liang} et~al.,}{{Liang}
  et~al.}{2019}]{Liang2019}
{Liang} L.,  et~al., 2019, \mn@doi [\mnras] {10.1093/mnras/stz2134}, \href
  {https://ui.adsabs.harvard.edu/abs/2019MNRAS.tmp.2072L} {p.~2072}

\bibitem[\protect\citeauthoryear{{Lilly} et~al.,}{{Lilly}
  et~al.}{2007}]{Lilly2007zbright}
{Lilly} S.~J.,  et~al., 2007, \mn@doi [The Astrophysical Journal Supplement
  Series] {10.1086/516589}, \href
  {https://ui.adsabs.harvard.edu/\#abs/2007ApJS..172...70L} {172, 70}

\bibitem[\protect\citeauthoryear{{Lilly}, {Carollo}, {Pipino}, {Renzini}  \&
  {Peng}}{{Lilly} et~al.}{2013}]{Lilly2013}
{Lilly} S.~J.,  {Carollo} C.~M.,  {Pipino} A.,  {Renzini} A.,   {Peng} Y.,
  2013, \mn@doi [\apj] {10.1088/0004-637X/772/2/119}, \href
  {https://ui.adsabs.harvard.edu/abs/2013ApJ...772..119L} {772, 119}

\bibitem[\protect\citeauthoryear{{Madau} \& {Dickinson}}{{Madau} \&
  {Dickinson}}{2014}]{Madau2014}
{Madau} P.,  {Dickinson} M.,  2014, \mn@doi [Annual Review of Astronomy and
  Astrophysics] {10.1146/annurev-astro-081811-125615}, \href
  {https://ui.adsabs.harvard.edu/\#abs/2014ARA&A..52..415M} {52, 415}

\bibitem[\protect\citeauthoryear{{Magnelli} et~al.,}{{Magnelli}
  et~al.}{2014}]{Magnelli2014}
{Magnelli} B.,  et~al., 2014, \mn@doi [\aap] {10.1051/0004-6361/201322217},
  \href {https://ui.adsabs.harvard.edu/abs/2014A&A...561A..86M} {561, A86}

\bibitem[\protect\citeauthoryear{{Ma{\l}ek} et~al.,}{{Ma{\l}ek}
  et~al.}{2018}]{Malek2018}
{Ma{\l}ek} K.,  et~al., 2018, \mn@doi [\aap] {10.1051/0004-6361/201833131},
  \href {http://adsabs.harvard.edu/abs/2018A%26A...620A..50M} {620, A50}

\bibitem[\protect\citeauthoryear{{Malek}, {Buat}, {Burgarella}, {Roehlly},
  {Shirley}  \& {the HELP team}}{{Malek} et~al.}{2019}]{Malek2019short}
{Malek} K.,  {Buat} V.,  {Burgarella} D.,  {Roehlly} Y.,  {Shirley} R.,   {the
  HELP team} 2019, arXiv e-prints, \href
  {https://ui.adsabs.harvard.edu/abs/2019arXiv190412498M} {p. arXiv:1904.12498}

\bibitem[\protect\citeauthoryear{{Marchesi} et~al.,}{{Marchesi}
  et~al.}{2016}]{Marchesi2016}
{Marchesi} S.,  et~al., 2016, \mn@doi [\apj] {10.3847/0004-637X/817/1/34},
  \href {https://ui.adsabs.harvard.edu/abs/2016ApJ...817...34M} {817, 34}

\bibitem[\protect\citeauthoryear{{McCracken} et~al.,}{{McCracken}
  et~al.}{2010}]{McCracken2010}
{McCracken} H.~J.,  et~al., 2010, \mn@doi [\apj] {10.1088/0004-637X/708/1/202},
  \href {https://ui.adsabs.harvard.edu/abs/2010ApJ...708..202M} {708, 202}

\bibitem[\protect\citeauthoryear{{McCracken} et~al.,}{{McCracken}
  et~al.}{2012}]{McCracken2012}
{McCracken} H.~J.,  et~al., 2012, \mn@doi [\aap] {10.1051/0004-6361/201219507},
  \href {https://ui.adsabs.harvard.edu/\#abs/2012A&A...544A.156M} {544, A156}

\bibitem[\protect\citeauthoryear{{Micha{\l}owski} et~al.,}{{Micha{\l}owski}
  et~al.}{2017}]{Michalowski2017}
{Micha{\l}owski} M.~J.,  et~al., 2017, \mn@doi [\mnras] {10.1093/mnras/stx861},
  \href {https://ui.adsabs.harvard.edu/abs/2017MNRAS.469..492M} {469, 492}

\bibitem[\protect\citeauthoryear{{Nguyen} et~al.,}{{Nguyen}
  et~al.}{2010}]{Nguyen2010}
{Nguyen} H.~T.,  et~al., 2010, \mn@doi [\aap] {10.1051/0004-6361/201014680},
  \href {https://ui.adsabs.harvard.edu/\#abs/2010A&A...518L...5N} {518, L5}

\bibitem[\protect\citeauthoryear{{Noeske} et~al.,}{{Noeske}
  et~al.}{2007}]{Noeske2007}
{Noeske} K.~G.,  et~al., 2007, \mn@doi [\apj] {10.1086/517926}, \href
  {https://ui.adsabs.harvard.edu/abs/2007ApJ...660L..43N} {660, L43}

\bibitem[\protect\citeauthoryear{{Noll}, {Burgarella}, {Giovannoli}, {Buat},
  {Marcillac}  \& {Mu{\~n}oz-Mateos}}{{Noll} et~al.}{2009}]{Noll2009}
{Noll} S.,  {Burgarella} D.,  {Giovannoli} E.,  {Buat} V.,  {Marcillac} D.,
  {Mu{\~n}oz-Mateos} J.~C.,  2009, \mn@doi [\aap]
  {10.1051/0004-6361/200912497}, \href
  {https://ui.adsabs.harvard.edu/\#abs/2009A&A...507.1793N} {507, 1793}

\bibitem[\protect\citeauthoryear{{Pearson} et~al.,}{{Pearson}
  et~al.}{2013}]{Pearson2013}
{Pearson} E.~A.,  et~al., 2013, \mn@doi [\mnras] {10.1093/mnras/stt1369}, \href
  {https://ui.adsabs.harvard.edu/abs/2013MNRAS.435.2753P} {435, 2753}

\bibitem[\protect\citeauthoryear{{Peng} \& {Maiolino}}{{Peng} \&
  {Maiolino}}{2014}]{Peng2014}
{Peng} Y.-j.,  {Maiolino} R.,  2014, \mn@doi [\mnras] {10.1093/mnras/stu1288},
  \href {https://ui.adsabs.harvard.edu/abs/2014MNRAS.443.3643P} {443, 3643}

\bibitem[\protect\citeauthoryear{{Peng} et~al.,}{{Peng}
  et~al.}{2010}]{Peng2010}
{Peng} Y.-j.,  et~al., 2010, \mn@doi [\apj] {10.1088/0004-637X/721/1/193},
  \href {https://ui.adsabs.harvard.edu/abs/2010ApJ...721..193P} {721, 193}

\bibitem[\protect\citeauthoryear{{Pilbratt} et~al.,}{{Pilbratt}
  et~al.}{2010}]{Pilbratt2010}
{Pilbratt} G.~L.,  et~al., 2010, \mn@doi [\aap] {10.1051/0004-6361/201014759},
  \href {https://ui.adsabs.harvard.edu/\#abs/2010A&A...518L...1P} {518, L1}

\bibitem[\protect\citeauthoryear{{Planck Collaboration} et~al.,}{{Planck
  Collaboration} et~al.}{2016}]{Planck2015}
{Planck Collaboration} et~al., 2016, \mn@doi [\aap]
  {10.1051/0004-6361/201525830}, \href
  {https://ui.adsabs.harvard.edu/abs/2016A&A...594A..13P} {594, A13}

\bibitem[\protect\citeauthoryear{{Price-Whelan} et~al.,}{{Price-Whelan}
  et~al.}{2018}]{Astropy2018}
{Price-Whelan} A.~M.,  et~al., 2018, \mn@doi [\aj] {10.3847/1538-3881/aabc4f},
  \href {https://ui.adsabs.harvard.edu/#abs/2018AJ....156..123T} {156, 123}

\bibitem[\protect\citeauthoryear{{R{\'e}my-Ruyer} et~al.,}{{R{\'e}my-Ruyer}
  et~al.}{2014}]{RemyRuyer2014}
{R{\'e}my-Ruyer} A.,  et~al., 2014, \mn@doi [\aap]
  {10.1051/0004-6361/201322803}, \href
  {https://ui.adsabs.harvard.edu/abs/2014A&A...563A..31R} {563, A31}

\bibitem[\protect\citeauthoryear{{Riechers} et~al.,}{{Riechers}
  et~al.}{2013}]{Riechers2013}
{Riechers} D.~A.,  et~al., 2013, \mn@doi [\nat] {10.1038/nature12050}, \href
  {https://ui.adsabs.harvard.edu/abs/2013Natur.496..329R} {496, 329}

\bibitem[\protect\citeauthoryear{{Rodighiero} et~al.,}{{Rodighiero}
  et~al.}{2011}]{Rodighiero2011}
{Rodighiero} G.,  et~al., 2011, \mn@doi [\apj] {10.1088/2041-8205/739/2/L40},
  \href {https://ui.adsabs.harvard.edu/abs/2011ApJ...739L..40R} {739, L40}

\bibitem[\protect\citeauthoryear{{Saintonge} et~al.,}{{Saintonge}
  et~al.}{2011}]{Saintonge2011}
{Saintonge} A.,  et~al., 2011, \mn@doi [\mnras]
  {10.1111/j.1365-2966.2011.18823.x}, \href
  {https://ui.adsabs.harvard.edu/abs/2011MNRAS.415...61S} {415, 61}

\bibitem[\protect\citeauthoryear{{Saintonge} et~al.,}{{Saintonge}
  et~al.}{2012}]{Saintonge2012}
{Saintonge} A.,  et~al., 2012, \mn@doi [\apj] {10.1088/0004-637X/758/2/73},
  \href {https://ui.adsabs.harvard.edu/abs/2012ApJ...758...73S} {758, 73}

\bibitem[\protect\citeauthoryear{{Saintonge} et~al.,}{{Saintonge}
  et~al.}{2016}]{Saintonge2016}
{Saintonge} A.,  et~al., 2016, \mn@doi [\mnras] {10.1093/mnras/stw1715}, \href
  {https://ui.adsabs.harvard.edu/abs/2016MNRAS.462.1749S} {462, 1749}

\bibitem[\protect\citeauthoryear{{Saintonge} et~al.,}{{Saintonge}
  et~al.}{2017}]{Saintonge2017}
{Saintonge} A.,  et~al., 2017, \mn@doi [\apjs] {10.3847/1538-4365/aa97e0},
  \href {https://ui.adsabs.harvard.edu/abs/2017ApJS..233...22S} {233, 22}

\bibitem[\protect\citeauthoryear{{Sandstrom} et~al.,}{{Sandstrom}
  et~al.}{2013}]{Sandstrom2013}
{Sandstrom} K.~M.,  et~al., 2013, \mn@doi [\apj] {10.1088/0004-637X/777/1/5},
  \href {https://ui.adsabs.harvard.edu/abs/2013ApJ...777....5S} {777, 5}

\bibitem[\protect\citeauthoryear{{Santini} et~al.,}{{Santini}
  et~al.}{2009}]{Santini2009}
{Santini} P.,  et~al., 2009, \mn@doi [\aap] {10.1051/0004-6361/200811434},
  \href {https://ui.adsabs.harvard.edu/abs/2009A&A...504..751S} {504, 751}

\bibitem[\protect\citeauthoryear{{Santini} et~al.,}{{Santini}
  et~al.}{2014}]{Santini2014}
{Santini} P.,  et~al., 2014, \mn@doi [\aap] {10.1051/0004-6361/201322835},
  \href {https://ui.adsabs.harvard.edu/abs/2014A&A...562A..30S} {562, A30}

\bibitem[\protect\citeauthoryear{{Sargent} et~al.,}{{Sargent}
  et~al.}{2014}]{Sargent2014}
{Sargent} M.~T.,  et~al., 2014, \mn@doi [\apj] {10.1088/0004-637X/793/1/19},
  \href {https://ui.adsabs.harvard.edu/abs/2014ApJ...793...19S} {793, 19}

\bibitem[\protect\citeauthoryear{{Sawicki}}{{Sawicki}}{2012}]{Sawicki2012}
{Sawicki} M.,  2012, \mn@doi [\mnras] {10.1111/j.1365-2966.2012.20452.x}, \href
  {https://ui.adsabs.harvard.edu/abs/2012MNRAS.421.2187S} {421, 2187}

\bibitem[\protect\citeauthoryear{{Schreiber} et~al.,}{{Schreiber}
  et~al.}{2015}]{Schreiber2015}
{Schreiber} C.,  et~al., 2015, \mn@doi [\aap] {10.1051/0004-6361/201425017},
  \href {https://ui.adsabs.harvard.edu/abs/2015A&A...575A..74S} {575, A74}

\bibitem[\protect\citeauthoryear{{Scoville} et~al.,}{{Scoville}
  et~al.}{2007}]{Scoville2007}
{Scoville} N.,  et~al., 2007, \mn@doi [The Astrophysical Journal Supplement
  Series] {10.1086/516585}, \href
  {https://ui.adsabs.harvard.edu/abs/2007ApJS..172....1S} {172, 1}

\bibitem[\protect\citeauthoryear{{Scoville}, {Sheth}, {Aussel}, {Manohar}  \&
  {ALMA Cycle 0 Teams}}{{Scoville} et~al.}{2013}]{Scoville2013}
{Scoville} N.,  {Sheth} K.,  {Aussel} H.,  {Manohar} S.,   {ALMA Cycle 0 Teams}
  2013, in {Kawabe} R.,  {Kuno} N.,   {Yamamoto} S.,  eds,  Astronomical
  Society of the Pacific Conference Series Vol. 476, New Trends in Radio
  Astronomy in the ALMA Era: The 30th Anniversary of Nobeyama Radio
  Observatory. p.~1

\bibitem[\protect\citeauthoryear{{Scoville} et~al.,}{{Scoville}
  et~al.}{2014}]{Scoville2014}
{Scoville} N.,  et~al., 2014, \mn@doi [\apj] {10.1088/0004-637X/783/2/84},
  \href {https://ui.adsabs.harvard.edu/abs/2014ApJ...783...84S} {783, 84}

\bibitem[\protect\citeauthoryear{{Scoville} et~al.,}{{Scoville}
  et~al.}{2016}]{Scoville2016}
{Scoville} N.,  et~al., 2016, \mn@doi [\apj] {10.3847/0004-637X/820/2/83},
  \href {http://adsabs.harvard.edu/abs/2016ApJ...820...83S} {820, 83}

\bibitem[\protect\citeauthoryear{{Scoville} et~al.,}{{Scoville}
  et~al.}{2017}]{Scoville2017}
{Scoville} N.,  et~al., 2017, \mn@doi [\apj] {10.3847/1538-4357/aa61a0}, \href
  {http://adsabs.harvard.edu/abs/2017ApJ...837..150S} {837, 150}

\bibitem[\protect\citeauthoryear{{Seymour} et~al.,}{{Seymour}
  et~al.}{2008}]{Seymour2008}
{Seymour} N.,  et~al., 2008, \mn@doi [\mnras]
  {10.1111/j.1365-2966.2008.13166.x}, \href
  {https://ui.adsabs.harvard.edu/abs/2008MNRAS.386.1695S} {386, 1695}

\bibitem[\protect\citeauthoryear{{Shirley} et~al.,}{{Shirley}
  et~al.}{2019}]{Shirley2019}
{Shirley} R.,  et~al., 2019, arXiv e-prints, \href
  {https://ui.adsabs.harvard.edu/abs/2019arXiv190904003S} {p. arXiv:1909.04003}

\bibitem[\protect\citeauthoryear{{Simpson} et~al.,}{{Simpson}
  et~al.}{2019}]{Simpson2019}
{Simpson} J.~M.,  et~al., 2019, \mn@doi [\apj] {10.3847/1538-4357/ab23ff},
  \href {https://ui.adsabs.harvard.edu/abs/2019ApJ...880...43S} {880, 43}

\bibitem[\protect\citeauthoryear{{Solomon} \& {Vanden Bout}}{{Solomon} \&
  {Vanden Bout}}{2005}]{Solomon2005}
{Solomon} P.~M.,  {Vanden Bout} P.~A.,  2005, \mn@doi [\araa]
  {10.1146/annurev.astro.43.051804.102221}, \href
  {https://ui.adsabs.harvard.edu/abs/2005ARA&A..43..677S} {43, 677}

\bibitem[\protect\citeauthoryear{{Sorba} \& {Sawicki}}{{Sorba} \&
  {Sawicki}}{2015}]{Sorba2015}
{Sorba} R.,  {Sawicki} M.,  2015, \mn@doi [\mnras] {10.1093/mnras/stv1235},
  \href {https://ui.adsabs.harvard.edu/abs/2015MNRAS.452..235S} {452, 235}

\bibitem[\protect\citeauthoryear{{Sorba} \& {Sawicki}}{{Sorba} \&
  {Sawicki}}{2018}]{Sorba2018}
{Sorba} R.,  {Sawicki} M.,  2018, \mn@doi [\mnras] {10.1093/mnras/sty186},
  \href {https://ui.adsabs.harvard.edu/abs/2018MNRAS.476.1532S} {476, 1532}

\bibitem[\protect\citeauthoryear{{Stach} et~al.,}{{Stach}
  et~al.}{2019}]{Stach2019}
{Stach} S.~M.,  et~al., 2019, \mn@doi [\mnras] {10.1093/mnras/stz1536}, \href
  {https://ui.adsabs.harvard.edu/abs/2019MNRAS.487.4648S} {487, 4648}

\bibitem[\protect\citeauthoryear{{Stanway}, {McMahon}  \& {Bunker}}{{Stanway}
  et~al.}{2005}]{Stanway2005}
{Stanway} E.~R.,  {McMahon} R.~G.,   {Bunker} A.~J.,  2005, \mn@doi [\mnras]
  {10.1111/j.1365-2966.2005.08977.x}, \href
  {https://ui.adsabs.harvard.edu/abs/2005MNRAS.359.1184S} {359, 1184}

\bibitem[\protect\citeauthoryear{{Steinhardt} et~al.,}{{Steinhardt}
  et~al.}{2014}]{Steinhardt2014}
{Steinhardt} C.~L.,  et~al., 2014, \mn@doi [\apj]
  {10.1088/2041-8205/791/2/L25}, \href
  {https://ui.adsabs.harvard.edu/abs/2014ApJ...791L..25S} {791, L25}

\bibitem[\protect\citeauthoryear{{Suh} et~al.,}{{Suh} et~al.}{2019}]{Suh2019}
{Suh} H.,  et~al., 2019, \mn@doi [\apj] {10.3847/1538-4357/ab01fb}, \href
  {https://ui.adsabs.harvard.edu/abs/2019ApJ...872..168S} {872, 168}

\bibitem[\protect\citeauthoryear{{Tacconi} et~al.,}{{Tacconi}
  et~al.}{2010}]{Tacconi2010}
{Tacconi} L.~J.,  et~al., 2010, \mn@doi [\nat] {10.1038/nature08773}, \href
  {https://ui.adsabs.harvard.edu/abs/2010Natur.463..781T} {463, 781}

\bibitem[\protect\citeauthoryear{{Tacconi} et~al.,}{{Tacconi}
  et~al.}{2013}]{Tacconi2013}
{Tacconi} L.~J.,  et~al., 2013, \mn@doi [\apj] {10.1088/0004-637X/768/1/74},
  \href {https://ui.adsabs.harvard.edu/abs/2013ApJ...768...74T} {768, 74}

\bibitem[\protect\citeauthoryear{{Tacconi} et~al.,}{{Tacconi}
  et~al.}{2018}]{Tacconi2018}
{Tacconi} L.~J.,  et~al., 2018, \mn@doi [\apj] {10.3847/1538-4357/aaa4b4},
  \href {http://adsabs.harvard.edu/abs/2018ApJ...853..179T} {853, 179}

\bibitem[\protect\citeauthoryear{{Taniguchi} et~al.,}{{Taniguchi}
  et~al.}{2007}]{Taniguchi2007}
{Taniguchi} Y.,  et~al., 2007, \mn@doi [\apjs] {10.1086/516596}, \href
  {https://ui.adsabs.harvard.edu/abs/2007ApJS..172....9T} {172, 9}

\bibitem[\protect\citeauthoryear{{Taniguchi} et~al.,}{{Taniguchi}
  et~al.}{2015}]{Taniguchi2015}
{Taniguchi} Y.,  et~al., 2015, \mn@doi [\pasj] {10.1093/pasj/psv106}, \href
  {https://ui.adsabs.harvard.edu/abs/2015PASJ...67..104T} {67, 104}

\bibitem[\protect\citeauthoryear{{Taylor}}{{Taylor}}{2005}]{Taylor2005}
{Taylor} M.~B.,  2005, in {Shopbell} P.,  {Britton} M.,   {Ebert} R.,  eds,
  Astronomical Society of the Pacific Conference Series Vol. 347, Astronomical
  Data Analysis Software and Systems XIV. p.~29

\bibitem[\protect\citeauthoryear{{Tomczak} et~al.,}{{Tomczak}
  et~al.}{2016}]{Tomczak2016}
{Tomczak} A.~R.,  et~al., 2016, \mn@doi [\apj] {10.3847/0004-637X/817/2/118},
  \href {https://ui.adsabs.harvard.edu/abs/2016ApJ...817..118T} {817, 118}

\bibitem[\protect\citeauthoryear{{Vaccari}}{{Vaccari}}{2016}]{Vaccari2016}
{Vaccari} M.,  2016, in {Napolitano} N.~R.,  {Longo} G.,  {Marconi} M.,
  {Paolillo} M.,   {Iodice} E.,  eds,  Vol. 42, The Universe of Digital Sky
  Surveys. p.~71 (\mn@eprint {arXiv} {1508.06444}),
  \mn@doi{10.1007/978-3-319-19330-4_10}

\bibitem[\protect\citeauthoryear{{Viero} et~al.,}{{Viero}
  et~al.}{2013}]{Viero2013}
{Viero} M.~P.,  et~al., 2013, \mn@doi [\apj] {10.1088/0004-637X/779/1/32},
  \href {http://adsabs.harvard.edu/abs/2013ApJ...779...32V} {779, 32}

\bibitem[\protect\citeauthoryear{{Whitaker}, {van Dokkum}, {Brammer}  \&
  {Franx}}{{Whitaker} et~al.}{2012}]{Whitaker2012}
{Whitaker} K.~E.,  {van Dokkum} P.~G.,  {Brammer} G.,   {Franx} M.,  2012,
  \mn@doi [\apj] {10.1088/2041-8205/754/2/L29}, \href
  {https://ui.adsabs.harvard.edu/abs/2012ApJ...754L..29W} {754, L29}

\bibitem[\protect\citeauthoryear{{Wilkins}, {Bunker}, {Stanway}, {Lorenzoni}
  \& {Caruana}}{{Wilkins} et~al.}{2011}]{Wilkins2011}
{Wilkins} S.~M.,  {Bunker} A.~J.,  {Stanway} E.,  {Lorenzoni} S.,   {Caruana}
  J.,  2011, \mn@doi [\mnras] {10.1111/j.1365-2966.2011.19315.x}, \href
  {https://ui.adsabs.harvard.edu/abs/2011MNRAS.417..717W} {417, 717}

\bibitem[\protect\citeauthoryear{{Wright}}{{Wright}}{2006}]{Wright2006}
{Wright} E.~L.,  2006, \mn@doi [\pasp] {10.1086/510102}, \href
  {https://ui.adsabs.harvard.edu/abs/2006PASP..118.1711W} {118, 1711}

\bibitem[\protect\citeauthoryear{{Wright} et~al.,}{{Wright}
  et~al.}{2016}]{Wright2016}
{Wright} A.~H.,  et~al., 2016, \mn@doi [\mnras] {10.1093/mnras/stw832}, \href
  {https://ui.adsabs.harvard.edu/\#abs/2016MNRAS.460..765W} {460, 765}

\bibitem[\protect\citeauthoryear{{da Cunha}, {Charlot}  \& {Elbaz}}{{da Cunha}
  et~al.}{2008}]{daCunha2008}
{da Cunha} E.,  {Charlot} S.,   {Elbaz} D.,  2008, \mn@doi [\mnras]
  {10.1111/j.1365-2966.2008.13535.x}, \href
  {https://ui.adsabs.harvard.edu/\#abs/2008MNRAS.388.1595D} {388, 1595}

\bibitem[\protect\citeauthoryear{{van der Wel} et~al.,}{{van der Wel}
  et~al.}{2014}]{vanderWel2014}
{van der Wel} A.,  et~al., 2014, \mn@doi [\apj] {10.1088/0004-637X/788/1/28},
  \href {https://ui.adsabs.harvard.edu/\#abs/2014ApJ...788...28V} {788, 28}

\makeatother
\end{thebibliography}


\appendix
\section{Cross check of COSMOS source catalogues} \label{App:Acrosscheck}
Here, we compare the Driver/{\sc magphys} source catalogue to the HELP/CIGALE source catalogue, test the consistency of the stellar mass estimates resulting from the two SED fitting programmes, and subsequently justify our chosen source list used in the stacking analysis. A description of the Driver/{\sc magphys} source catalogue can be found in the main text (Section \ref{Sec:DriverMAGPHYS}).

\subsection{HELP/CIGALE catalogue of region}
CIGALE (\citealt{Noll2009}, \citealt{Boquien2019}) is an SED fitting routine wherein SED models are built using several modular components in a similar way to {\sc magphys}, but also includes different dust attenuation curves, active galactic nuclei (AGN) emission, and  radio emission (\citealt{Ciesla2016}, \citealt{Hunt2019}).

Several modules for a given component can be considered in the fitting process to try and help disentangle different physical implications of similarly looking SEDs. Much like {\sc magphys}, CIGALE also makes use of the concept of energy-balance; the energy absorbed by dust in the UV-NIR is re-emitted in the MIR-FIR. The parameters returned for the physical systems fitted by CIGALE are chosen by the user. Values of physical parameters returned by CIGALE are either the best-fit values (from the best-fitting SED), or likelihood-weighted means and likelihood-weighted standard deviations. The likelihood is taken to be $e^{-\chi^2 / 2}$.  We use the likelihood-weighted values in this analysis, and refer the interested reader to \citet{Boquien2019} for further details. 

In this work, we make use of photometry and CIGALE data from the {\it Herschel} Extragalactic Legacy Project (HELP) database (\citealt{Vaccari2016}; \citealt{Malek2018}; \citealt{Malek2019short}; \citealt{Shirley2019}; Oliver et al. {\it in prep.}). HELP provides a homogenized, multi-wavelength database of all the fields observed by {\it Herschel}, covering 1270 square degrees over 23 different fields \citep{Shirley2019}, including the COSMOS field. As the {\it Herschel} maps suffer from source confusion \citep{Nguyen2010}, XID+, a probabilistic de-blending tool that uses Bayesian techniques to assign FIR fluxes to sources based on NIR and MIR positional prior catalogues \citep{Hurley2017}, was used to assign FIR fluxes to sources. 

The HELP-COSMOS database includes 33 photometric bands, per source, that are suitable for SED fitting. To avoid over-dense photometry data causing forced SED fits, and to ensure that the deepest data are used, where there is photometry in similar bands for a given object, priority is given to the deepest observations. We refer the reader to \citet{Malek2018} for details, but briefly, the 19 bands used in the CIGALE COSMOS fits are: {\it ugriz}, N921 (a narrow band filter on Suburu/HSC), {\it y}JK, IRAC1234, MIPS24, PACS100/160, SPIRE250/350/500. Again, note that not every source may have fluxes for all filters.

Unlike {\sc magphys}, CIGALE offers the user a choice for the input parameters. We refer the reader to \citet{Malek2018} for details, but briefly, the HELP consortium uses the Single Stellar Population (SSP) model of \cite{BruzualCharlot2003}, assuming a \cite{Chabrier2003} IMF. They follow the dust attenuation curve of \cite{CharlotFall2000} and dust emission is based on \cite{DraineLi2007}, with AGN based on \cite{Fritz2006}. Star Formation Histories (SFHs) with delays and additional, optional, bursts are also implemented.

The HELP photometry catalogue for the COSMOS field is based on the COSMOS2015 catalogue from \cite{Laigle2016} (see Section \ref{Sec:COSMOS2015} for a description of the COSMOS2015 catalogue). 
The CIGALE catalogue we use was compiled by fitting every source within the HELP photometric catalogue for the COSMOS field that has at least four `optical' and `NIR' fluxes, where `optical' bands are defined as {\it ugrizy} and N921, and `NIR' bands are {\it J} and {\it K} (\citealt{Malek2018}; \citealt{Shirley2019})

Photometric redshifts are calculated as part of the HELP pipeline, using a Bayesian combination approach, which combines popular photometric redshift estimator templates to achieve the best estimate of the redshift, see \citet{Duncan2018a} and \citet{Duncan2018b}. Spectroscopic redshifts are used, where possible, and are sourced from various different surveys compiled by the HELP consortium including: SDSS \citep{SDSS13}, PRIMUS \citep{Cool2013}, zBRIGHT \citep{Lilly2007zbright} and GAMA \citep{Davies2015}. In total, CIGALE fits are available for 639,873 sources with photometric redshifts and 39,890 sources with spectroscopic redshifts.

\subsection{Cross check and final selection of COSMOS source catalogue}

As a consistency check, we cross-match the results from {\sc magphys} and CIGALE for the COSMOS field, and compare the calculated stellar masses ($M_*$) for the resulting population. We make use of TOPCAT (Tools for OPerations on Catalogues And Tables, \citealt{Taylor2005}), selecting sources that are matched within 0.1$^{\prime \prime}$ on the sky. 

There are several RA and Dec options within the {\sc magphys} catalogue - we choose to cross-match to the CIGALE catalogue using the RA and Dec values from the COSMOS2015 catalogue \citep{Laigle2016}. For sources with photometric redshifts in the CIGALE catalogue, the separation between matching sources is well below typical astrometric uncertainties. 
We find 114,546 matches in this way. Similarly for the sources with spectroscopic redshifts in the CIGALE catalogue, we find most sources are separated by a radius between 0.07960$^{\prime \prime}$ and 0.07965$^{\prime \prime}$. We find 26,749 matches for sources with spectroscopic redshifts in the CIGALE catalogue. This brings our total number of matches between {\sc magphys} and CIGALE to 141,295.

Although we are confident in the accuracy of the cross-matching, there are still likely to be some spurious matches in our resultant catalogue. We estimate the number of false matches as:
\begin{equation}
    {\rm spurious \ matches} = \frac{N_M \pi r^2 N_C}{A}
\end{equation}
where $N_M$ is the number of sources in the {\sc magphys} catalogue, $r$ is the matching radius (in arcsec), $N_C$ is the number of sources in the CIGALE catalogue within the area covered by the {\sc magphys} catalogue, and $A$ is the area covered by the {\sc magphys} catalogue (in arcsec$^2$). For sources in the CIGALE catalogue with photometric redshifts, we estimate that there may be 136 spurious matches with the {\sc magphys} catalogue, using a matching radius of 0.1$^{\prime \prime}$. For sources in the CIGALE catalogue with spectroscopic redshifts, we estimate that there may be 13 spurious matches. This is a false-positive rate of 0.1\% and $<$0.05\%, respectively, and thus are  are unlikely to have a significant effect on our analysis. 

\begin{figure}
	\includegraphics[width=\columnwidth]{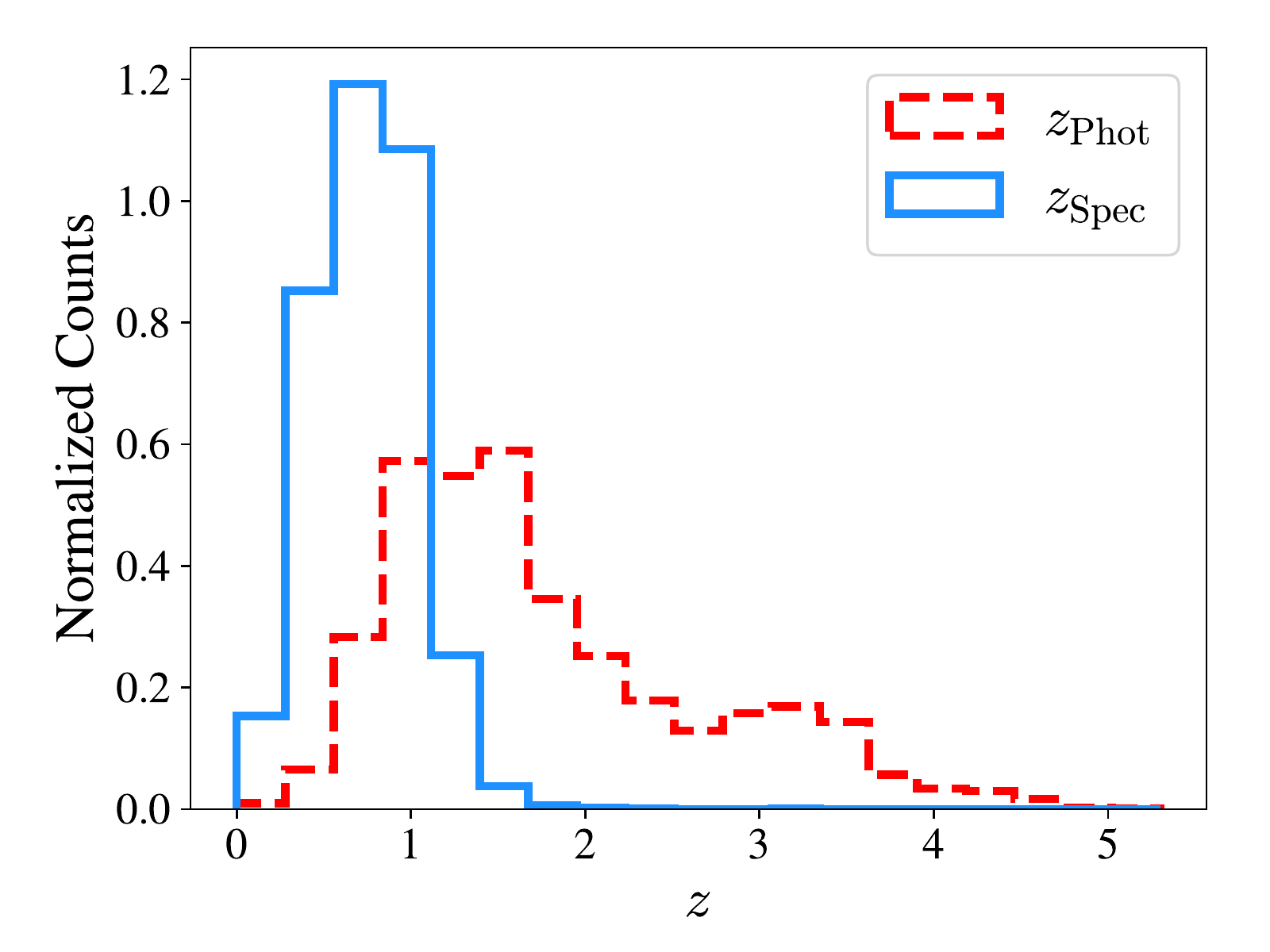}
    \caption{The redshift distribution of a filtered sub-sample of 23,164 galaxies in the COSMOS field that have both {\sc magphys} and CIGALE fits, $\Delta z < 0.02(1+z)$ and log($M_*/M_{\odot}) \geq 9.5$. The blue solid line indicates the distribution of sources with CIGALE spectroscopic redshifts. The red dashed line indicates the distribution of sources with CIGALE photometric redshifts.}
    \label{fig:zdistrib_Mstar_compare}
\end{figure}

Since both SED fitting routines use the galaxy redshift as an input, we also filter this `matched' selection of galaxies to exclude sources with redshift differences between the two catalogues of $\Delta z > 0.02 (1+z)$. This is based on the maximum quoted photometric redshift error in the COSMOS2015 catalogue \citep{Laigle2016}, as such is likely a conservative estimate in the uncertainty in $z$. We choose to use the {\sc magphys} redshifts in this selection criteria. We further filter for sources that have poor fits. To do this, we determine a $\chi^2$ cut to apply to the {\sc magphys} and CIGALE catalogues by fitting continuous probability distributions to the $\chi^2$ values, where the mean of the distribution provides our $\chi^2$ cut value. We determine a threshold of $\chi^2_{\rm thr,MAG} < 1.93$ for {\sc magphys} and $\chi^2_{\rm thr,CIG} < 5.60$ for CIGALE (see Appendix \ref{App:Achi2} for details). The difference in the $\chi^2$ values determined for the two different SED fitting codes is related to the different number of free parameters and the way of calculating the final $\chi^2$, which is non-trivial. Finally, we filter for sources that have log($M_*/M_{\odot}$) < 9.5 in both catalogues, as we do not consider galaxies below this stellar mass in this study (see Section \ref{Sec:stackinggas}). This leaves us with 23,164 sources out to a maximum photometric redshift of $z=5.4$, and a maximum spectroscopic redshift of $z=3.1$, for which to compare stellar masses (Figure \ref{fig:zdistrib_Mstar_compare}).

\begin{figure}
	\includegraphics[width=\columnwidth]{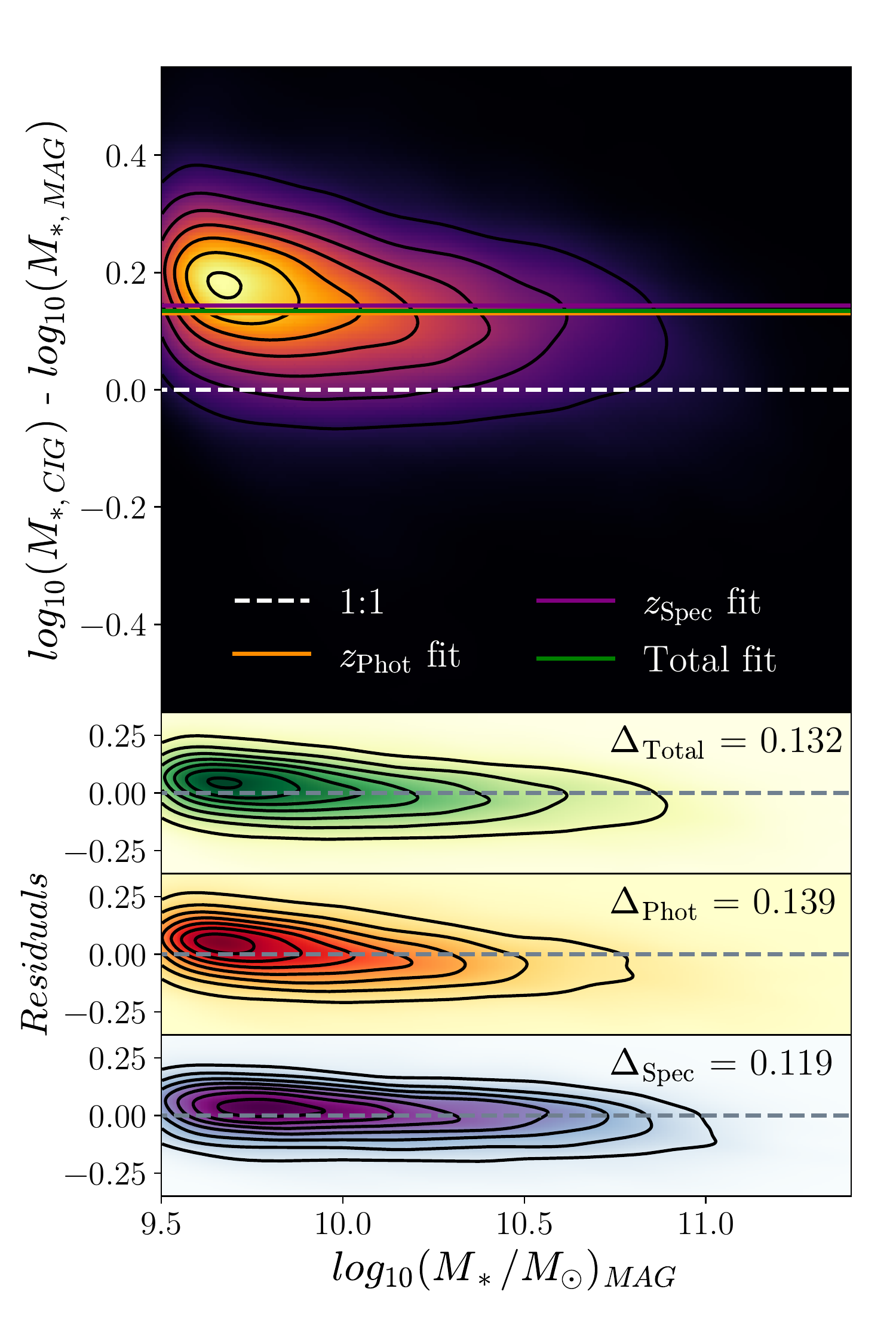}
    \caption{A comparison of the stellar masses for a filtered sub-sample of galaxies in the COSMOS field that have both {\sc magphys} and CIGALE fits. The sources displayed here have log$(M_*/M_{\odot})$ > 9.5 and a redshift difference between the two catalogues of $\Delta z < 0.02(1+z)$. \textit{Upper:} Distribution of the matched sources. White dashed line is the 1:1 line. Green, orange and purple lines are best fits to different populations, with a fixed gradient of zero. \textit{Orange line}: sources with photometric redshifts in the CIGALE dataset. \textit{Purple line}: sources with spectroscopic redshifts in the CIGALE dataset. \textit{Green line}: all matched sources. \textit{Lower three plots:} residuals of best fit line for different populations and corresponding dispersion values, $\Delta$. \textit{Lower-top:} all matched sources. \textit{Lower-middle:} sources with photometric redshifts in the CIGALE catalogue. \textit{Lower-bottom:} sources with spectroscopic redshifts in the CIGALE catalogue. }
    \label{fig:Mstar_compare}
\end{figure}

Figure \ref{fig:Mstar_compare} compares the ratio of stellar masses resulting from {\sc magphys} and CIGALE. The ratio of stellar masses indicate a systematic offset between CIGALE and {\sc magphys}, with a maximum dispersion value of 0.139 dex (the best fit line offsets are displayed in Table \ref{tab:line_offsets}). The offsets are similar in magnitude to the dispersions of the different populations of sources, implying that the offset in results for the different SED fitting programmes is small (at most 25\,per\,cent)\footnote{We also examined sources where $\Delta z > 0.02(1+z)$. The best fit line offset for all sources is 0.113, around 20\,per\,cent lower than the sources with $\Delta z < 0.02(1+z)$, but with a stronger evolution with stellar mass. The offset is more pronounced at low stellar masses ($<10^{10}M_{\odot}$). Sources with larger $\Delta z$ showed a larger dispersion; 0.167 for all sources, compared to 0.132 for the sources with smaller $\Delta z$ (an increase of around 25\,per\,cent).}. We note that this offset is lower than the systematic underestimation of stellar masses from SED fitting due to outshining, where bright young stars can mask underlying older stellar populations \citep{Sorba2015, Sorba2018, Abdurrouf2018}. Therefore the choice of source catalogue is not likely to be our largest source of error, and Figure \ref{fig:Mstar_compare} demonstrates that choosing stellar masses from the CIGALE catalogue would not change our conclusions. 

\begin{table}
	\centering
	\caption{Line offsets for the best fit lines fit to the {\sc magphys} and CIGALE ratio of stellar masses displayed in Figure \ref{fig:Mstar_compare}. The lines are fixed to have a gradient of 0. {\it Total fit}: all of the sources with $\Delta z < 0.02(1+z)$, log$(M_*/M_{\odot}) > 9.5$, and with $\chi^2_{\rm thr, MAG} < 1.93$ and $\chi^2_{\rm thr, CIG} < 5.60$. $z_{\rm Phot}$ {\it fit}: similar to `total fit', but only for sources that have photometric redshifts in the CIGALE catalogue. $z_{\rm Spec}$ {\it fit}: similar to `total fit', but only for sources that have spectroscopic redshifts in the CIGALE catalogue.}
	\label{tab:line_offsets}
	\begin{tabular}{cc} 
		\hline
		Name & Offset (dex) \\
		\hline
		Total fit & 0.135\\
		$z_{\rm Phot}$ fit & 0.130\\
		$z_{\rm Spec}$ fit & 0.143\\
		\hline
	\end{tabular}
\end{table}

\section{Determining the Threshold of Good SED Fits} \label{App:Achi2}
To determine the $\chi^2$ thresholds for our {\sc magphys} and CIGALE sources, we first filter for catastrophic fitting failures, to allow the fitting functions to converge to a result - for {\sc magphys}, we do not consider sources with $\chi^2 > 40$, and for CIGALE, we do not consider sources with $\chi^2 > 80$. This leaves us with 141,088 {\sc magphys} sources and 138,215 CIGALE sources to determine the $\chi^2$ thresholds. 

\begin{figure}
	\includegraphics[width=\columnwidth]{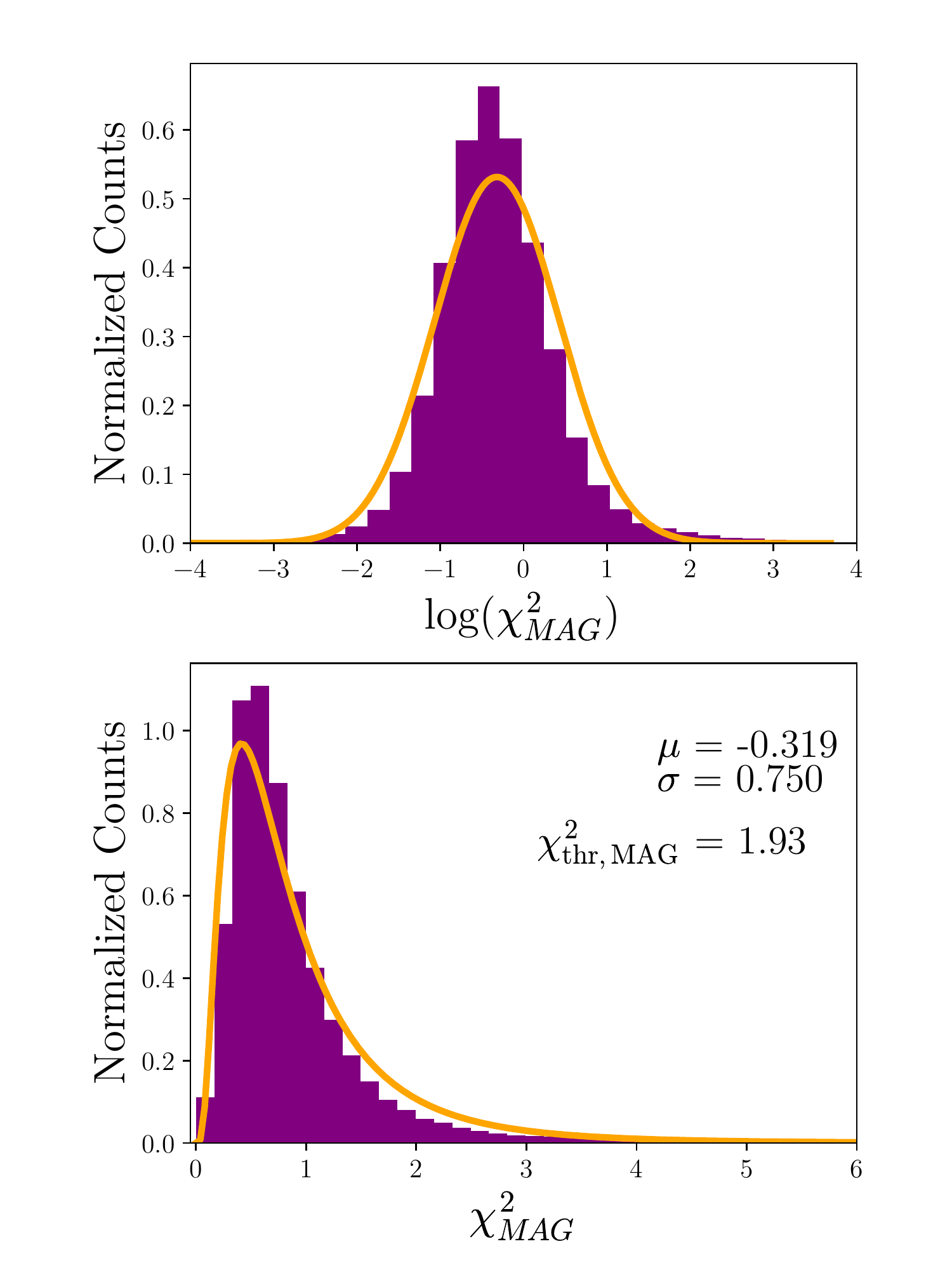}
    \caption{The {\sc magphys} $\chi^2$ distribution of a filtered sub-sample of galaxies in the COSMOS field that have both {\sc magphys} and CIGALE fits. The sources displayed here have $\Delta z < 0.02(1+z)$ between the {\sc magphys} and CIGALE catalogues. {\it Upper}: logged {\sc magphys} $\chi^2$ values clearly display a Gaussian distribution. Purple is a histogram of the logged {\sc magphys} $\chi^2$ values; orange is a Gaussian fit to the data. {\it Lower}: a lognormal fit to the {\sc magphys} $\chi^2$ distributions, with calculated mean, $\mu$, standard deviation, $\sigma$, and subsequently calculated {\sc magphys} $\chi^2$ threshold, $\chi^2_{\rm thr,MAG}$ displayed.}
    \label{fig:mag_chi2}
\end{figure}

When logged, the {\sc magphys} $\chi^2$ values clearly display a Gaussian distribution (Figure \ref{fig:mag_chi2}). A lognormal distribution was fitted to the {\sc magphys} $\chi^2$ values (Figure \ref{fig:mag_chi2}), giving a threshold $\chi^2_{\rm thr,MAG} = 1.93$. 

\begin{figure}
	\includegraphics[width=\columnwidth]{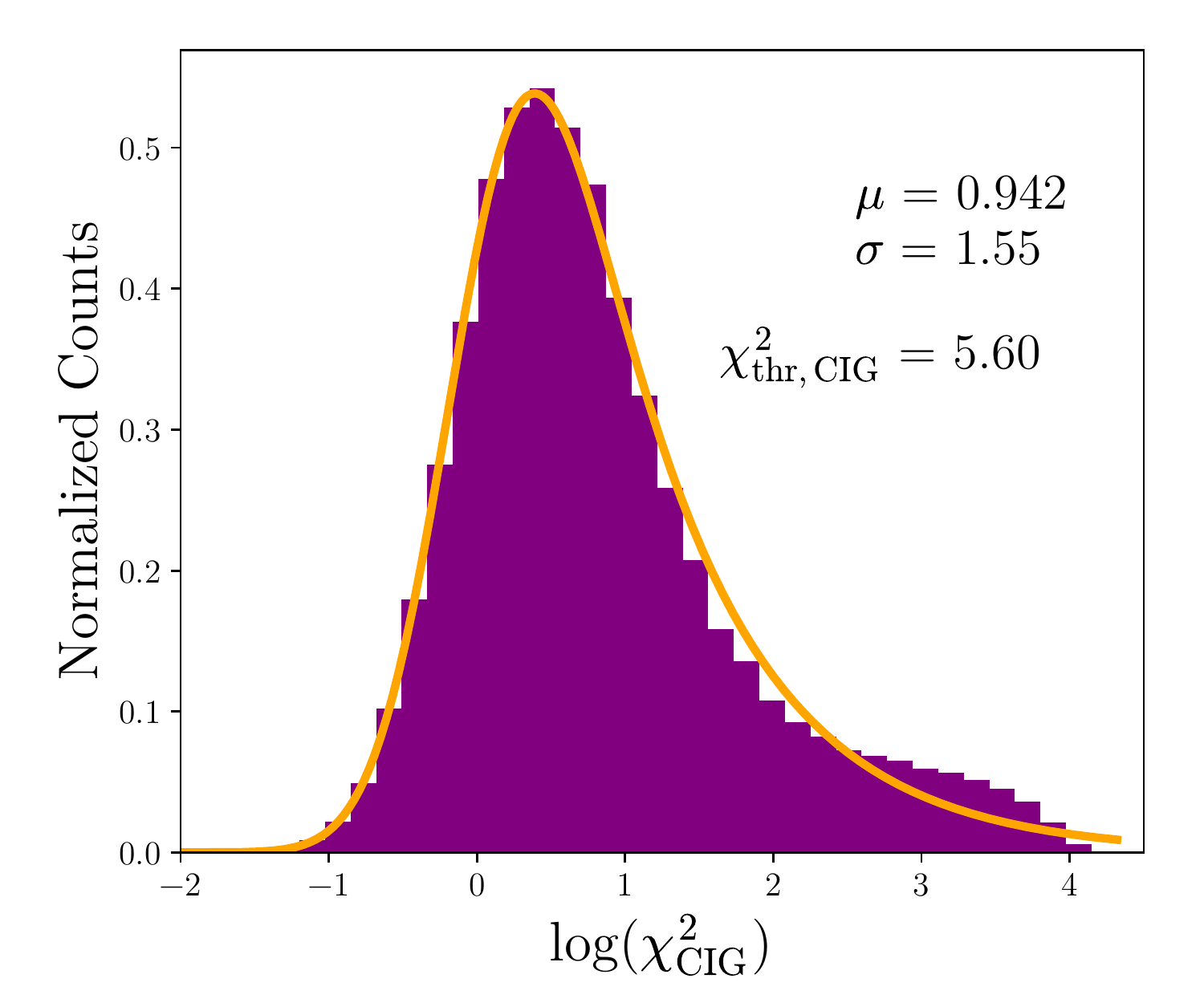}
    \caption{The CIGALE $\chi^2$ distribution of a filtered sub-sample of galaxies in the COSMOS field that have both {\sc magphys} and CIGALE fits. The sources displayed here have $\Delta z < 0.02(1+z)$ between the {\sc magphys} and CIGALE catalogues. Logged CIGALE $\chi^2$ values clearly do not display a Gaussian distribution (purple histogram); as such, a lognormal distribution cannot be fitted to the data. Instead, an exponentially modified Gaussian distribition is fitted to the logged $\chi^2$ values (orange line). The calculated mean, $\mu$, standard deviation, $\sigma$, and subsequently calculated CIGALE $\chi^2$ threshold, $\chi^2_{\rm thr,CIG}$ are displayed.}
    \label{fig:cig_chi2}
\end{figure}

When logged, the CIGALE $\chi^2$ values did not display a Gaussian distribution. The distribution displayed by the data was similar to than of an exponentially modified Gaussian (Figure \ref{fig:cig_chi2}). Subsequently, this function was fit to the logged CIGALE $\chi^2$ values, giving a threshold $\chi^2_{\rm thr,CIG} = 5.60$. 

\section{Sources in {\sc magphys} catalogue without matches in COSMOS2015}
\begin{figure}
	\includegraphics[width=\columnwidth]{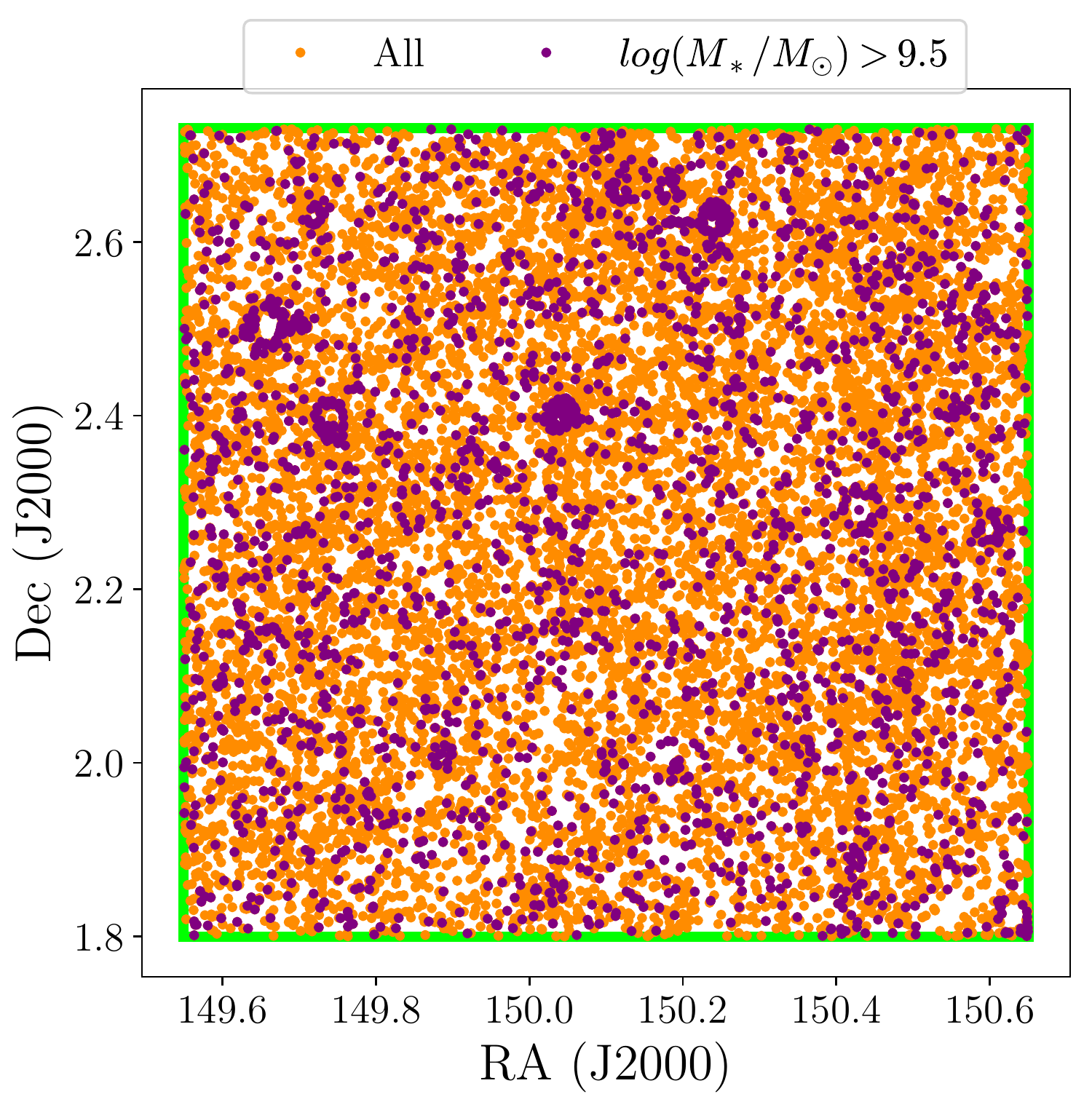}
    \caption{The distribution of sources in the {\sc magphys} catalogue but not in the COSMOS2015 catalogue. Orange markers are all the sources; purple markers are sources with log($M_*/M_{\odot}$) > 9.5. The green box is the same as that shown in Figure \ref{fig:SCUBA2_map} i.e. the overall extent of the {\sc magphys} sources.}
    \label{fig:radec_notin_COSMOS2015}
\end{figure}

\begin{figure}
	\includegraphics[width=\columnwidth]{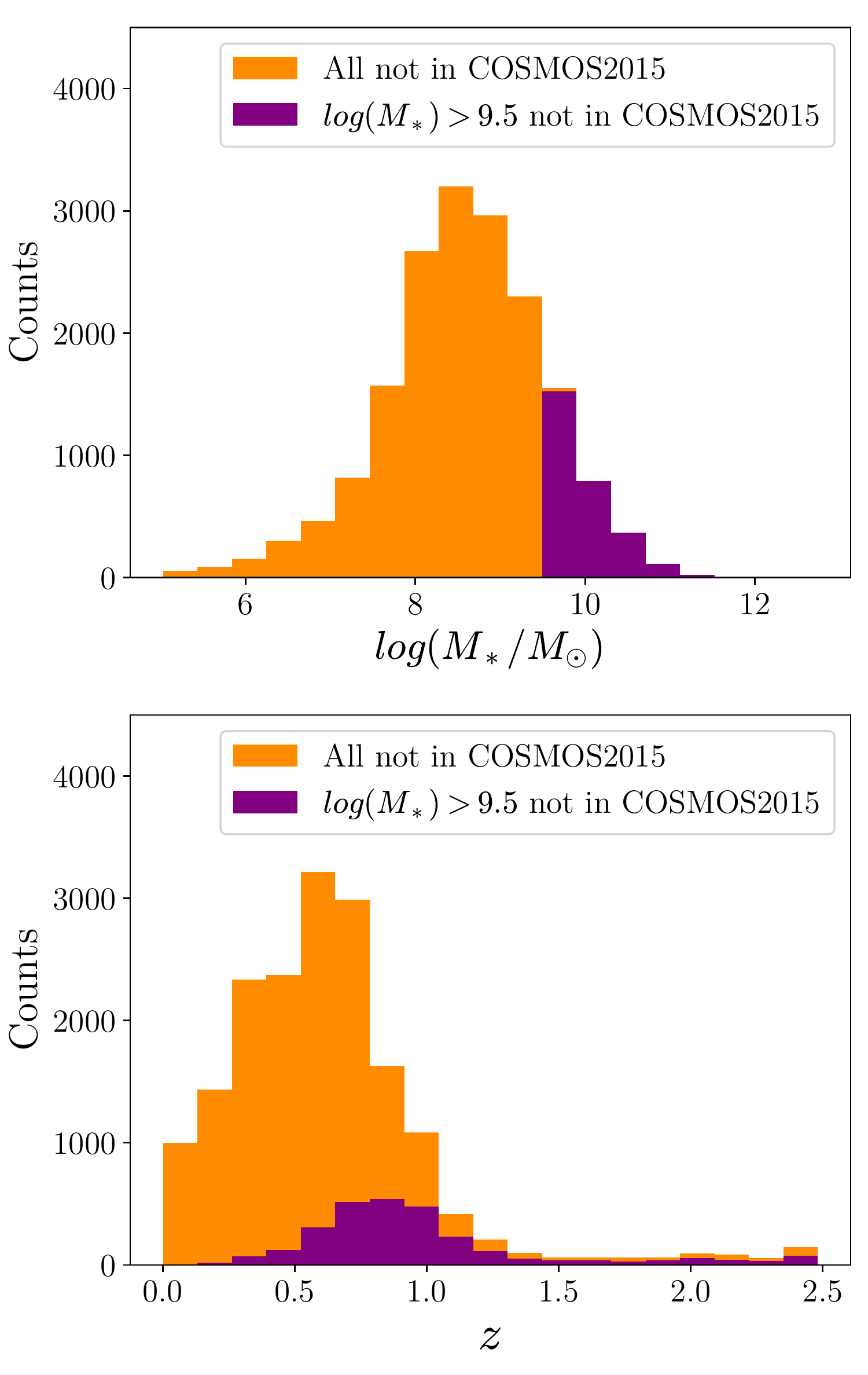}
    \caption{The stellar mass and redshift distribution of sources in the {\sc magphys} catalogue but not in the COSMOS2015 catalogue. Orange marks all the sources; purple marks sources with log$(M_*/M_{\odot})$ > 9.5.}
    \label{fig:mstarz_notin_COSMOS2015}
\end{figure}

Figure \ref{fig:radec_notin_COSMOS2015} shows the distribution of sources in the {\sc magphys} catalogue without matches in COSMOS2015. Figure \ref{fig:mstarz_notin_COSMOS2015} illustrates the stellar mass and redshift distribution of sources removed from the original {\sc magphys} catalogue. Note here that only sources with log$({M_*/M_{\odot}})$ > 5 are considered, as suggested by \cite{Driver2018}.

\section{Stacking Data}
\begin{landscape}
\begin{table}
	\centering
	\caption{Gas mass fractions and relevant ancillary data. Red values indicate calculations using 3$\sigma$ upper limits. Gas mass fractions are not calculated for SIMSTACK results where 3$\sigma$ upper limits are used for IVW fluxes.}
	\label{tab:bin_numbers}
	\begin{tabular}{ccccccccccc} 
		\hline
		log(M$_*/M_{\odot}$) & Bootstrapped &\multirow{2}{*}{<$z_{\rm bin}$>} & \multirow{2}{*}{\# sources} & IVW Stacked & IVW ISM & \multirow{2}{*}{IVW $f_{\rm ISM}$} & SIMSTACK Stacked & SIMSTACK ISM & SIMSTACK \\ 
		bin & log(M$_*/M_{\odot})_{\rm Med}$  & & & Flux (mJy) & Mass ($10^{10}M_{\odot}$) & & Flux (mJy) & Mass ($10^{10}M_{\odot}$) & $f_{\rm ISM}$\\
		\hline
         & 9.62 & 0.28 & 1489 & 0.087 $\substack{ +0.034 \\ -0.034 }$ & 0.116 $\substack{ +0.034 \\ -0.034 }$ & 0.217 $\substack{ +0.086 \\ -0.088 }$ & 0.088 & 0.117 & 0.219 \\ 
         & 9.62 & 0.82 & 4579 & 0.051 $\substack{ +0.018 \\ -0.021 }$ & 0.199 $\substack{ +0.018 \\ -0.021 }$ & 0.323 $\substack{ +0.122 \\ -0.138 }$ & 0.024 & 0.096 & 0.187 \\ 
         & 9.62 & 1.36 & 3574 & 0.094 $\substack{ +0.022 \\ -0.021 }$ & 0.460 $\substack{ +0.022 \\ -0.021 }$ & 0.524 $\substack{ +0.138 \\ -0.133 }$ & 0.033 & 0.164 & 0.281 \\ 
         & 9.62 & 1.90 & 2165 & 0.130 $\substack{ +0.027 \\ -0.029 }$ & 0.683 $\substack{ +0.027 \\ -0.029 }$ & 0.620 $\substack{ +0.151 \\ -0.163 }$ & 0.170 & 0.895 & 0.681 \\ 
        9.50 - 9.75 & 9.62 & 2.44 & 1654 & 0.060 $\substack{ +0.032 \\ -0.031 }$ & 0.328 $\substack{ +0.032 \\ -0.031 }$ & 0.438 $\substack{ +0.250 \\ -0.247 }$ & 0.053 & 0.286 & 0.405 \\ 
         & 9.63 & 2.98 & 1246 & 0.106 $\substack{ +0.035 \\ -0.035 }$ & 0.595 $\substack{ +0.035 \\ -0.035 }$ & 0.585 $\substack{ +0.224 \\ -0.223 }$ & 0.053 & 0.298 & 0.414 \\ 
         & 9.63 & 3.52 & 603 & 0.122 $\substack{ +0.054 \\ -0.050 }$ & 0.723 $\substack{ +0.054 \\ -0.050 }$ & 0.629 $\substack{ +0.325 \\ -0.301 }$ & 0.056 & 0.329 & 0.435 \\ 
         & 9.64 & 4.05 & 236 & -0.000 $\substack{ +0.090 \\ -0.079 }$ & \textcolor{red}{1.696} & \textcolor{red}{0.796} & 0.069 & - & - \\ 
         & 9.65 & 4.59 & 69 & 0.054 $\substack{ +0.167 \\ -0.150 }$ & 0.367 $\substack{ +0.167 \\ -0.150 }$ & 0.449 $\substack{ +1.533 \\ -1.373 }$ & 0.218 & 1.495 & 0.769 \\ 
         & 9.68 & 5.13 & 12 & 0.442 $\substack{ +0.371 \\ -0.367 }$ & 3.337 $\substack{ +0.371 \\ -0.367 }$ & 0.875 $\substack{ +0.976 \\ -0.967 }$ & 0.349 & 2.634 & 0.847 \\ 
        \hline
         & 9.87 & 0.82 & 4052 & 0.131 $\substack{ +0.020 \\ -0.019 }$ & 0.513 $\substack{ +0.020 \\ -0.019 }$ & 0.409 $\substack{ +0.069 \\ -0.066 }$ & 0.139 & 0.544 & 0.424 \\ 
         & 9.87 & 1.36 & 2957 & 0.155 $\substack{ +0.021 \\ -0.024 }$ & 0.761 $\substack{ +0.021 \\ -0.024 }$ & 0.507 $\substack{ +0.079 \\ -0.090 }$ & 0.142 & 0.698 & 0.486 \\ 
         & 9.87 & 1.90 & 1857 & 0.178 $\substack{ +0.030 \\ -0.029 }$ & 0.932 $\substack{ +0.030 \\ -0.029 }$ & 0.556 $\substack{ +0.106 \\ -0.105 }$ & 0.127 & 0.666 & 0.472 \\ 
         & 9.87 & 2.44 & 1148 & 0.201 $\substack{ +0.039 \\ -0.037 }$ & 1.092 $\substack{ +0.039 \\ -0.037 }$ & 0.597 $\substack{ +0.133 \\ -0.129 }$ & 0.232 & 1.259 & 0.631 \\ 
        9.75 - 10.00 & 9.87 & 2.98 & 958 & 0.212 $\substack{ +0.039 \\ -0.041 }$ & 1.194 $\substack{ +0.039 \\ -0.041 }$ & 0.616 $\substack{ +0.133 \\ -0.140 }$ & 0.282 & 1.588 & 0.681 \\ 
         & 9.87 & 3.52 & 664 & 0.101 $\substack{ +0.049 \\ -0.045 }$ & 0.596 $\substack{ +0.049 \\ -0.045 }$ & 0.444 $\substack{ +0.234 \\ -0.215 }$ & 0.206 & 1.217 & 0.620 \\ 
         & 9.87 & 4.05 & 205 & 0.167 $\substack{ +0.087 \\ -0.080 }$ & 1.052 $\substack{ +0.087 \\ -0.080 }$ & 0.584 $\substack{ +0.353 \\ -0.323 }$ & 0.035 & 0.220 & 0.227 \\ 
         & 9.88 & 4.59 & 118 & 0.300 $\substack{ +0.120 \\ -0.107 }$ & 2.054 $\substack{ +0.120 \\ -0.107 }$ & 0.730 $\substack{ +0.360 \\ -0.322 }$ & 0.199 & 1.359 & 0.642 \\ 
         & 9.86 & 5.13 & 14 & -0.386 $\substack{ +0.309 \\ -0.313 }$ & \textcolor{red}{7.017} & \textcolor{red}{0.907} & -0.276 & - & - \\  
        \hline
         & 10.12 & 0.82 & 3531 & 0.132 $\substack{ +0.020 \\ -0.022 }$ & 0.518 $\substack{ +0.020 \\ -0.022 }$ & 0.282 $\substack{ +0.044 \\ -0.048 }$ & 0.093 & 0.366 & 0.217 \\ 
         & 10.12 & 1.36 & 2292 & 0.192 $\substack{ +0.027 \\ -0.028 }$ & 0.943 $\substack{ +0.027 \\ -0.028 }$ & 0.418 $\substack{ +0.063 \\ -0.065 }$ & 0.111 & 0.545 & 0.293 \\ 
         & 10.12 & 1.90 & 1375 & 0.326 $\substack{ +0.031 \\ -0.039 }$ & 1.710 $\substack{ +0.031 \\ -0.039 }$ & 0.567 $\substack{ +0.061 \\ -0.077 }$ & 0.309 & 1.625 & 0.554 \\ 
         & 10.11 & 2.44 & 773 & 0.366 $\substack{ +0.047 \\ -0.046 }$ & 1.989 $\substack{ +0.047 \\ -0.046 }$ & 0.605 $\substack{ +0.091 \\ -0.089 }$ & 0.356 & 1.934 & 0.598 \\ 
        10.00 - 10.25 & 10.12 & 2.98 & 615 & 0.389 $\substack{ +0.050 \\ -0.050 }$ & 2.187 $\substack{ +0.050 \\ -0.050 }$ & 0.627 $\substack{ +0.095 \\ -0.096 }$ & 0.369 & 2.073 & 0.614 \\ 
         & 10.11 & 3.52 & 415 & 0.209 $\substack{ +0.062 \\ -0.067 }$ & 1.235 $\substack{ +0.062 \\ -0.067 }$ & 0.490 $\substack{ +0.161 \\ -0.175 }$ & 0.062 & 0.365 & 0.221 \\ 
         & 10.12 & 4.05 & 138 & 0.251 $\substack{ +0.118 \\ -0.105 }$ & 1.583 $\substack{ +0.118 \\ -0.105 }$ & 0.547 $\substack{ +0.293 \\ -0.260 }$ & 0.083 & 0.521 & 0.285 \\ 
         & 10.13 & 4.59 & 80 & 0.216 $\substack{ +0.135 \\ -0.134 }$ & 1.477 $\substack{ +0.135 \\ -0.134 }$ & 0.525 $\substack{ +0.370 \\ -0.368 }$ & 0.223 & 1.523 & 0.533 \\ 
         & 10.12 & 5.13 & 18 & 0.387 $\substack{ +0.299 \\ -0.260 }$ & 2.928 $\substack{ +0.299 \\ -0.260 }$ & 0.687 $\substack{ +0.644 \\ -0.561 }$ & 0.215 & 1.624 & 0.550 \\ 
        \hline
         & 10.37 & 0.28 & 1130 & 0.141 $\substack{ +0.039 \\ -0.041 }$ & 0.186 $\substack{ +0.039 \\ -0.041 }$ & 0.074 $\substack{ +0.021 \\ -0.021 }$ & 0.053 & 0.070 & 0.029 \\ 
         & 10.37 & 0.82 & 3387 & 0.144 $\substack{ +0.023 \\ -0.023 }$ & 0.563 $\substack{ +0.023 \\ -0.023 }$ & 0.195 $\substack{ +0.032 \\ -0.031 }$ & 0.074 & 0.289 & 0.110 \\ 
         & 10.37 & 1.36 & 2053 & 0.211 $\substack{ +0.029 \\ -0.026 }$ & 1.038 $\substack{ +0.029 \\ -0.026 }$ & 0.306 $\substack{ +0.045 \\ -0.040 }$ & 0.124 & 0.611 & 0.206 \\ 
         & 10.36 & 1.90 & 964 & 0.377 $\substack{ +0.041 \\ -0.041 }$ & 1.979 $\substack{ +0.041 \\ -0.041 }$ & 0.463 $\substack{ +0.055 \\ -0.056 }$ & 0.403 & 2.114 & 0.480 \\ 
        10.25 - 10.50 & 10.35 & 2.44 & 452 & 0.504 $\substack{ +0.062 \\ -0.061 }$ & 2.739 $\substack{ +0.062 \\ -0.061 }$ & 0.550 $\substack{ +0.077 \\ -0.076 }$ & 0.547 & 2.971 & 0.570 \\ 
         & 10.36 & 2.98 & 334 & 0.489 $\substack{ +0.072 \\ -0.071 }$ & 2.753 $\substack{ +0.072 \\ -0.071 }$ & 0.548 $\substack{ +0.092 \\ -0.091 }$ & 0.446 & 2.508 & 0.525 \\ 
         & 10.35 & 3.52 & 216 & 0.565 $\substack{ +0.091 \\ -0.085 }$ & 3.337 $\substack{ +0.091 \\ -0.085 }$ & 0.600 $\substack{ +0.112 \\ -0.105 }$ & 0.589 & 3.480 & 0.610 \\ 
         & 10.36 & 4.05 & 85 & 0.524 $\substack{ +0.151 \\ -0.128 }$ & 3.299 $\substack{ +0.151 \\ -0.128 }$ & 0.592 $\substack{ +0.198 \\ -0.168 }$ & 0.526 & 3.314 & 0.593 \\ 
         & 10.34 & 4.59 & 51 & 0.201 $\substack{ +0.177 \\ -0.169 }$ & 1.375 $\substack{ +0.177 \\ -0.169 }$ & 0.386 $\substack{ +0.365 \\ -0.348 }$ & -0.010 & -0.071 & -0.034 \\ 
         & 10.36 & 5.13 & 13 & 0.514 $\substack{ +0.336 \\ -0.293 }$ & 3.887 $\substack{ +0.336 \\ -0.293 }$ & 0.631 $\substack{ +0.488 \\ -0.426 }$ & 0.477 & 3.604 & 0.613 \\ 
         \hline
         & 10.62 & 0.28 & 1069 & 0.199 $\substack{ +0.035 \\ -0.040 }$ & 0.264 $\substack{ +0.035 \\ -0.040 }$ & 0.060 $\substack{ +0.011 \\ -0.012 }$ & 0.230 & 0.305 & 0.068 \\ 
         & 10.61 & 0.82 & 2895 & 0.225 $\substack{ +0.023 \\ -0.024 }$ & 0.884 $\substack{ +0.023 \\ -0.024 }$ & 0.177 $\substack{ +0.019 \\ -0.019 }$ & 0.200 & 0.786 & 0.161 \\ 
        10.50 - 10.75 & 10.62 & 1.36 & 2057 & 0.386 $\substack{ +0.027 \\ -0.027 }$ & 1.893 $\substack{ +0.027 \\ -0.027 }$ & 0.313 $\substack{ +0.023 \\ -0.023 }$ & 0.349 & 1.712 & 0.292 \\ 
for          & 10.61 & 1.90 & 678 & 0.593 $\substack{ +0.046 \\ -0.049 }$ & 3.114 $\substack{ +0.046 \\ -0.049 }$ & 0.431 $\substack{ +0.037 \\ -0.039 }$ & 0.549 & 2.885 & 0.413 \\ 
         & 10.60 & 2.44 & 196 & 0.670 $\substack{ +0.093 \\ -0.093 }$ & 3.641 $\substack{ +0.093 \\ -0.093 }$ & 0.476 $\substack{ +0.073 \\ -0.073 }$ & 0.640 & 3.478 & 0.464 \\  
        \hline
	\end{tabular}
\end{table}
\addtocounter{table}{-1}
\end{landscape}

\begin{landscape}
\begin{table}
	\centering
	\caption{\textit{Continued}}
	\begin{tabular}{ccccccccccc} 
		\hline
		log(M$_*/M_{\odot}$) & Bootstrapped &\multirow{2}{*}{<$z_{\rm bin}$>} & \multirow{2}{*}{\# sources} & IVW Stacked & IVW ISM & \multirow{2}{*}{IVW $f_{\rm ISM}$} & SIMSTACK Stacked & SIMSTACK ISM & SIMSTACK \\ 
		bin & log(M$_*/M_{\odot})_{\rm Med}$  & & & Flux (mJy) & Mass ($10^{10}M_{\odot}$) & & Flux (mJy) & Mass ($10^{10}M_{\odot}$) & $f_{\rm ISM}$\\
		\hline
         & 10.59 & 2.98 & 122 & 0.843 $\substack{ +0.108 \\ -0.113 }$ & 4.740 $\substack{ +0.108 \\ -0.113 }$ & 0.550 $\substack{ +0.080 \\ -0.084 }$ & 0.948 & 5.334 & 0.579 \\ 
        10.50 - 10.75 & 10.60 & 3.52 & 66 & 0.897 $\substack{ +0.151 \\ -0.156 }$ & 5.296 $\substack{ +0.151 \\ -0.156 }$ & 0.571 $\substack{ +0.110 \\ -0.114 }$ & 0.563 & 3.325 & 0.455 \\ 
         & 10.63 & 4.05 & 31 & 1.123 $\substack{ +0.224 \\ -0.206 }$ & 7.076 $\substack{ +0.224 \\ -0.206 }$ & 0.625 $\substack{ +0.147 \\ -0.135 }$ & 1.384 & 8.722 & 0.673 \\ 
         & 10.59 & 4.59 & 18 & 0.428 $\substack{ +0.316 \\ -0.285 }$ & 2.927 $\substack{ +0.316 \\ -0.285 }$ & 0.430 $\substack{ +0.347 \\ -0.313 }$ & 0.756 & 5.176 & 0.572 \\
        \hline
         & 10.86 & 0.28 & 783 & 0.321 $\substack{ +0.045 \\ -0.043 }$ & 0.425 $\substack{ +0.045 \\ -0.043 }$ & 0.056 $\substack{ +0.008 \\ -0.008 }$ & 0.302 & 0.400 & 0.053 \\ 
         & 10.85 & 0.82 & 2033 & 0.310 $\substack{ +0.028 \\ -0.029 }$ & 1.214 $\substack{ +0.028 \\ -0.029 }$ & 0.145 $\substack{ +0.013 \\ -0.014 }$ & 0.289 & 1.131 & 0.137 \\ 
         & 10.86 & 1.36 & 1653 & 0.436 $\substack{ +0.032 \\ -0.029 }$ & 2.142 $\substack{ +0.032 \\ -0.029 }$ & 0.228 $\substack{ +0.017 \\ -0.016 }$ & 0.370 & 1.816 & 0.200 \\ 
        10.75 - 11.00 & 10.87 & 1.90 & 512 & 0.888 $\substack{ +0.054 \\ -0.059 }$ & 4.663 $\substack{ +0.054 \\ -0.059 }$ & 0.388 $\substack{ +0.025 \\ -0.028 }$ & 0.876 & 4.601 & 0.384 \\ 
         & 10.86 & 2.44 & 97 & 1.490 $\substack{ +0.134 \\ -0.121 }$ & 8.093 $\substack{ +0.134 \\ -0.121 }$ & 0.530 $\substack{ +0.054 \\ -0.049 }$ & 1.630 & 8.850 & 0.553 \\ 
         & 10.88 & 2.98 & 39 & 2.276 $\substack{ +0.203 \\ -0.193 }$ & 12.800 $\substack{ +0.203 \\ -0.193 }$ & 0.630 $\substack{ +0.066 \\ -0.063 }$ & 2.425 & 13.640 & 0.645 \\ 
         & 10.83 & 3.52 & 23 & 0.872 $\substack{ +0.274 \\ -0.276 }$ & 5.146 $\substack{ +0.274 \\ -0.276 }$ & 0.431 $\substack{ +0.147 \\ -0.148 }$ & 0.841 & 4.965 & 0.422 \\ 
         & 10.82 & 4.05 & 10 & -0.095 $\substack{ +0.401 \\ -0.386 }$ & \textcolor{red}{7.589} & \textcolor{red}{0.532} & 0.155 & - & - \\ 
        \hline
         & 11.10 & 0.28 & 413 & 0.194 $\substack{ +0.063 \\ -0.060 }$ & 0.258 $\substack{ +0.063 \\ -0.060 }$ & 0.020 $\substack{ +0.007 \\ -0.006 }$ & 0.234 & 0.310 & 0.024 \\ 
         & 11.09 & 0.82 & 864 & 0.375 $\substack{ +0.039 \\ -0.043 }$ & 1.470 $\substack{ +0.039 \\ -0.043 }$ & 0.106 $\substack{ +0.011 \\ -0.012 }$ & 0.324 & 1.271 & 0.093 \\ 
        11.00 - 11.25 & 11.09 & 1.36 & 784 & 0.706 $\substack{ +0.045 \\ -0.042 }$ & 3.466 $\substack{ +0.045 \\ -0.042 }$ & 0.219 $\substack{ +0.014 \\ -0.013 }$ & 0.692 & 3.399 & 0.216 \\ 
         & 11.11 & 1.90 & 293 & 1.339 $\substack{ +0.080 \\ -0.067 }$ & 7.035 $\substack{ +0.080 \\ -0.067 }$ & 0.352 $\substack{ +0.022 \\ -0.019 }$ & 1.306 & 6.856 & 0.346 \\ 
         & 11.12 & 2.44 & 58 & 1.867 $\substack{ +0.168 \\ -0.154 }$ & 10.138 $\substack{ +0.168 \\ -0.154 }$ & 0.437 $\substack{ +0.043 \\ -0.039 }$ & 1.777 & 9.653 & 0.425 \\ 
         & 11.07 & 2.98 & 13 & 2.514 $\substack{ +0.318 \\ -0.331 }$ & 14.140 $\substack{ +0.318 \\ -0.331 }$ & 0.544 $\substack{ +0.078 \\ -0.081 }$ & 2.401 & 13.503 & 0.532 \\ 
        \hline
         & 11.33 & 0.28 & 145 & 0.019 $\substack{ +0.108 \\ -0.103 }$ & 0.025 $\substack{ +0.108 \\ -0.103 }$ & 0.001 $\substack{ +0.007 \\ -0.006 }$ & 0.199 & 0.264 & 0.012 \\ 
         & 11.32 & 0.82 & 204 & 0.501 $\substack{ +0.088 \\ -0.085 }$ & 1.966 $\substack{ +0.088 \\ -0.085 }$ & 0.085 $\substack{ +0.015 \\ -0.014 }$ & 0.361 & 1.414 & 0.063 \\ 
        11.25 - 11.50 & 11.32 & 1.36 & 217 & 1.242 $\substack{ +0.083 \\ -0.080 }$ & 6.095 $\substack{ +0.083 \\ -0.080 }$ & 0.226 $\substack{ +0.016 \\ -0.015 }$ & 1.316 & 6.459 & 0.237 \\ 
         & 11.33 & 1.90 & 87 & 1.594 $\substack{ +0.142 \\ -0.137 }$ & 8.373 $\substack{ +0.142 \\ -0.137 }$ & 0.281 $\substack{ +0.026 \\ -0.025 }$ & 1.279 & 6.716 & 0.239 \\ 
         & 11.32 & 2.44 & 26 & 3.157 $\substack{ +0.234 \\ -0.216 }$ & 17.143 $\substack{ +0.234 \\ -0.216 }$ & 0.449 $\substack{ +0.037 \\ -0.034 }$ & 3.657 & 19.863 & 0.486 \\ 
        \hline
         & 11.57 & 0.28 & 28 & -0.317 $\substack{ +0.233 \\ -0.232 }$ & \textcolor{red}{0.928} & \textcolor{red}{0.024} & -0.270 & - & - \\ 
        11.50 - 11.75 & 11.62 & 0.82 & 35 & 0.012 $\substack{ +0.194 \\ -0.203 }$ & 0.045 $\substack{ +0.194 \\ -0.203 }$ & 0.001 $\substack{ +0.018 \\ -0.019 }$ & -0.297 & -1.162 & -0.029 \\ 
         & 11.55 & 1.36 & 33 & 1.243 $\substack{ +0.217 \\ -0.220 }$ & 6.100 $\substack{ +0.217 \\ -0.220 }$ & 0.148 $\substack{ +0.026 \\ -0.026 }$ & 1.589 & 7.802 & 0.182 \\ 
         & 11.56 & 1.90 & 15 & 1.875 $\substack{ +0.321 \\ -0.293 }$ & 9.850 $\substack{ +0.321 \\ -0.293 }$ & 0.213 $\substack{ +0.037 \\ -0.034 }$ & 1.849 & 9.712 & 0.211 \\ 
        \hline
	\end{tabular}
\end{table}
\addtocounter{table}{-1}
\end{landscape}

\section{Distributions of central pixel values} \label{App:MChists}
\begin{figure*}
	\includegraphics[width=\textwidth]{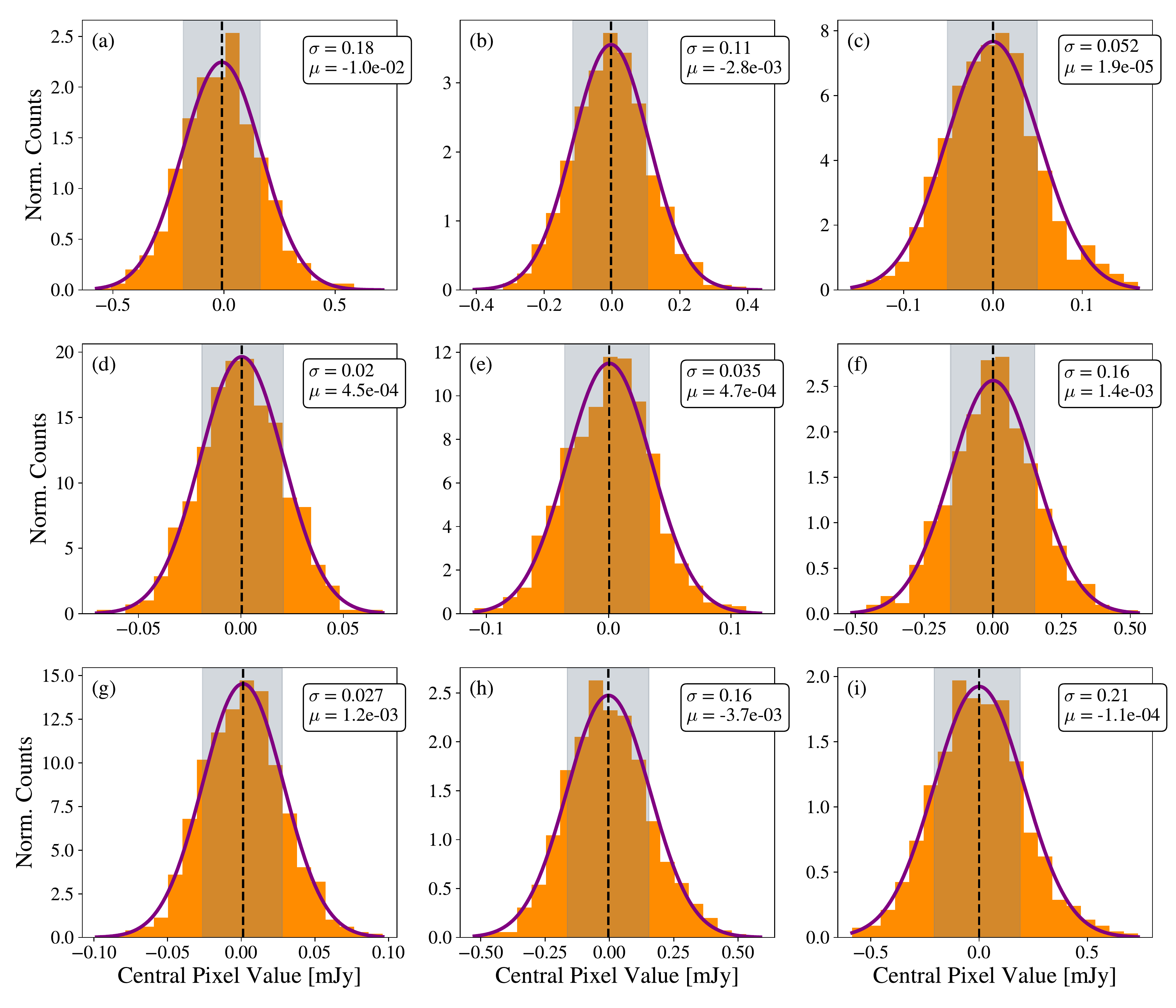}
    \caption{Normalised histograms of the central pixel values determined using 1000 random source catalogues and MC methodologies for a random selection of $(M_*-z)$ bins. The black dashed vertical line marks the mean of the distribution. The grey shaded region marks the extent of the 16th- and 84th-percentiles. The standard deviation, $\sigma$, and mean, $\mu$, are denoted in the upper right corner. The purple line is a Gaussian fit to the data, calculated using the stated mean and standard deviation. Details of the bins displayed in the figure can be seen in Table \ref{tab:MCsims_hists}; the letter in the upper left corner of each subplot is the cross-reference for this table. For further details on the displayed bins, see Table \ref{tab:bin_numbers}.}
    \label{fig:MCsims_hists}
\end{figure*}

Figure \ref{fig:MCsims_hists} shows illustrative examples of the central pixel values determined for a selection of bins, using the 1000 random source catalogues generated for the MC simulation on the location of sources within the S2COSMOS map. The central pixel values of the stacked stamps generated using the random sources broadly display a Gaussian distribution centred around zero. This is to be expected; the average of the S2COSMOS map is zero, and these sources are not expected to have any true 850$\mu$m flux associated with them. Therefore, on average, stacking on these random sources should return a null result. Since the central pixel values are approximately Gaussian distributed, for a given bin, the width of the distribution is an estimate of the error on our fluxes determined using our sample of {\sc magphys} galaxies. We make use of the 16th- and 84th- percentiles of these distributions of central pixel values as our flux errors.

\begin{table}
	\centering
	\caption{Details of the ($M_*-z$) bins displayed in Figure \ref{fig:MCsims_hists}. Additional information can be found in Table \ref{tab:bin_numbers}.}
	\label{tab:MCsims_hists}
	\begin{tabular}{ccc} 
		\hline
		Letter & $<z_{\rm bin}>$ & log$M_*$ bin \\
		\hline
		(a) & 4.59 & 10.25 - 10.50 \\
		(b) & 2.98 & 10.50 - 10.75 \\
		(c) & 2.98 & 10.00 - 10.25 \\
		(d) & 0.82 & 9.75 - 10.00 \\
		(e) & 1.90 & 10.00 - 10.25 \\
		(f) & 3.51 & 10.50 - 10.75 \\
		(g) & 1.36 & 10.00 - 10.25 \\
		(h) & 4.59 & 9.50 - 9.75 \\
		(i) & 0.82 & 11.50 - 11.75 \\
		\hline
	\end{tabular}
\end{table}

\bsp	
\label{lastpage}
\end{document}